\UseRawInputEncoding
\documentclass[fleqn,usenatbib]{mnras}
\usepackage{graphicx}
\usepackage{float}
\usepackage{wrapfig}
\usepackage{amsmath}
\usepackage{amssymb}

\usepackage{bm}
\usepackage[font=small,labelfont=bf]{caption}
\usepackage{citesort}
\usepackage{threeparttable}
\usepackage{comment}
\usepackage[dvipsnames]{xcolor}
\usepackage{pdflscape}
\usepackage{adjustbox}

\usepackage[normalem]{ulem}  

\newcommand{\SB}[1]{{\color{ForestGreen}SB: #1}}

\newcommand{\HI}{H\,{\sc i}}

\newcommand{\CI}{C\,{\sc i}}

\usepackage[textwidth=1in]{todonotes}
\usepackage{multicol, cuted}

\def\apj{Astroph. J.}
\def\apjl{Astroph. J. Lett.}
\def\apjs{Astroph. J. Suppl.}
\def\mnras{MNRAS}
\def\nat{Nature}
\def\aap{Astron. Astroph.}

\def\ssr{Space Sci. Rev.}
\def\araa{Ann. Rev. Astron. Astroph.}
\def\aj{Astron. J.}
\def\pasp{Pub. Astron. Soc. Pacific}

\title[HD at high redshift: cosmic-ray ionization rate in the diffuse ISM]{
HD molecules at high redshift: cosmic-ray ionization rate in the diffuse interstellar medium}

\author[D. N.~Kosenko et al.]{D. N.~Kosenko,$^{1}$\thanks{E-mail: kosenkodn@yandex.ru}
S. A.~Balashev,$^{1}$\thanks{E-mail: s.balashev@gmail.com} 
P.~Noterdaeme$^2$, J.-K.~Krogager$^2$, \newauthor 
R.~Srianand$^3$ and C.~Ledoux$^4$
\\
$^{1}$Ioffe Institute, 26 Politeknicheskaya st., St.\ Petersburg, 194021, Russia\\
$^{2}$Institut d'Astrophysique de Paris, UMR\,7095, CNRS-SU, 98bis boulevard Arago, 75014 Paris, France\\
$^{3}$Inter-University Centre for Astronomy and Astrophysics, Post Bag 4, Ganeshkhind, 411 007, Pune, India\\
$^{4}$European Southern Observatory, Alonso de C\'ordova 3107, Vitacura, Casilla 19001, Santiago, Chile \label{eso}
}

\date{Accepted XXX. Received YYY; in original form ZZZ}

\pubyear{2021}

\begin{document}

\label{firstpage}
\pagerange{\pageref{firstpage}--\pageref{lastpage}}
\maketitle

\begin{abstract}
We present a systematic study of deuterated molecular hydrogen (HD) 
at high redshift, detected in absorption in the spectra of quasars. 
We present four new 
identifications of HD lines associated with known $\rm H_2$-bearing
Damped Lyman-$\alpha$ systems.
In addition, we measure upper limits on the $\rm HD$ column density 
in twelve recently identified $\rm H_2$-bearing DLAs. 
%
We find that the new $\rm HD$ detections have similar 
$N({\rm HD})/N(\rm H_2)$ ratios
as previously found, 
further strengthening a marked difference with 
measurements 
through the Galaxy. This 
is likely due to 
differences in physical conditions and metallicity between 
the local and the high-redshift interstellar media.
Using the measured $N({\rm HD})/N({\rm H_2})$ ratios 
together with priors on the UV flux ($\chi$) and number densities ($n$), obtained from analysis of $\rm H_2$ and associated \CI\ lines, we are able to 
constrain the cosmic-ray ionization rate (CRIR, $\zeta$)
for the new $\rm HD$ detections and for eight known HD-bearing systems where priors on $n$ and $\chi$ are available. We find significant dispersion in
$\zeta$, from a few $\times 10^{-18}$\,s$^{-1}$ to a few
$\times 10^{-15}$\,s$^{-1}$. We also find that $\zeta$ strongly correlates with $\chi$ -- showing almost quadratic dependence, slightly correlates with $Z$, and does not correlate with $n$, which probably reflects a physical connection
between cosmic rays and star-forming regions.


\end{abstract}


\section{Introduction}

The formation and evolution of galaxies is intimately linked to their interstellar medium (ISM). Indeed, the ISM provides the fuel for star formation and in turn, the physical and chemical properties of the ISM are affected by stars (through UV radiation, cosmic rays, winds, enrichment by metals and dust, mechanical energy injection, etc.). The ISM presents several phases: 
the cold dense phases (cold neutral medium, CNM, itself including molecular phases) that may eventually collapse to form stars, the warmer, less dense phases (warm neutral and ionized medium, WNM and WIM, respectively), and the hot ionized medium (HIM) \citep{Field1969, McKee1977}. 
These phases are well studied in the local Universe via analysis of emission over a large range of wavelengths in the electromagnetic spectrum from X-ray \citep{Snowden1997} to radio \citep{Heiles2003a}, but their description is still limited at high redshift, due to flux dimming at cosmological distances and significantly coarser spatial resolution available for emission-line studies of most tracers of the ISM. 

This problem can be overcome by absorption-line spectroscopy.
Both the WNM and CNM at high redshift are detectable in the spectra of background quasars and $\gamma$-ray burst (GRB) afterglows as Damped Lyman-$\alpha$ systems (DLAs) -- absorption-line systems with the highest column densities of neutral hydrogen, \ion{H}{I} ($\log N({\rm \ion{H}{I}}) > 20.3$\footnote{Here and in what follows, $N$ is the column density in cm$^{-2}$.}) and a collection of associated metal lines \citep[for a review, see][]{Wolfe2005}.
Most DLAs actually represent WNM \citep{Srianand2005, Neeleman2015}, while CNM is much more rarely detected \citep[in a few percent of DLAs,
see e.g.,][]{Balashev2018}. 

One of the main tracers of CNM is molecular hydrogen (H$_2$), the most abundant molecule in the Universe.
Using UV absorption lines of H$_2$ in the Lyman and Werner bands, one can probe diffuse and translucent molecular clouds along the line of sight \citep{Ledoux2003, Noterdaeme2008, Noterdaeme2010, Balashev2017, Ranjan2018}.
If the H$_2$ column density is large enough, the less abundant isotopologue, HD, can also be detected \citep{Varshalovich2001}. 
%
To date, HD lines have been detected only in twelve intervening systems among $\sim40$ 
confirmed H$_2$-bearing DLAs at high redshift ($>0$) 
\citep{Noterdaeme2008, Balashev2010, Tumlinson2010, Ivanchik2010, Noterdaeme2010, Klimenko2015, Ivanchik2015, Klimenko2016, Noterdaeme2017,  Balashev2017, Rawlins2018, Kosenko2018}. This number remains limited 
since the detailed analysis of H$_2$ and HD lines can be done only in high-resolution quasar spectra, which require observations with 
the largest optical telescopes. Also, as mentioned before, the incident rate of the cold ISM in DLAs at high $z$ is quite low. Hence, blind searches for HD/H$_2$ are very inefficient \citep{Jorgenson2014}.
Notwithstanding, in recent years several efficient techniques were proposed to pre-select saturated $\rm H_2$ lines in DLAs where HD is then easier to detect \citep[][]{Balashev2014, Ledoux2015, Noterdaeme2018}.


Some of the high-redshift $N({\rm HD})/2N(\rm H_2)$ 
measurements lie close to the primordial isotopic (D/H)$_p$ ratio, triggering discussion on whether the molecular isotopic ratio could serve as a proxy for D/H, in particular at high column densities where the cloud is thought to be fully molecularized \citep[e.g.,][]{Ivanchik2010}. However, models
suggest that the HD/H$_2$ ratio varies significantly with depth into the clouds \citep{LePetit2002, Liszt2015, Balashev2020} since HD and H$_2$ have different main 
formation mechanisms: H$_2$ is forming mainly on the surface of dust grains, while HD is mostly formed via fast ion-molecular reactions.
At the same time, destruction of both HD and H$_2$ mainly occurs via photo-dissociation by UV photons.\footnote{Photo-dissociation is the main destruction process for molecules but there can be additional reactions such as destruction by cosmic rays or reversed reaction (Eq.~\ref{H2+D+}) that lead to a non-unity molecular fraction even in the fully self-shielded part of the clouds.}
This implies that the HD/H$_2$ ratio is sensitive to a combination of physical conditions, and that the HD/H$_2$ ratio can differ from the isotopic ratio even at high column densities in self-shielded regions \citep{Balashev2020}.
Moreover, under some conditions the D/HD transition may take place earlier than the H/H$_2$ transition \citep{Balashev2020}, which leads to HD/2H$_2$ $>$ D/H \citep{Tumlinson2010, Noterdaeme2017}, and therefore HD/H$_2$ may not be used as a lower limit for the isotopic ratio.

From the known HD-bearing systems, it was found that the relative HD/H$_2$ abundance tends to systematically be higher at high redshift than in the Galaxy \citep{Snow2008, Balashev2010, Tumlinson2010, Ivanchik2015}. This discrepancy cannot be solely explained by the progressive destruction of deuterium, since the astration of D through stellar evolution is expected to be small \citep{Dvorkin2016}. Therefore, the most probable explanation is to be sought in differences in physical conditions between the ISM of the Galaxy and that of distant galaxies. Indeed, models of ISM chemistry show that the HD/H$_2$ ratio is sensitive to the physical conditions in the ISM  -- UV flux, cosmic-ray ionization rate (CRIR), metallicity, number density, and cloud depth (\citealt{LePetit2002, Cirkovic2006, Liszt2015, Balashev2020}).
Among these parameters, 
the cosmic-ray ionization rate seems to play a major role, being extremely important for the ISM chemistry. 
Indeed, cosmic rays are 
an important source of heating and the main ionizing source and therefore drives almost all the chemistry in the ISM. In the case of $\rm HD$, cosmic rays promote the main channel of its formation as follows:
\begin{equation}
\label{eq:HD_equations}
{\rm H} \xrightarrow{\rm CR} {\rm H^+} \xrightarrow{ \rm D } {\rm D^+} \xrightarrow{\rm H_2} {\rm HD}
\end{equation}

Therefore, HD can, in principle, be used to constrain the CRIR \citep[e.g.,][]{Balashev2020}. Such independent constraint would be extremely valuable, given the still loose constraints on CRIR in both the local Universe \citep[see, e.g.,][]{Hartquist1978a, vanDishoeck1986, Federman1996, Indriolo2007, Neufeld2017, Gonzalez_Alfonso2013, Gonzalez_Alfonso2018, vanderTak2016} and at high redshift \citep{Muller2016, Shaw2016, Indriolo2018}. 
Additionally, a $\rm HD$-based method to constrain CRIR has an important advantage compared to other, widely-used methods based on oxygen-bearing molecules: the abundance of $\rm HD$ is found to increase relative to H$_2$ when the metallicity decreases \citep[mostly due to chemistry;][]{Liszt2015, Balashev2020}. 
Therefore, $\rm HD$ allows us to probe the CNM in lower metallicity environments.



Motivated by this emerging new possibility of using $\rm HD$ as a probe of CRIR and the lack of known HD detections to date,
we performed a systematic search for $\rm HD$ in recently-published and archival H$_2$-bearing DLAs at high redshift. 
We report four new HD detections. Additionally, we refit HD, $\rm H_2$, and \ion{C}{i} in a few systems to obtain confidence upper limits on HD olumn density and to get consistent constraints on its physical parameters. Finally, in one DLA we find $\rm H_2$ that has not been reported before (while we only obtain an upper limit on $\rm HD$). 
This paper is organized as follows: in Sect.~\ref{sec:data} we present the sample in which we searched for HD lines. Sect.~\ref{sec:analysis} describes the data analysis and details on individual systems. The measurements of $\rm HD/H_2$ abundances are summarized in Sect.~\ref{sec:results} and used in Sect.~\ref{sec:phys_cond} to constrain the physical conditions in the absorbing medium. In Sect.~\ref{sec:discussion}, we discuss some implications on the derived CRIR and limitations of the model. Lastly, in Sect.~\ref{sec:conclusion} we offer our concluding remarks. 

\section{Data}
\label{sec:data}
To search for HD absorption lines in high-$z$ DLAs, we used quasar spectra obtained at medium and high resolving power with X-shooter ($R\sim6000$; \citealt{Vernet2011}) and the Ultraviolet and Visual Echelle Spectrograph  (UVES, $R\sim50\,000$; \citealt{Dekker2000}) on the Very Large Telescope (VLT) as well as the High Resolution Echelle Spectrograph (HIRES, $R\sim50\,000$; \citealt{Voht1994}) on the Keck telescope.

Most of the data come from X-shooter and include the spectra of quasars with recently reported 
high-$z$ H$_2$-bearing DLAs from \citet{Noterdaeme2018, Ranjan2018, Balashev2019, Ranjan2020}.
The detailed description of the observations and data reduction is presented in the above-mentioned papers. Typically, these quasars were observed with 1-4 exposures, each about one hour long.

The UVES data include the system at $z=3.09$ towards J\,1311+2225, recently reported by \citet{Noterdaeme2018}, where \ion{C}{I} together with H$_2$ and CO molecules were detected, the well-known three ESDLA systems at $z=2.402$ towards HE0027$-$1836 \citep{Noterdaeme2007} (for which further data was obtained by \citet{Rahmani2013}, leading to an improved quality spectrum), at $z=3.85$ towards J\,0816+1446 \citep{Guimaraes2012}, at $z=2.48$ towards J\,2140$-$0321 \cite{Noterdaeme2010}, and the DLA system towards J\,2340-0053 where HD was reported independently and almost simultaneously by \citet{Kosenko2018} and \citet{Rawlins2018}. For this latter system, 
we refitted HD together with $\rm H_2$ and \ion{C}{I} lines to obtain self-consistent priors on physical parameters that are used to derive the CRIR. 
For these systems we used the spectra from original publications or the SQUAD UVES database \citep{Murphy2019}, or from KODIAQ DR2 database \citep{OMeara2017}. 

We also looked at all other 
known H$_2$-bearing DLAs at high $z$ to search for HD absorption lines that were not detected or considered in the original studies. 
\citet{Kosenko2018} reported a new H$_2$-bearing system at $z=2.067$ towards Q\,0812+3208. Unfortunately, only the weakest HD transition (L0-0 band) is covered 
by the HIRES spectrum (see details by \citealt[][]{Balashev2010}), so that we were only able to place an upper limit on $N$(HD). 


A summary of the H$_2$-bearing DLAs analysed in this paper is provided in Table~\ref{tab:qso}. In Table~\ref{tab:known_qso}, we provide information on previously known high-$z$ HD/H$_2$-bearing systems known to date, that were used later to derive physical conditions. 
\begin{center}
\begin{table*}
\centering
\caption{H$_2$-bearing DLA systems searched for HD.}
\label{tab:qso}
\begin{tabular}{lccccccc}
 \hline
 Quasar &$z_{\rm em}$ & $z_{\rm abs}$ & $\log N(\rm HI)$ & [X/H]$^a$ & X & $\log N(\rm H_2)$ &  References$^b$\\
 \hline
 \multicolumn{8}{c}{VLT/X-shooter data:} \\
 J\,0136+0440 & 2.78 & 2.779 &  $20.73\pm0.01$ & $-0.58\pm0.03$ & S & $18.65^{+0.06}_{-0.07}$ & 1 \\
 J\,0858+1749 & 2.65 & 2.625 &  $20.40\pm0.01$ & $-0.63\pm0.02$ & S &  $19.72^{+0.01}_{-0.02}$ & 1\\
 J\,0906+0548 & 2.79 & 2.567 &  $20.13\pm0.01$ & $-0.18^{+0.05}_{-0.08}$ & S &  $18.88\pm0.02$  & 1\\
 J\,0917+0154 & 2.18 & 2.107 &  $20.75\pm0.04$ & $0.17\pm0.07$ & Zn &  $20.11\pm0.06$ & 2, 3\\
 J\,0946+1216 & 2.66 & 2.607 &  $21.15\pm0.02$ & $-0.48\pm0.01$ & S & $19.97^{+0.01}_{-0.02}$ & 1\\
 J\,1143+1420 & 2.58 & 2.323 & $21.64\pm0.06$ & $-0.80\pm0.06$ & Zn
& $18.3\pm0.1$ & 4 \\
J\,1146+0743 & 3.03 & 2.840 &  $21.54\pm0.01$ & $-0.57\pm0.02$ & Zn & $18.82^{+0.03}_{-0.02}$ & 1\\
 J\,1236+0010 & 3.02 & 3.033 &  $20.78\pm0.01$ & $-0.58^{+0.04}_{-0.03}$ & S & $19.76\pm0.01$ & 1\\
 J\,1513+0352 & 2.68 & 2.46 &  $21.83\pm0.01$ & $-0.84\pm 0.23$ & Zn & $21.31\pm0.01$ & 5\\
 J\,2232+1242 & 2.30 & 2.230 & $21.75\pm0.03$ & $-1.48\pm0.05$ & Zn & $18.56\pm0.02$ & 4 \\
 J\,2347$-$0051 & 2.62 & 2.588 &  $20.47\pm0.01$ & $-0.60^{+0.06}_{-0.09}$ & S & $19.44\pm0.01$ & 1 \\
 
 \hline 
 \multicolumn{8}{c}{High-resolution (Keck/HIRES and VLT/UVES) data:} \\
 HE0027$-$1836 & 2.56 & 2.402 & $21.75\pm0.10$ & $-1.63\pm0.10$ & Zn & $17.43\pm0.02$ & 4, 6 \\
 J\,0812+3208  & 2.70 & 2.067 &  $21.50\pm0.20$ & $-1.83\pm0.20$ & Si & $19.28\pm0.01^{c}$ & 7, 8 \\
 J\,0816+1446  & 3.85 & 3.287 &  $22.00\pm0.10$ & $-1.10\pm0.10$ & Zn & $18.48\pm0.02^{c}$ & 9 \\
 J\,1311+2225 & 3.14 & 3.093 & $20.62\pm0.10$ & $-0.34^{+0.13}_{-0.14}$$^c$ & Zn & $19.69\pm0.01^{c}$ & 2 \\
 J\,2140$-$0321 & 2.48 & 2.339 & $22.41\pm0.03$ & $-1.52\pm0.08$ & Zn & $20.13\pm0.07$ & 4, 10 \\
 \hline
\end{tabular}
\begin{tablenotes}
    \item $(a)$ Metallicity with respect to solar \citep{Asplund2009}: $[{\rm X}/{\rm H}] = \log({\rm X/H}) - \log({\rm X/H})_{\odot}$.
	\item $(b)$ References: 
	(1) \citet{Balashev2019}, 
	(2) \citet{Noterdaeme2018},
	(3) \citet{Zou2018}, 
	(4) \citet{Ranjan2020}, 
	(5) \citet{Ranjan2018}, 
	(6) \citet{Noterdaeme2007},
	(7) \citet{Kosenko2018},
	(8) \citet{Jorgenson2010},
	(9) \citet{Guimaraes2012},
	(10) \citet{Noterdaeme2015}.
	\item $(c)$ This work.
\end{tablenotes}
\end{table*}
\end{center}

\begin{center}
\begin{table*}
\centering
\caption{Known HD-bearing DLA systems.}
\label{tab:known_qso}
\begin{tabular}{lcccccccc}
 \hline
 Quasar &$z_{\rm em}$ & $z_{\rm abs}$ & $\log N(\rm HI)$ & [X/H]$^a$ & X & $\log N(\rm H_2)$ & $N({\rm HD})$ &  References$^b$\\
 \hline 
  J\,0000+0048 & 3.03 & 2.5255 & $20.8\pm 0.1$ & $0.46\pm0.45$ & Zn & $20.43\pm0.02$ & $16.64^{+0.16}_{-0.18}$ & 1 \\
  B\,0120$-$28 & 0.434 & 0.18562 & $20.50\pm0.10$ & $-1.19^{+0.15}_{-0.21}$ & S & $20.00\pm0.10$ & $14.82\pm0.15$ & 2  \\
  Q\,0528$-$2505 & 2.77 & 2.81112 & $21.35\pm0.10$ & $-0.68\pm0.02$ & Zn & $17.85\pm0.02$ & $13.33\pm0.02$ & 3, 4 \\
  J\,0643$-$5041 & 3.09 & 2.658601 & $21.03\pm0.08$ & $-0.91\pm0.09$ & Zn & $18.54\pm0.01$ & $13.65\pm0.07$ & 5  \\
 J\,0812+3208 & 2.7 & 2.626443 & $21.35\pm0.10$ & $-0.81\pm0.10$ & Zn & $19.93\pm0.04$ & $15.71\pm0.07$ & 6, 7 \\
              &     & 2.626276 &                & $-0.81\pm0.10$ & Zn & $18.82\pm0.37$ & $12.98\pm0.22$ & 6, 7 \\
 J\,0843+0221 & 2.92 & 2.786 & $21.82\pm0.11$ & $-1.52^{+0.08}_{-0.10}$ & Zn & $21.21\pm0.02$ & $17.35^{+0.15}_{-0.34}$ & 8 \\
 J\,1232+0815 & 2.57 & 2.3377 & $20.90^{+0.08}_{-0.10}$ & $-1.32\pm0.12$ & S & $19.57^{+0.10}_{-0.13}$ & $15.53^{+0.17}_{-0.12}$ & 9, 10 \\   
 J\,1237+0647 & 2.78 & 2.68959 & $20.00\pm0.15$ & $0.34\pm0.12$ & Zn & $19.20\pm0.13$ & $14.48\pm0.05$ & 11   \\
 J\,1331+170 & 2.08 & 1.77637 & $21.18\pm0.04$ & $-1.22\pm0.10$ & Zn & $19.43\pm0.10$ & $14.83\pm0.15$ & 6, 12 \\
             &  & 1.77670 & & $-1.22\pm0.10$ & Zn & $19.39\pm0.11$ & $14.61\pm0.20$ & 6, 12 \\
 J\,1439+1117 & 2.58 & 2.41837 & $20.10\pm0.10$ & $0.16\pm0.11$ & Zn & $19.38\pm0.10$ & $14.87\pm0.03$ & 13, 14 \\ 
 J\,2100$-$0641 & 3.14 & 3.09149 & $21.05\pm0.15$ & $-0.73\pm0.15$ & Si & $18.76\pm0.04$ & $13.83\pm0.06$ & 15, 16 \\
 J\,2123$-$0050 & 2.261 & 2.0593 & $19.18\pm0.15$ & $-0.19\pm0.10$ & S & $17.94\pm0.01$ & $13.87\pm0.06$ & 17 \\
 J\,2340$-$0053 & 2.083 & 2.05 &
 $20.35\pm 0.05$ & $-0.52\pm 0.06$
 & S & 18.62$^{+0.02}_{-0.01}$ $^{c}$ & $14.11\pm0.06$ $^{c}$ & 18  \\
 \hline
 \end{tabular}
\begin{tablenotes}
    \item $(a)$ Metallicity  with respect to solar \citep{Asplund2009}: $[{\rm X}/{\rm H}] = \log({\rm X/H}) - \log({\rm X/H})_{\odot}$.
	 \item $(b)$ References: (1) \citet{Noterdaeme2017}, (2) \citet{Oliveira2014}, (3) \citet{Klimenko2015}, (4) \citet{Balashev2020b}, (5) \citet{Albornoz2014}, 
	 (6)  \citet{Balashev2010}, (7) \citet{Jorgenson2009}, (8) \citet{Balashev2017}, (9) \citet{Ivanchik2010}, (10) \citet{Balashev2011}, (11) \citet{Noterdaeme2010}, (12) \citet{Carswell2011}, (13) \citet{Srianand2008}, (14) \citet{Noterdaeme2008}, (15) \citet{Ivanchik2015}, (16) \citet{Jorgenson2010}, (17) \citet{Klimenko2016}, (18) \citet{Rawlins2018}.
	\item $(c)$ This work.
\end{tablenotes}

\end{table*}
\end{center}

\section{Analysis}
\label{sec:analysis}

We analyzed the absorption lines using multi-component Voigt profile\footnote{The Voigt profile is a convolution of Lorenzian and Gaussian functions, arising from natural broadening and thermal/turbulent motions of the gas, respectively.} fitting.
The unabsorbed continuum was typically constructed by-eye using spline interpolation constrained by the regions free from any evident absorption lines \citep[see e.g.][]{Balashev2019}. 
The lines were fitted simultaneously and the spectral pixels that were used to constrain the model were selected by eye to avoid blends (mainly with Ly-$\alpha$ forest lines). The best value and interval estimates on the fitting parameters (Doppler parameter, $b$, column density, $N$ and redshift, $z$) were obtained with a Bayesian approach, using standard $\chi^2$ likelihood to compare the data and the model. To sample the posterior distribution function of the parameters we used Monte Carlo Markov Chain (MCMC) \citep[see e.g.][]{Balashev2017} with affine-invariant sampling \citep{Goodman2010}. By default the priors on most parameters were assumed to be flat (for $b$, $\log N$ and $z$). 
However,
for most X-shooter spectra, the resolution is not high enough to accurately resolve the velocity structure and some HD lines can be in the saturated regime. In these cases, we found that column densities and Doppler parameters can be highly degenerated, resulting in uncertain constraints. Therefore, we used priors on the number of components, their redshifts and Doppler parameters from the analysis of H$_2$ or \ion{C}{I} absorption lines  \citep[see e.g.][]{Balashev2019}. This is adequate, since H$_2$ is usually constrained by a large number of lines ($\sim50-100$) and \ion{C}{I} is fitted in the region out of Ly$\alpha$ forest. 
We used mostly components where the column density of H$_2$ exceeds $\log N (\rm H_2) \gtrsim 18$, since for lower H$_2$ columns, the expected HD column densities will be much lower than what the data can constrain, 
i.e. even upper limits will be uninformative.

Moreover, we found that in X-Shooter spectra, the continuum placement for some HD lines is non-trivial. 
We estimated the resulting uncertainty independently using the following procedure. We performed a large number ($\sim500$) of realizations, where we randomly shifted the continuum level for each line. The values of the shifts were drawn from a normal distribution with dispersion corresponding to the mean uncertainty of spectral pixels at the positions of absorption lines. For each realization, we also randomly drew an HD Doppler parameter using constraints obtained from H$_2$. The redshift uncertainty from H$_2$ (or \ion{C}{I}) in most cases is quite low and has only marginal effect on the results. 
We then fitted each realization $i$ with fixed $b$ and $z$ 
and obtained the best fit column density $N^i$(HD). We obtained the final HD column density measurement from the distribution of $N^i$(HD). 
We found that the uncertainties on HD column densities increase in most cases by a factor of $\sim 2$ compared to MCMC fit with fixed continuum,  meaning that the continuum placement uncertainty contributes significantly to the total $N($HD) uncertainty budget at medium resolution. 


We summarize the results of fitting HD lines in Table~\ref{tab:fit_results} 
and provide specific comments on each system as follows:
\subsection{VLT/X-shooter data:}

\subsubsection{J\,0136$+$0440}

We only tentatively detected HD absorption lines at the expected positions based on the redshift of the main H$_2$ component ($z=2.779430$) with column density $\log N(\rm H_2)=18.64^{+0.06}_{-0.08}$ and Doppler parameter $b=7.7^{+2.4}_{-1.9}$\,km\,s$^{-1}$. Therefore, fixing $z$ and using priors on Doppler parameter from H$_2$ analysis, we placed only an upper limit to the HD column density in this component, $\log N({\rm HD})<14.5$. The fits to the unblended HD absorption lines are shown in Fig.~\ref{fig:J0136}. Here and in the following figures we show only those HD absorption lines that are not totally blended with other absorption lines (from Ly$\alpha$ forest and/or H$_2$ and metal lines from corresponding DLA).

\subsubsection{J\,0858$+$1749}

We detected HD absorption lines at the position of H$_2$ component ($z=2.62524$) that has $\log N(\rm H_2)=19.72^{+0.01}_{-0.02}$ and 
$b=7.9^{+0.4}_{-0.4}$\,km\,s$^{-1}$. To fit HD lines we fixed $z$ and used $b$ as a prior from H$_2$ analysis. Using the HD\,L8-0R(0) line and red wings of HD\,L4-0R(0), HD\,L7-0R(0), HD\,L11-0R(0) and HD\,L12-0R(0) absorption lines (see Fig.~\ref{fig:J0858}), we constrained  $\log N({\rm HD}) = 14.87^{+0.06}_{-0.09}$. 

\subsubsection{J\,0906$+$0548}

We only tentatively detected HD absorption lines at the position of the main H$_2$ component ($z=2.56918$) that has 
$\log N(\rm H_2)=18.87\pm0.02$ and 
$b=6.8^{+0.1}_{-0.1}$\,km\,s$^{-1}$. Although we did find HD lines at the expected positions, all of them are partially or fully blended with other absorption lines (see Fig.~\ref{fig:J0906}). Therefore, using $z$ and priors on $b$ obtained from H$_2$ analysis, we were only able to place an upper limit to the HD column density in this component to be $\log N({\rm HD})<14.7$. 

\subsubsection{J\,0917$+$0154}

This system was selected by \citet{Ledoux2015} in their search for cold gas at high redshift through \ion{C}{I} lines. The detection and analysis of H$_2$ was presented by \citet{Noterdaeme2018} (they reported total column density $N(\rm H_2) =20.11\pm0.06$) and the metal lines were studied by \citet{Zou2018}.
Unfortunately, due to low resolution and relatively high velocity extent of H$_2$ lines, almost all HD lines are blended, including usually avaliable L3-0R0, L4-0R0 and W0-0R0 lines. The only not blended line L0-0R0 has a very low oscillator strength and therefore we were able to put only very conservative upper limit on HD column density using priors on the redshifts and Doppler parameters for three components fit obtained from the refitting jointly \CI\ and H$_2$ absorption lines. The fit to \ion{C}{I} and HD lines are shown in 
Fig.~\ref{fig:J0917} and H$_2$ lines profiles are presented in Fig.~\ref{fig:J0917_H2}. The detailed fit result is given in Table~\ref{tab:J0917}.


\subsubsection{J\,0946$+$1216}

The detection of HD at 
the position of the main H$_2$ component ($z=2.60642$,  $\log N(\rm H_2)=19.96^{+0.01}_{-0.02}$,  $b=9.8^{+0.8}_{-0.3}$\,km\,s$^{-1}$) for this system is also tentative. Unfortunately, the spectrum is very noisy and significantly contaminated by highly saturated H$_2$ lines and intervening Ly$\alpha$ forest absorption. 
Therefore we fixed $z$ and used Doppler parameter from H$_2$ analysis as a prior. Hence we were only able to obtain relatively loose constraint on the HD column density in this component to be $\log N({\rm HD})<15.2$, see 
Fig.~\ref{fig:J0946}.

\subsubsection{J\,1143$+$1420}

This extremely saturated DLA at $z = 2.3228054$ was previously analysed by \citet{Ranjan2020} and H$_2$ column density was found to be $\log N({\rm H_2}) = 18.3\pm0.1$. We looked for HD lines associated with H$_2$, and we were able to place an upper limit on HD column density. We used fixed $z$ and priors on Doppler parameter from H$_2$ analysis, and got $N({\rm HD}) < 15$.
The fit to HD lines is shown in Fig.~\ref{fig:J1143}.
\subsubsection{J\,1146$+$0743}

We do not detected HD absorption lines at the position of both H$_2$ components ($z=2.84163$ and 2.83946 with $N(\rm H_2) = 18.76\pm0.01$ and $17.94^{+0.11}_{-0.13}$ respectively). Therefore we constrained $\log N({\rm HD})<14.4$ and $\log N({\rm HD})<14.5$ for the red and blue components, respectively, using a combination of HD\,L3-0R(0), HD\,L8-0R(0), HD\,W0-0R(0), HD\,W1-0R(0), HD\,L11-0R(0) and HD\,L12-0R(0) lines and priors on $b$ and fixed $z$ from H$_2$ analysis.  The spectrum at the expected positions of HD absorption lines is shown in Fig.~\ref{fig:J1146}. 

\subsubsection{J\,1236$+$0010}

We do not detect HD absorption lines at the position of H$_2$ component of DLA ($z=3.03292$, $\log N(\rm H_2)=19.76\pm0.01$, 
$b=2.3^{+0.2}_{-0.2}$\,km\,s$^{-1}$). To fit HD lines we fixed $z$ and used Doppler parameter of  H$_2$ as a prior. Using HD\,L0-0R(0), HD\,L3-0R(0), HD\,L4-0R(0), HD\,L5-0R(0), HD\,W0-0R(0), HD\,L11-0R(0) and HD\,L14-0R(0) lines (see Fig.~\ref{fig:J1236}) we put quite loose constraint on HD column density to be 
$\log N(\rm HD) \lesssim 16.1$
since lines are found to be in the intermediate 
regime. 

\subsubsection{J\,1513$+$0352}

The extremely saturated DLA at $z = 2.463598$ towards J\,1513$+$0352   was found in SDSS database by \citet{Noterdaeme2014}. Detailed analysis of system by \citet{Ranjan2018} using X-shooter spectrum revealed a very high H$_2$ column density: $\log N({\rm H_2}) = 21.31\pm0.01$  (actually the highest value reported to the date at high-$z$). 
We detected HD L0-0R0, HD L5-0R0 and HD L7-0R0 absorption lines in this system. However, because of the H$_2$ lines were damped, they did not 
constrain the 
Doppler parameters. We therefore used instead the value obtained from associated \ion{C}{I} as a prior for HD and obtained 
$\log N({\rm HD}) = 17.42^{+0.64}_{-1.09}$. 
This makes it the DLA with one of the highest HD column density as well. However, since the absorption lines are in the saturated regime and resolution is moderate, the uncertainty on $N(\rm HD)$ remains quite large.  
The fit to the HD absorption lines is shown in Fig.~\ref{fig:J1513}.

\subsubsection{J\,2232$+$1242}

We do not detect HD absorption lines in  H$_2$-bearing DLA ($z = 2.2279378$, $N({\rm H_2}) = 18.56\pm0.02$, \citealt{Ranjan2020}) towards J\,2232$+$1242. Using redshift and prior on Doppler parameter from H$_2$ fit we obtain the upper limit on HD column density to be $\log N(\rm HD) < 13.8$ (see Fig.~\ref{fig:J2232}).

\subsubsection{J\,2347$+$0051}

We detect HD absorption lines at the position of H$_2$ ($z=2.58797$, $\log N(\rm H_2)=19.44\pm0.01$, $b=6.2^{+0.2}_{-0.2}$\,km\,s$^{-1}$). Using HD\,L3-0R(0), HD\,L5-0R(0), HD\,L7-0R(0), HD\,L13-0R(0) and HD\,L15-0R(0) lines (see Fig.~\ref{fig:J2347}), we measured HD column density to be $\log N({\rm HD})=14.33^{+0.18}_{-0.16}$ (to fit HD lines we fixed $z$ and used prior on $b$ from H$_2$ analysis).  

\subsection{KECK/HIRES and VLT/UVES data:}
\label{sect:high}
\subsubsection{HE\,0027$-$1836}

The extremely saturated DLA system at $z = 2.4018258$ have been studied by \citet{Noterdaeme2007, Rahmani2013}. H$_2$ was identified in this DLA with column density  $\log N({\rm H_2}) = 17.43$. Searching for HD absorption lines at the redshift of H$_2$ absorption lines, we obtained an upper limit on the HD column density $\log N(\rm HD) < 13.6$ (we fixed $z$ and used Doppler parameter as a prior from H$_2$ analysis) due to inconsistency of L2-0R0 and L3-0R0 lines with the fit in the spectrum (see Fig.~\ref{fig:HE0027_HD}). 

\subsubsection{J\,0812$+$3208}

The spectrum towards J\,0812$+$3208 features two DLAs at $z=2.626491$ and $z=2.06779$ \citep{Prochaska2003}. \citet{Jorgenson2010} detected absorption lines from \ion{C}{I} fine-structure levels in both of them. Associated HD/H$_2$ absorption lines at $z=2.626491$ were studied in details by several authors \citep{Jorgenson2009, Balashev2010, Tumlinson2010}, however, no significant attention have been paid to the system at $z=2.06678$. Knowing that \ion{C}{i} is an excellent tracer of H$_2$ in ISM \citep{Noterdaeme2018}, we searched for H$_2$ and HD molecules in this system as well. 
We used the Keck/HIRES spectrum whose 
reduction is detailed in \cite{Balashev2010}. We detected H$_2$ absorption lines from $J\le4$ rotational levels, which we fitted 
using a one component model, with tied redshifts and Doppler parameter rotational levels for all levels. Indeed, H$_2$ lines are 
located at the blue end of the spectrum, covering only one-two unblended H$_2$ lines from each rotational level. The fit results is given in Table~\ref{table:J0812} and line profiles are shown in Fig.~\ref{fig:J0812}. Using relative population of $J=1$ and $J=0$ levels, we found the excitation temperature to be $T_{01}=67^{+4}_{-3}$\,K. 

Unfortunately, only two HD lines (L0-0R0 and L1-0R0) were covered in this spectra and only the weakest HD L0-0R0 line from this system was unblended (see Fig.~\ref{fig:J0812}). Thus, we estimated only an upper limit to the HD column density, fixing the redshift and Doppler parameter from H$_2$, and obtained $\log N(\rm HD)<14.4$.

\subsubsection{J\,0816$+$1446}
The multicomponent $\rm H_2$-bearing DLA system towards J\,0816$+$1446 was identified by \citet{Guimaraes2012}. This system have quite large redshift and hence is significantly blended with Ly$\alpha$ forest lines. \citet{Guimaraes2012} reported H$_2$ in two components, with one at $z=3.28742$ indicates a significantly high H$_2$ column density, $\log N(\rm H_2) = 18.66\pm0.27$ to be searched for $\rm HD$. We refit H$_2$ absorption lines at $z=3.28742$ with three subcomponents, since it provides a better fit, and measured the total $\log N(\rm H_2) = 18.51\pm0.04$ in agreement with \citet[][]{Guimaraes2012}. Unfortunately, all $\rm HD$ lines are blended and therefore using fixed z and Doppler parameter from H$_2$ analysis we were able to obtain an upper limits on the HD column densities $\log N (\rm HD) \lesssim 15$ from the L4-0 R(0) line (fit results are presented in Table~\ref{tab:J0816} and Fig.~\ref{fig:J0816_HD}).

\subsubsection{J\,1311$+$2225}

This multicomponent $\rm H_2$-bearing DLA system was selected through \ion{C}{i} 
by \citet{Ledoux2015}. \citet{Noterdaeme2018} reported $\log N({\rm H_2}) = 19.69\pm 0.01$ in this system, using single component model, 
but they noted that four components for H$_2$ lines can be distinguished.  
We refitted $\rm H_2$ and \ion{C}{I} lines in this system using four-component model. First we fit \ion{C}{I} absorption lines from three fine-structure levels, where we tied Doppler parameters for each component. Then we performed a four-component fit to the H$_2$ lines, where the selection of initial guess of components was based on \ion{C}{I} result. 
For $\rm H_2$, we tied Doppler parameters only between $J=0$ and $J=1$ levels, while Doppler parameters for other rotational levels were allowed to vary independently. However, since the components are significantly blended among themselves and the data is quite noisy, we added two penalty functions to the likelihood. The first one is set to artificially suppress situations where the Doppler parameter of the some $J$ level would be lower than that of the $J-1$ level. This is well motivated physically and observationally, since the increase of the Doppler parameters for the higher $\rm H_2$ rotational levels has been established in many $\rm H_2$ absorption systems \citep[see e.g.][]{Lacour2005, Noterdaeme2007, Balashev2009}. The other penalty is to keep a reasonable excitation diagram of $\rm H_2$: we penalized models with $T_{J-1,J} > T_{J,J+1}$\footnote{where $T_{J,J+1}$ is
the excitation temperature between 
$J$ and $J+1$ levels}. This is also reasonably motivated by both observations and modelling \cite[see e.g.][]{Klimenko2020}.
Therefore we get total H$_2$ column density to be $N({\rm H_2}) = 19.59\pm 0.01$, which is a bit lower than the value $19.69\pm 0.01$ reported previously \citep{Noterdaeme2018}.
The fitting results are shown in Table~\ref{tab:J1311_results} and \ion{C}{I} and H$_2$ profiles in Figs.~\ref{fig:J1311_CI}, \ref{fig:J1311_H2_low}, \ref{fig:J1311_H2_j2}, \ref{fig:J1311_H2_j3}, \ref{fig:J1311_H2_j45}.

We also estimated metallicity in this system. Unfortunately, very few metal lines, that are usually used to obtain metallicity, were covered in this spectrum, and almost all covered lines are blended. Therefore to obtain metallicity we used \ion{Zn}{II}\,2062 line. We fitted this line, assuming 4 components in the positions of \ion{C}{I} components, and obtained \ion{Zn}{II} total column density to be $12.84^{+0.09}_{-0.11}$, therefore the metallicity is $-0.34^{+0.13}_{-0.14}$ relative to solar. The fit to \ion{Zn}{II} absorption line is shown in Fig.~\ref{fig:J1311_ZnII}.

We again used a four-component model to analyse HD, associated with \ion{C}{I} components. We found that component 3 for HD is shifted in comparison with \ion{C}{I} lines. However, the component 3 in \CI\ have quite large Doppler parameter, that indicates that there is velocity structure within this component, which meanwhile we can not resolve due to low quality of the spectrum and mutual blending from other components. 
So for HD we did not use the H$_2$ and \ion{C}{I} priors on redshifts (except weak component 1, where only upper limit on HD column density could be placed) and Doppler parameters.
After MCMC procedure we found HD to be detected in the component 2, 3 and 4, and the redshifts of the components are well agree within uncertainties (see Table~\ref{tab:J1311_results}). Component 1 is too weak, so we could only place an upper limit on $N({\rm HD})$ there. 
The fit to the HD lines is shown in Fig.~\ref{fig:J1311_HD} and HD column densities reported in Table~\ref{tab:J1311_results}.

\subsubsection{J\,2140$-$0321}

H$_2$ absorption lines were previously found and analysed by \citet{Noterdaeme2015, Ranjan2020} at $z = 2.339$ and H$_2$ column density was found to be quite large $\log N(\rm H_2) = 20.13$.  To fit HD absorption lines we used together the spectra, obtained by X-shooter and UVES. However, since the UVES spectrum is very noisy, and X-shooter is low-resolution hence it is not appropriate for HD analysis.  Therefore we were able only to place upper limit on HD column density to be $\log N(\rm HD) < 14.6$ using the priors on the Doppler parameters and the redshifts obtained from H$_2$ analysis \citep{Noterdaeme2015}, see Fig.~\ref{fig:J2140_HD}.

\subsubsection{J\,2340$-$0053}

\ion{C}{I} and H$_2$ absorption lines in the DLA at $z\approx2.055$ towards J\,2340$-$0053 were first reported by \citet{Jorgenson2010}. These authors found \ion{C}{i} in nine components, while they fitted H$_2$ using a six components model. This spectrum was recently reanalysed by \citet{Rawlins2018} with a seven components for both \ion{C}{I} and H$_2$ and found their redshifts to be consistent with each other. HD absorption lines, associated with H$_2$ were later independently detected by \citet{Kosenko2018} and \citet{Rawlins2018}. In this paper, we present detailed reanalysis of HD, H$_2$ and \ion{C}{I} absorption lines. 

Using the reduced 1D-spectrum of J\,2340$-$0053 from the KODIAQ database \citep{OMeara2017}, we refitted \ion{C}{I}, H$_2$ and HD absorption lines with seven component model using the same methodology as in the previous section for J\,1311$+$2225. We fit \ion{C}{I} lines first, taking into account the partial coverage of the background emission line region by \ion{C}{I} line at $\sim$1560\AA\ reported by \citet{Bergeron2017}. We fit the covering factors as an independent parameter following the methodology from \citet{Balashev2011}. We found that a fit with three independent covering factors for each of the three main components provides a better fit, than using a single covering factor for all components. 
We then used the \ion{C}{I} fit as first guess to the redshifts of the $\rm H_2$ lines. Unfortunately, three central components are significantly blended with each other in almost all $\rm H_2$ absorption lines from $J=0$, 1, 2 and 3 rotational levels. Therefore we used redshifts determined during \ion{C}{I} fit as priors, and as for J1311$+$2225, we used penalty functions during $\rm H_2$ analysis to reproduce physically reasonable constraints. 
We obtained the total H$_2$ column density to be $\log N({\rm H_2}) = 18.57\pm 0.02$, which is higher than reported by \citep[][$\log N({\rm H_2}) = 17.99\pm0.05$]{Rawlins2018}. 
The difference is partly due to the fact that 
\citet{Rawlins2018} tied all H$_2$ Doppler parameters for $J > 0$ to H$_2$ $J = 0$, while we tied only H$_2$ $J = 1$ and allowed increasing $b$-values for other levels. 
The fitting results are shown in Table~\ref{tab:J2340_results} and \ion{C}{I} and H$_2$ profiles in Figs~\ref{fig:J2340_CI}, \ref{fig:J2340_H2_J01}, \ref{fig:J2340_H2_J23}, \ref{fig:J2340_H2_J45}.
 
We fit HD absorption lines at the positions of these components using the priors on the Doppler parameters from the fit of $J=0$ and $J=1$ rotational levels of $\rm H_2$. However, the exact $b$-values affect little the results since the $\rm HD$ absorption lines are optically thin. 
The obtained total HD column density is $\log N({\rm HD}) = 14.11\pm 0.06$, which is is a bit lower than found by \citet{Rawlins2018} ($\log N({\rm HD}) = 14.28\pm0.08$).
The fit to the HD lines is reported in Table~\ref{tab:J2340_results} and shown in Fig.~\ref{fig:J2340_HD}.


\begin{table*}
\centering
\caption{Results from the analysis of HD lines.}
\label{tab:fit_results}
\begin{tabular}{lccccccl}
 \hline
 Quasar &$z$ & $b$ (km\,s$^{-1}$) & $\log N(\rm HD)^a$& $\log N(\rm H_2)$ & $N({\rm HD})/2N({\rm H}_2)$\\
 \hline
 \multicolumn{6}{c}{X-shooter data:}\\
 J\,0136$+$0440 & 2.779430 & $7.7^{+2.4}_{-1.9}$ & $< 14.5$ & $18.64^{+0.06}_{-0.08}$ & $< 3.6 \times 10^{-5}$ \\
               
 J\,0858$+$1749 & 2.625241 & $7.9^{+0.4}_{-0.4}$ & $14.87^{+0.06}_{-0.09}$ & $19.72^{+0.01}_{-0.02}$ & $\left(7.1^{+1.1}_{-1.4}\right)\times 10^{-6}$  \\

 J\,0906$+$0548 & 2.569180 & $6.8^{+0.1}_{-0.1}$ & $< 14.7$ & $18.87^{+0.02}_{-0.02}$ & $< 3.4\times 10^{-5}$ \\
 
 J\,0917+0154$(b)$ & $2.10586$ & $5.2^{+1.1}_{-1.8}$ & $<12$  & $17.96^{+0.82}_{-0.16}$  & $<5.5\times10^{-7}$ \\
             & $2.10624$ & $6.4^{+1.5}_{-2.4}$ & $<15.9$ & $18.4^{+1.0}_{-0.3}$ & $<1.6\times10^{-3}$ \\
             & $2.106812$ & $4.7^{+1.1}_{-1.3}$ & $<18.1$ & $20.09^{+0.07}_{-0.08}$ & $<5.1\times10^{-3}$ \\
    
 J\,0946$+$1216 & 2.606406 & $9.8^{+0.8}_{-0.3}$ & $<15.2$ & $19.96^{+0.01}_{-0.02}$ & $< 9.0\times 10^{-6}$  \\
 
J\,1143+1420 & 2.3228054 & $2.2^{+2.0}_{-0.6}$ & $< 15$ & $18.3^{+0.1}_{-0.1}$ & $< 2.5\times10^{-4}$ \\
 			   
 J\,1146$+$0743 & 2.839459 & $7.6^{+0.1}_{-0.4}$ & $< 14.5$ & $17.94^{+0.11}_{-0.13}$ & $< 1.8\times 10^{-4}$ \\
 &  2.841629 & $11.4^{+0.5}_{-0.7}$ & $< 14.4$ & $18.76^{+0.01}_{-0.01}$ & $< 2.2\times 10^{-5}$ \\
 				
 J\,1236$+$0010 & 3.03292 & $2.3^{+0.2}_{-0.2}$ & $< 16.1$
 & $19.76^{+0.01}_{-0.01}$ & $< 1.1\times 10^{-4}$
 \\
  J\,1513$+$0352 & 2.463598 & $3.9^{+0.3}_{-0.3}$ & $17.42^{+0.64}_{-1.09}$ & $21.31^{+0.01}_{-0.01}$ & $\left(6.4^{+2.1}_{-5.9}\right)\times 10^{-5}$ \\
 
 J\,2232+1242 & 2.2279378 & $8.1^{+1.1}_{-1.2}$ & $<13.8$ & $18.56^{+0.02}_{-0.02}$ & $<8.7\times10^{-4}$ \\
 
 J\,2347$+$0051 & 2.587971 & $6.2^{+0.2}_{-0.2}$ & $14.33^{+0.18}_{-0.16}$ & $19.44^{+0.01}_{-0.01}$ & $\left(3.9^{+2.0}_{-1.2}\right)\times 10^{-6}$ \\

\hline
\multicolumn{6}{c}{High-resolution data:}\\

HE\,0027$-$1836 & 2.4018258 & $1.2^{+0.1}_{-0.2}$ & $< 13.6$ & $17.43^{+0.02}_{-0.02}$ & $<7.4\times10^{-5}$  \\ 
J\,0812$+$3208 & $2.066780(^{+1}_{-1})$ & $4.4^{+0.1}_{-0.1}$ & $<14.4$ & $19.26^{+0.02}_{-0.01}$ & $< 7.4\times 10^{-6}$ \\
J\,0816$+$1446$(b)$ & $3.287252(^{+3}_{-2})$ & $0.6^{+0.1}_{-0.1}$ & $<14.9$ & $16.97^{+0.09}_{-0.10}$ & $<4.3\times10^{-3}$ \\
&  $3.287399(^{+2}_{-3})$ & $1.5^{+0.1}_{-0.1}$ & $<14$ & $18.43^{+0.04}_{-0.03}$ & $<1.9\times10^{-5}$ \\
&  $3.287515(^{+2}_{-3})$ & $1.1^{+0.1}_{-0.1}$ & $<14.2$ & $17.60^{+0.10}_{-0.10}$ & $<2.0\times10^{-4}$ \\
J\,1311+2225 & $3.091410(^{+21}_{-14})$ & $8.0^{+4.6}_{-5.4}$ & $<12.8$ & $17.87^{+0.37}_{-0.33}$ & $<4.4\times 10^{-6}$ \\
& $3.0915397(^{+66}_{-77})$ & $5.4^{+0.8}_{-0.8}$ & $14.82^{+0.08}_{-0.08}$ & $19.52^{+0.02}_{-0.02}$ & $\left(1.0^{+0.3}_{-0.2}\right)\times 10^{-5}$ \\
& $3.091714(^{+28}_{-48})$ & $\lesssim 2.8$ & $14.30^{+0.37}_{-0.31}$ & $18.25^{+0.22}_{-0.39}$ & $
\left(5.6^{+13.7}_{-3.2}\right)\times 10^{-5}$ \\
& $3.091871(^{+11}_{-26})$ & $4.0^{+1.6}_{-1.2}$ & $14.27^{+0.10}_{-0.13}$ & $18.57^{+0.05}_{-0.09}$ & $\left( 2.5^{+0.9}_{-0.7}\right)\times 10^{-6}$ \\
& Total: &  & $15.02^{+0.11}_{-0.07}$ & $19.59^{+0.01}_{-0.01}$ & $\left(1.3^{+0.4}_{-0.2}\right)\times 10^{-5}$ \\
 
J\,2140$-$0321 & $2.33996(^{+3}_{-3})$ & $4.5^{+0.9}_{-0.7}$ & $<14.6$ & $20.13^{+0.07}_{-0.07}$ & $<1.5\times 10^{-6}$ \\
 
J\,2340$-$0053$^{b}$ & $2.0541703(^{+6}_{-4})$ & $2.5^{+0.1}_{-0.1}$ & $<13.5$  & $15.99^{+0.04}_{-0.04}$ & $<1.7\times 10^{-3}$ \\
& $2.0542913(^{+4}_{-9})$ & $1.7^{+0.1}_{-0.2}$ & $<12.7$ & $15.24^{+0.04}_{-0.03}$ & $<1.4\times 10^{-3}$\\
& $2.054528(^{+3}_{-3})$ & $3.0^{+0.1}_{-0.2}$ & $<13.8$ & $17.11^{+0.12}_{-0.14}$ & $<2.2\times 10^{-4}$\\
& $2.054610(^{+1}_{-1})$ & $1.0^{+0.1}_{-0.3}$ & $13.60^{+0.15}_{-0.14}$ & $18.27^{+0.06}_{-0.06}$ & $\left(1.1^{+0.5}_{-0.3}\right)\times 10^{-5}$\\
& $2.054723(^{+3}_{-3})$ & $3.1^{+0.1}_{-0.1}$ & $13.84^{+0.05}_{-0.05}$ & $18.14^{+0.04}_{-0.04}$ & $\left(2.5^{+0.4}_{-0.3}\right)\times 10^{-5}$\\
& $2.0549952(^{+5}_{-4})$ & $3.8^{+0.1}_{-0.1}$ & $<12.6$ & $16.43^{+0.03}_{-0.03}$ & $<7.1\times 10^{-5}$\\
& $2.0551398(^{+6}_{-4})$ & $1.8^{+0.1}_{-0.1}$ & $13.29^{+0.15}_{-0.21}$ & $17.43^{+0.04}_{-0.05}$ & $ \left(3.6^{+1.6}_{-0.3}\right)\times 10^{-5}$ \\
 & Total: &  & $14.11^{+0.06}_{-0.06}$ & $18.57^{+0.02}_{-0.02}$ & $\left(1.7^{+0.3}_{-0.2}\right)\times 10^{-5}$ \\
\hline
\end{tabular}
\begin{tablenotes}
\item $(a)$ The point and interval estimates were obtained from a 1D marginalized posterior distribution function, and correspond to its maximum and 0.683 (1$\sigma$) confidence interval, respectively. In case of tentative detection, the upper limits are constrained from the 1$\sigma$ one-sided confidence interval.
\item $(b)$ These system were re-fitted to get consistent results for HD, H$_2$, and \ion{C}{i} (see text).
\end{tablenotes}
\end{table*}

\section{Results}
\label{sec:results}

We summarize our new measurements of HD (and H$_2$) column densities and relative abundance of $N({\rm HD})/2N({\rm H}_2)$ in Table~\ref{tab:fit_results}. In total, 
we report four new detections of HD molecules in high-redshift DLAs (sometimes in several components) and place upper-limits for another twelve. 

\begin{figure*}
    \centering
    \includegraphics[width=\textwidth]{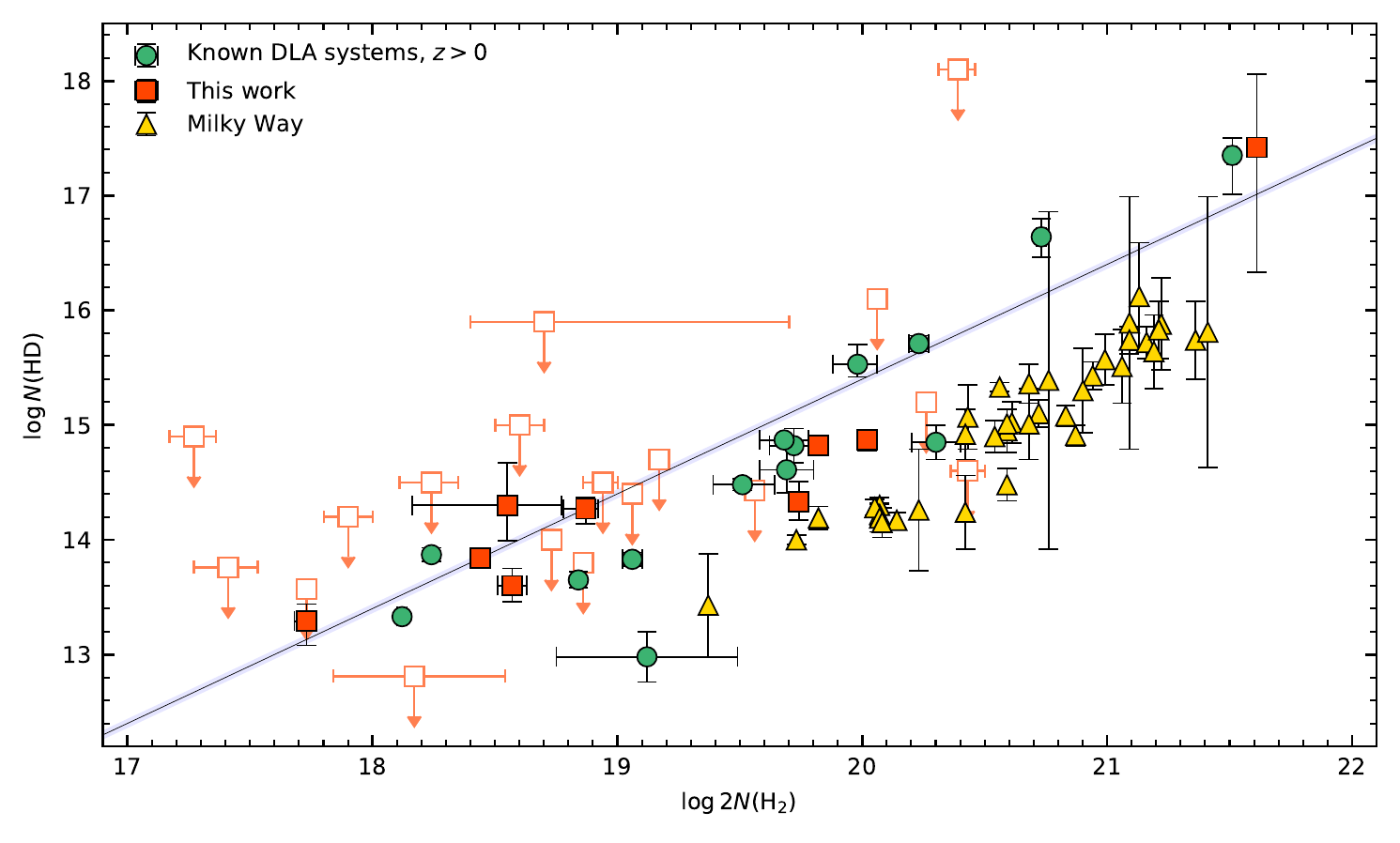}
    \caption{Relative abundance of HD and H$_2$ molecules. Green circles, red squares, and yellow triangles correspond to known HD-bearing systems at high redshift (for references, see text), new HD detections at high redshift and upper limits on HD column densities (filled and empty squares; this work), and measurements in the Galaxy \citep{Snow2008}, respectively. The solid blue line shows the D/H isotopic ratio estimated using standard BBN calculations and $\Omega_{\rm b}h^2$ from the \citet{Planck2018}.
    }
    \label{fig:HD_H2}
\end{figure*}

Fig.~\ref{fig:HD_H2} compares the HD and H$_2$ column densities in the Galaxy (\citealt{Snow2008}) and at high redshift (new measurements and values from Table~\ref{tab:known_qso}). 
We also compare the data to the primordial D/H isotopic ratio derived from updated Big Bang Nucleosynthesis (BBN) calculations \citep{Pitrou2018} and $\Omega_{\rm b} h^2$ from \citep{Planck2018}. 
One can see that the molecular ratios are well below the primordial isotopic ratio in the Galaxy, 
while distant measurements do not show such a tendency and instead indicate a systematically higher $\rm HD/H_2$ relative abundance than locally at specific $\log N(\rm H_2)$, and closer to the BBN value. 

Several processes can in principle affect the $\rm HD/H_2$ relative abundance, such as fractionation and astration of deuterium. The chemical fractionation of D (the process through which D efficiently replaces hydrogen in complex molecules, such as D$_2$, HDO, D$_2$O, NH$_2$D, NHD$_2$, ND$_3$, H$_2$D$^+$, DCO$^+$, etc.) should play a minor role, since complex molecules mainly reside in the cold dense medium with $n_{\rm H} \gtrsim 10^{5}$\,cm$^{-3}$ and $T\lesssim 25$\,K 
\citep[see e.g.][and references therein]{Kim2020}, which is not the case here as we probe more diffuse clouds. 
Measurements at high $z$ tend to probe lower metallicities than locally 
and hence represent gas that has been less processed in stars. Such gas should therefore be less affected by astration of deuterium 
than measurements in the Galaxy (i.e. $\sim$10 Gyr later). However, \citet{Dvorkin2016} showed that D/H is never reduced to less than 1/3 of its primordial ratio, i.e. astration cannot explain the observed discrepancy. 
On the other hand, the low metallicity affects the $\rm HD$ abundance from the chemical pathway \citep{Liszt2015,Balashev2020}. Indeed, as the metallicity decreases, both the dust abundance and electron fraction comes from carbon (here we consider diffuse in the diffuse ISM) decrease. 
This results in a drop of the radiative and $\rm H^+$ grain recombination rates, and hence the ionization fraction of H and D increases (see set of reactions leading to HD, \ref{eq:HD_equations}). This then results in an increase of the $\rm HD$ formation rate through the reaction:
\begin{equation}
\label{H2+D+}
    {\rm H_2} + \rm{D^{+}} \longrightarrow \rm{HD} + \rm{H^{+}}
\end{equation}

The enhanced $\rm HD$ formation rate consequently increases the $\rm HD$ abundance relative to $\rm H_2$. Interestingly, in certain physical conditions, this may lead to a D/HD transition occurring earlier (lower penetration depth) in ISM clouds than the H/H$_2$ transition \citep{Balashev2020}. 
The evident observational consequence of this is that $ N({\rm HD})/2 N({\rm H_2}) > {\rm D/H}$, while the opposite case was generally assumed \citep[e.g.][]{LePetit2002}, since naively $\rm HD$ is always significantly less self-shielded in the medium than $\rm H_2$. 
In conclusion, the typically lower metallicities at high $z$ can in principle explain the systematic difference in relative $\rm HD/H_2$ abundance between high-$z$ and Milky-Way measurements \citep[see also][]{Liszt2015}. 


\section{Physical conditions}
\label{sec:phys_cond}

The relative $\rm HD/H_2$ abundance depends not only on the metallicity, but also on the physical conditions in the medium -- number density, UV flux, and cosmic-ray ionization rate \citep{LePetit2002, Cirkovic2006, Liszt2015}. To describe this dependence, we used recently published simple semi-analytic description of the dependence of the HD/H$_2$ ratio on these parameters \citep{Balashev2020}.
This method includes solving the HD balance equation between formation and destruction processes in a plane-parallel, steady-state cloud and permits the determination of how $N({\rm HD})$ -- as a function of $N({\rm H_2})$ -- depends on the physical properties in the cloud, namely cosmic-ray ionization rate per hydrogen atom (CRIR, $\zeta$), UV field intensity (relative to Draine field, \citealt{Draine1978}, $\chi$), number density ($n = n_{\rm \ion{H}{I}} + 2n_{\rm H_2}$), and metallicity ($Z$). We assumed that the D/H isotopic ratio is $2.5\times10^{-5}$ for all systems, i.e., we neglected a possible astration of D, which is typically much smaller \citep{Dvorkin2016} than the uncertainties of this method (see below).

\begin{figure}
    \centering
    \includegraphics[width=0.45\textwidth]{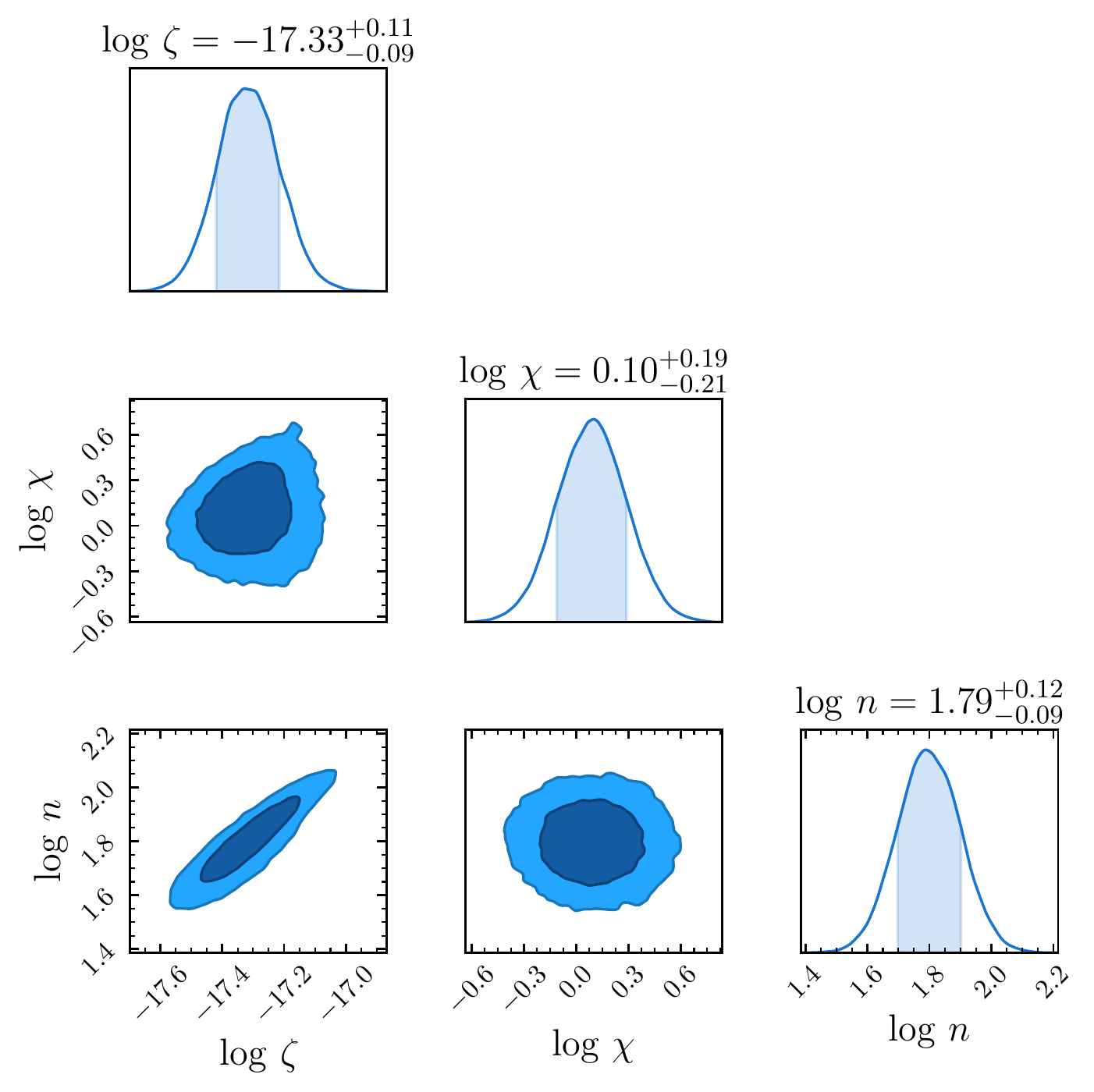}
    \caption{Posterior probability functions for CRIR ($\zeta$), UV field intensity ($\chi$), and number density ($n$) obtained from HD/H$_2$ fitting for the system at $z = 2.625241$ towards J\,0858$+$1749. The diagonal panels show 1D marginalized posterior function, non-diagonal show 2D posterior functions, where the dark- and light-blue regions correspond to 1$\sigma$ and 2$\sigma$ confidence levels, respectively. 
    \label{fig:J0858_MCMC}
    }
\end{figure}

To constrain the distributions of the physical parameters from the measured $N(\rm HD)$ and $N(\rm H_2)$, we followed a Bayesian approach using affine-invariant Markov Chain Monte Carlo (MCMC) sampling \citep{Foreman2013}. Because we don't have access to the total hydrogen $(N(\text{\HI})+N({\rm H_2}))$ in individual $\rm HD$-bearing components, the metallicity for each cloud 
was set to the overall metallicity in the corresponding DLA, as provided in Tables~\ref{tab:qso} and \ref{tab:known_qso}. 
For the intensity of the UV field as well as the number density, we used priors that have been estimated from the analysis of the relative population of H$_2$ rotational and \ion{C}{I} fine-structure levels \citep{Balashev2019, Klimenko2020}. 
This allows us to significantly reduce the constrained probability distribution function of CRIR, for which  we used a flat prior on $\log\zeta$. 
An example of the constrained 1D and 2D-posterior probability distribution functions of the parameters for the system towards J\,0858$+$1749 is given in Fig.~\ref{fig:J0858_MCMC}. 
The derived physical conditions for the sample are summarised in Table~\ref{tab:phys_params}, 
and plots of the marginalized posterior distribution functions for each component are shown in Figures~\ref{fig: MCMC_results1} -- \ref{fig: MCMC_results5}. 
We do not report results for J\,1513$+$0352 nor J\,1311$+$2225 (component 3), towards which we obtained a very loose constraint on $\zeta$, due to large uncertainties on the HD column densities. 
Note that the constraints on the number density, $n$, and UV flux, $\chi$, typically match the priors used. 


\begin{table}
 \caption{Constraints on physical conditions.}
 \label{tab:phys_params}
    \centering
    \begin{tabular}{lcccc}
    \hline 
      Quasar   & $\log \zeta$ & $\log\chi$ & $\log n$ & Ref.$^{\dagger}$  \\
      \hline 
        J\,0000$+$0048 & $\gtrsim -16.3$ & $0.0^{+0.3}_{-0.3}$ & $1.2^{+0.5}_{-0.4}$ & (2) \\
        Q\,0528$-$2505 & $-14.9^{+0.2}_{-0.1}$ & $1.1^{+0.1}_{-0.1}$ & $2.4^{+0.1}_{-0.1}$ & (3) \\
         J\,0812$+$3208, c1 & $-16.6^{+1.4}_{-0.5}$ & $-0.1^{+0.2}_{-0.1}$ & $2.4^{+0.2}_{-0.2}$ & (2) \\
        J\,0812$+$3208, c2 & $\lesssim -19.2$ & $-0.8^{+0.2}_{-0.2}$ & $0.8^{+0.3}_{-0.3}$ & (2) \\
        J\,0843$+$0221 & $-16.5^{+0.9}_{-1.1}$ & $2.0^{+0.1}_{-0.1}$ & $1.9^{+0.1}_{-0.1}$ & (2) \\
        J\,0858$+$1749 & $-17.3^{+0.1}_{-0.1}$ & $0.1^{+0.2}_{-0.2}$ & $1.8^{+0.1}_{-0.1}$ & (1) \\
        J\,1232$+$0815 & $-18.3^{+0.3}_{-0.3}$ & $-0.4^{+0.2}_{-0.2}$ & $1.6^{+0.1}_{-0.1}$ & (2) \\
        J\,1237$+$0647 & $-14.8^{+0.2}_{-0.2}$ & $1.1^{+0.1}_{-0.1}$ & $1.3^{+0.1}_{-0.1}$ & (2) \\
        J\,1311$+$2225, c2 & $-16.2^{+0.1}_{-0.1}$ & $1.1^{+0.1}_{-0.1}$ & $1.7^{+0.2}_{-0.2}$ & (4) \\
        J\,1311$+$2225, c3 & $-$ & $1.0^{+0.1}_{-0.1}$ & $1.9^{+0.1}_{-0.1}$ & (4) \\
        J\,1311$+$2225, c4 & $-15.1^{+0.2}_{-0.3}$ & $0.6^{+0.2}_{-0.2}$ & $2.1^{+0.3}_{-0.2}$ & (4) \\
        J\,1439$+$1118 & $-15.4^{+0.3}_{-0.2}$ & $0.8^{+0.2}_{-0.2}$ & $0.9^{+0.2}_{-0.2}$ & (2) \\
        J\,1513$+$0352 & $-$ & $0.6^{+0.3}_{-0.2}$ & $1.9^{+0.1}_{-0.2}$ & (2) \\
        
        J\,2100$-$0641 & $-17.2^{+0.3}_{-0.2}$ & $-0.3^{+0.3}_{-0.3}$ & $1.4^{+0.3}_{-0.3}$ & (2) \\
        
        J\,2340$-$0053, c4 & $-16.4^{+0.7}_{-0.7}$ & $-0.1^{+0.2}_{-0.3}$ & $0.6^{+0.3}_{-0.4}$ & (4) \\
        J\,2340$-$0053, c5 & $-14.8^{+0.2}_{-0.2}$ & $0.6^{+0.1}_{-0.1}$ & $1.2^{+0.1}_{-0.1}$ & (4) \\ 
        J\,2340$-$0053, c7 & $-15.4^{+0.8}_{-1.0}$ & $-0.2^{+0.2}_{-0.2}$ & $0.8^{+0.4}_{-0.4}$ & (4) \\
        J\,2347$+$0051 & $-17.6^{+0.6}_{-0.5}$ & $-0.4^{+0.4}_{-0.4}$ & $2.8^{+0.1}_{-0.1}$ & (1) \\
        \hline
    \end{tabular}
    \begin{tablenotes}
    \item $\dagger$  References used to obtain priors on $\chi$ and $n$: (1) \citet{Balashev2019}, (2) \citet{Klimenko2020}, (3) \citet{Balashev2020b}, (4) this work.
    \end{tablenotes}
\end{table}

\section{Discussion}
\label{sec:discussion}


We find the CRIRs to vary significantly from $\zeta \sim 10^{-18}$ 
to $10^{-15}$ s$^{-1}$, possibly reflecting  
a wide range of environments being probed by our sample. 
Indeed, DLA systems are selected owing to their absorption cross-section and likely probe 
the overall 
galaxy population, with a high fraction of low-mass galaxies at high redshift \citep[e.g.,][]{Cen2012}, in which the 
star-formation and cosmic-ray ionization rates are expected to vary significantly. 
Even though the $\rm HD/H_2$ absorption systems in our sample do not necessarily probe the immediate environments of star formation, as we will show below, the measured high CRIR values correlate with the relatively high UV fluxes that 
reach up to 10 times the Draine field. 

We find that the range of CRIR estimates are in line with other recent 
measurements 
both at high redshift \citep{Indriolo2018, Muller2016, Shaw2016} and in nearby galaxies \citep{vanderTak2016, Gonzalez_Alfonso2013, Gonzalez_Alfonso2018}, which also show quite large dispersion. This dispersion can be partly due to the use of various methods, or connected to a real physical dispersion of the CRIR. Indeed, the measurements in the Galaxy \citep[for a review see][and references therein]{Padovani2020} and in the lensed system at $z\sim 0.89$ towards PKS\,1830$-$211 \citep{Muller2016} show that this parameter can vary significantly between different sightlines even inside a given galaxy, mostly depending on the proximity to the CR accelerator. \citet{LePetit2016} also present 
evidence of CRIR enhancement 
in the center of the Galaxy relative to the disk. Finally, we think that the comparison of previous data with our measurement is likely not straightforward since different methods have been used, which probe various environments. Indeed, the aforementioned and most recent constraints on CRIR in local and high-$z$ galaxies have been based on oxygen-bearing species ($\rm OH^+$ and $\rm H_2O^+$). Since these have been analysed in quite luminous starburst galaxies with roughly solar metallicity and high star formation rates, they may sample rather high CRIR values compared to the overall galaxy population.

\begin{figure*}
    \centering
    \includegraphics[width=\textwidth]{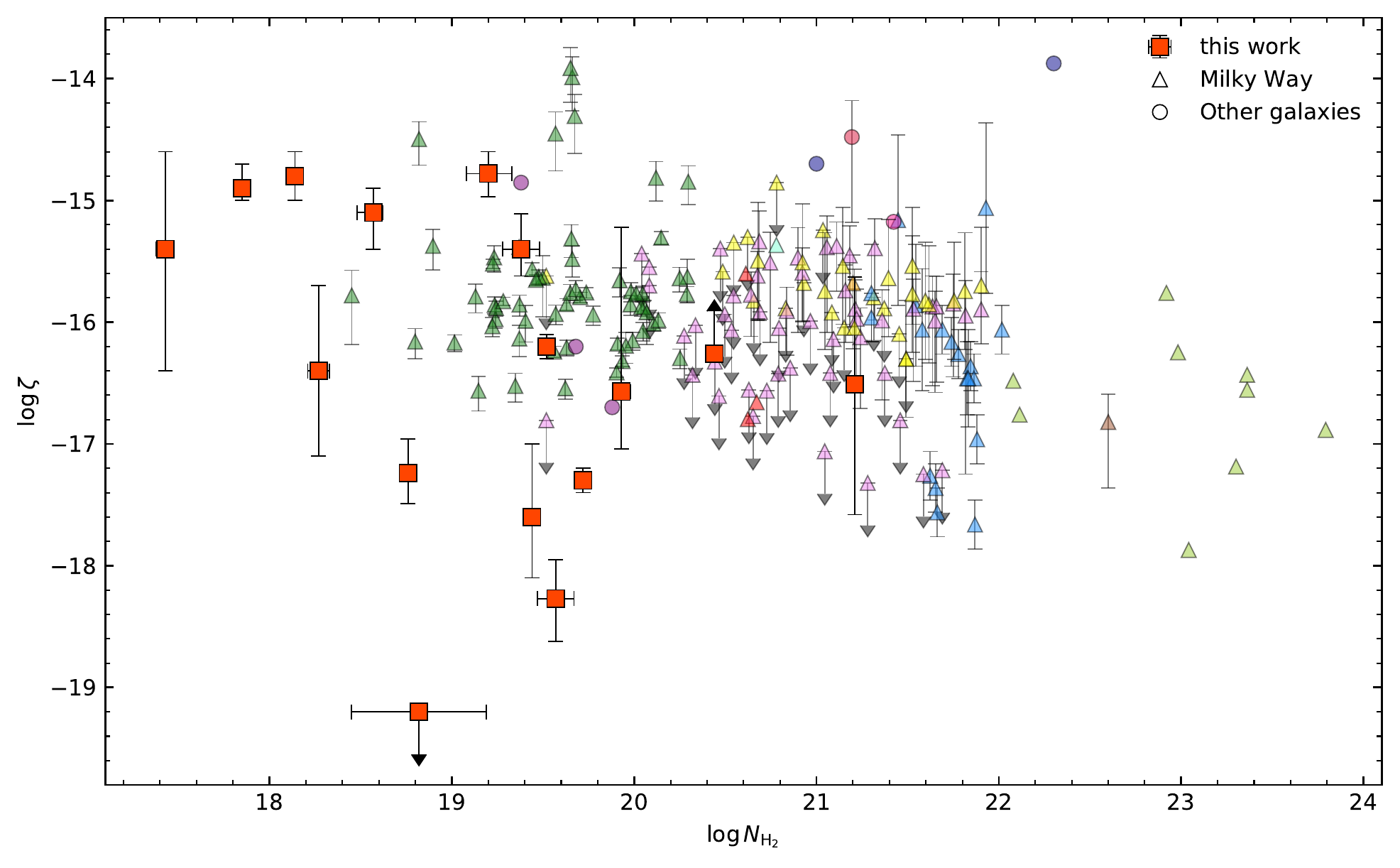}
    \caption{Estimated CRIR as a function of H$_2$ column density. Here, red squares are values obtained in this work, smaller squares are values obtained for other galaxies (blue - \citet{Muller2016}; violet - \citet{Shaw2016}; pink - \citet{Indriolo2018}), triangles are values measured in the Galaxy (yellow - \citet{Indriolo2007}; light green are protostellar envelopes \citep[for references, see table 6 from][]{Padovani2009}; blue - \citet{Caselli1998}; cyan - \citet{Shaw2008}; brown - \citet{Maret2007}; violet - \citet{Indriolo2012b}; dark green - \citet{Indriolo2015}).
    \label{fig:cr_H2_data}
    }
\end{figure*}

Figure~\ref{fig:cr_H2_data} 
shows and compares our measurements with literature ones in the [$\zeta$, $\log N(\rm H_2)$] plane.
An attenuation of the cosmic-ray ionization rate with increasing column density is theoretically expected \citep{Padovani2009}. 
However, we do not see strong evidence for a correlation between $\zeta$ and N(H$_2$) in our sample, probably because of the large dispersion (unweighted Pearson test gives correlation coefficient $r = -0.49$, with p-value $ 0.08$). 
In addition, we probe mostly diffuse clouds with low cloud depths (except Q\,$0843+0221$ which will be discussed later), which may be insufficient 
to attenuate the cosmic-ray flux. Additionally, the observed clouds should have quite large ($N(\rm H\,I)\gtrsim 10^{20}\,\rm cm^{-2}$) column densities of associated \HI, which is hard to constrain observationally, but which is also able to attenuate the CR flux, and therefore may provide an additional uncertainty in our calculations.  

Previous measurements at high redshift and in the Galaxy show that in case of a denser medium (e.g., dense cores, blue triangles, \citet{Caselli1998} and protostellar envelopes, light green triangles, for references see \citet{Padovani2009}) the cosmic-ray ionization rates tend to be slightly lower. 

\begin{figure*}
   \begin{minipage}{0.33\textwidth}
        \center{\includegraphics[width=1\linewidth]{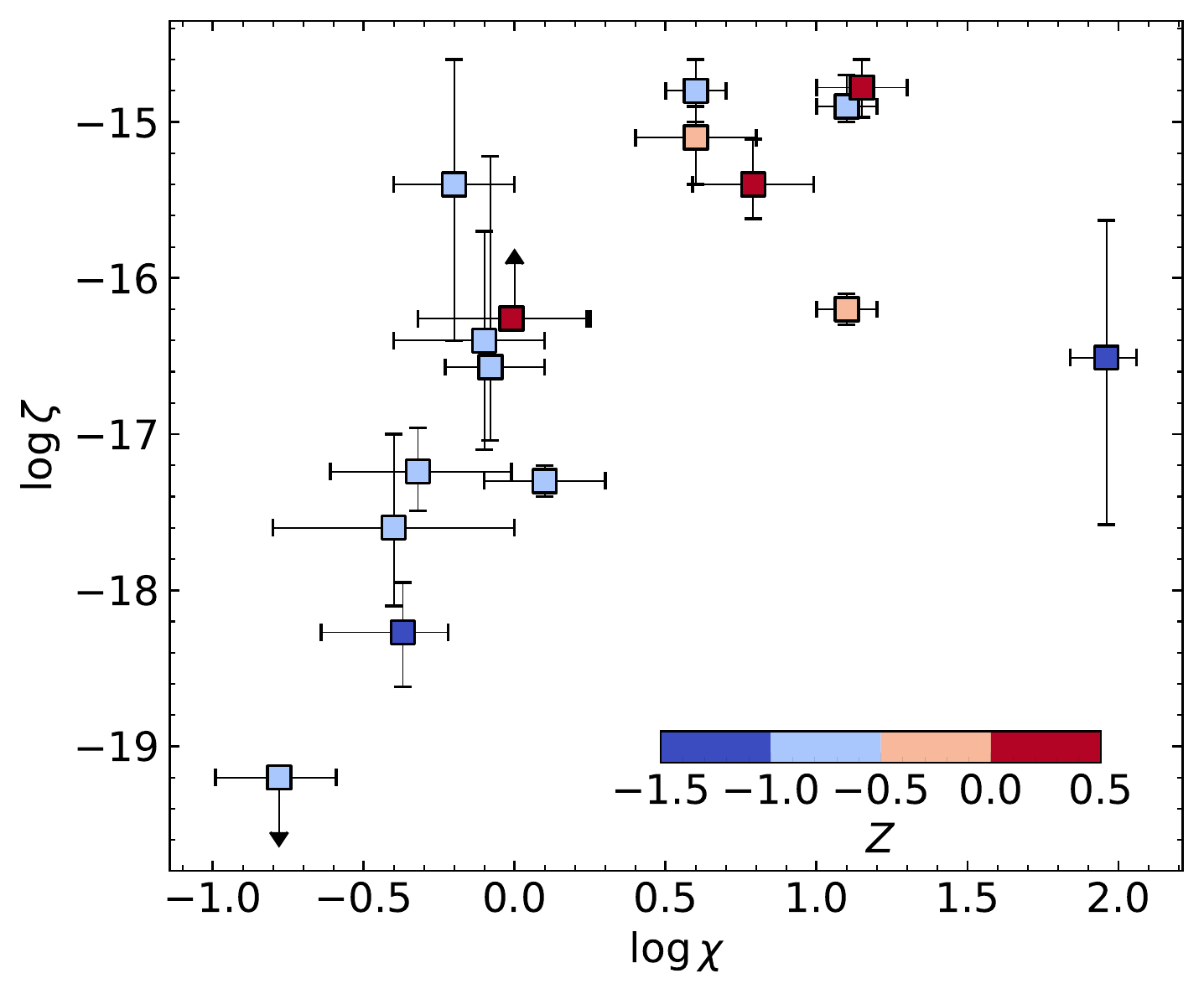}}
    \end{minipage}
    \hfill
    \begin{minipage}{0.33\textwidth}
        \center{\includegraphics[width=1\linewidth]{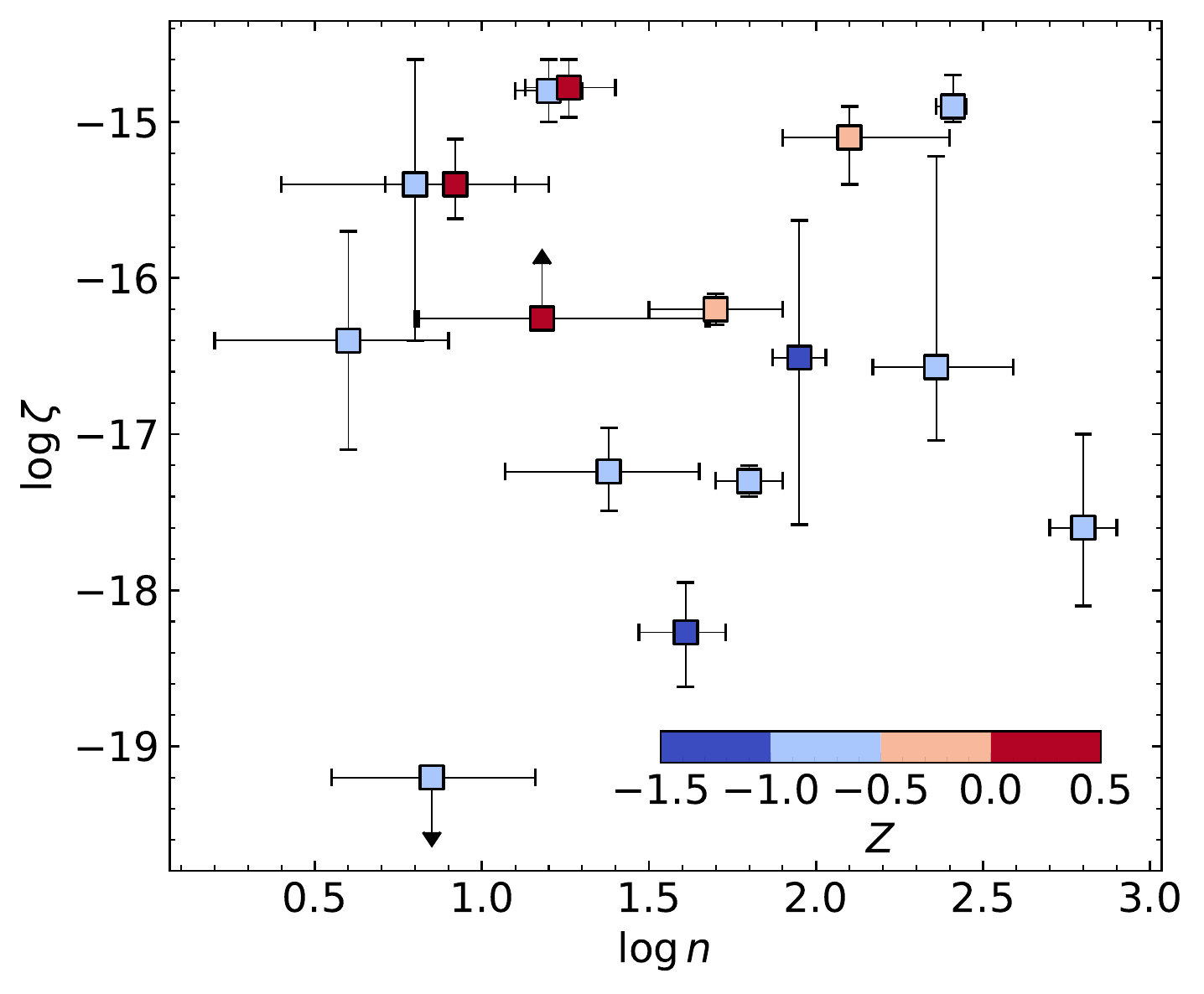}}
    \end{minipage}
    \begin{minipage}{0.33\textwidth}
    \center{\includegraphics[width=\linewidth]{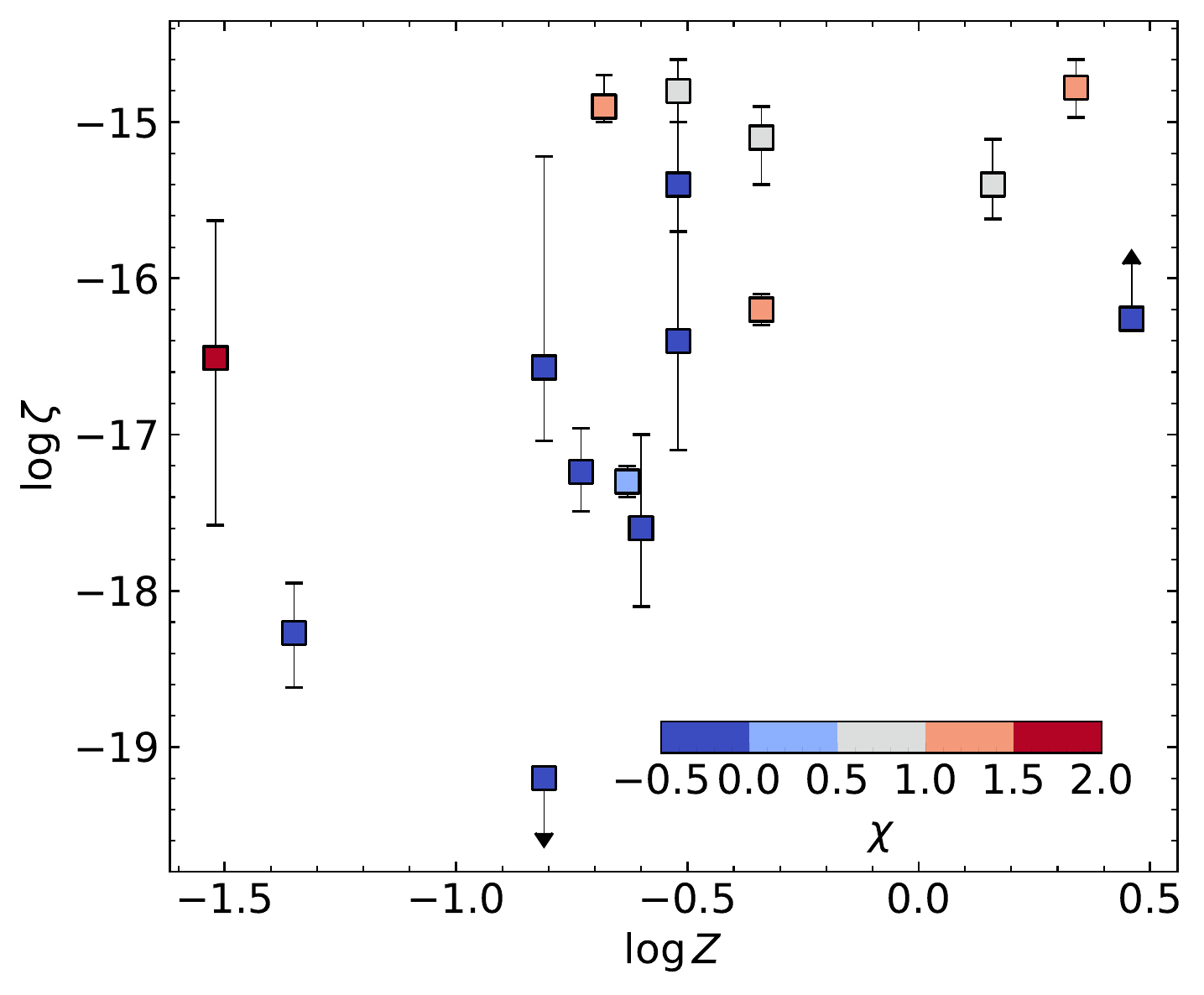}}
    \end{minipage}
    \caption{Estimated cosmic-ray ionization rate, $\zeta$, as a function of UV field strength (left panel), number density (middle panel), and metallicity (right panel), using the $\rm HD/H_2$ relative abundance measured in DLAs at high redshift. The points are color-coded by metallicity (left and middle panels) and UV field strength (right panel), with color bars provided within each panel.}
    \label{fig:cr_uv_n}
\end{figure*}

In Figure \ref{fig:cr_uv_n}, we 
investigate the dependence of $\zeta$ on
the intensity of the UV field, number density and metallicity in the medium. The CRIR is found to correlate strongly with 
the UV field intensity, while it does not correlate with number density and only 
slightly correlates with metallicity. 
%
Removing a possible outlier at $\log \chi\sim 2$ (corresponding to J\,$0843+0221$) and lower and upper limits from J\,0000$+$0048 and J\,0812$+$3208 (comp 2), we find 
a Pearson correlation coefficient between 
$\zeta$ and $\chi$ of $r=0.75$ (with p-value=0.002 to reject the null hypothesis that $\zeta$ and $\chi$ do not correlate). 
The outlier 
in this plot (J\,0843$+$0221) may be due to 
its very high H$_2$ column density, 
with 
suppression of the CRIR
or due to its exceptionally low metallicity.

Indeed, in our formalism, 
we assume CRIR, $\zeta$, to be constant throughout the cloud. However, it is expected that CRIR can be attenuated in the cloud at column densities $N(\rm H) \gtrsim 10^{20}\,\rm cm^{-2}$ \citep[see, e.g.,][]{Silsbee2019}.
Therefore, if the CRIR is attenuated inside the cloud, then the derived value of $\zeta$ is lower than the incident value.
That means that in principle, to draw accurate physical conclusions, cosmic-ray propagation effects at high column densities should be taken into account properly.
This would also require knowledge of the magnetic-field configuration in DLAs, which is not well probed by available observations.

Since the relative $\rm HD/H_2$ abundance depends in opposite ways on 
$\zeta$ and $\chi$ (i.e., $N({\rm HD})/N({\rm H_2})$ increases when $\zeta$ increases but also when $\chi$ decreases), the $\zeta$-$\chi$ correlation could be artificially introduced by issues in the measurements themselves.
However, the posterior distributions for individual systems (e.g., Fig~\ref{fig:J0858_MCMC}) indicate that the measurements of $\zeta$ may correlate more strongly with $n$ than with $\chi$ (if any). 
We also note that the HD/H$_2$ ratio depends on the number density and metallicity (with similar sensitivity on variation of $n$ as for $\chi$, and even higher sensitivity for metallicity (see \citealt{Balashev2020});
therefore, it is not evident why we should see a strong correlation between $\zeta$ and $\chi$, and a lack of correlation between $\zeta$ and $n$. This motivates us to assume that the $\zeta$-$\chi$ correlation has a real physical origin.

Indeed, we expect a common star-formation origin 
between cosmic rays and UV radiation. 
%
Furthermore, one can see that the slope of the $\log \zeta - \log \chi$ correlation is close to 2, i.e., CRIR increases quadratically with the strength of the UV field. This also may have reasonable explanation, since the low-energy cosmic rays ($\lesssim 100$ MeV, that mostly determine the ionization rate) may have complex propagation behaviour, related to the diffusion in the ISM magnetic fields. In addition, taking into account the energy losses \citep[see loss function for cosmic rays in][]{Padovani2018}, this may result in a local enhancement of the $\zeta / \chi$ ratio near the production sites and hence super-linear dependence of $\zeta$ on $\chi$, since UV photons escape much more easily from the star-forming regions.

\begin{figure}    
    \center{\includegraphics[width=1.0\linewidth]{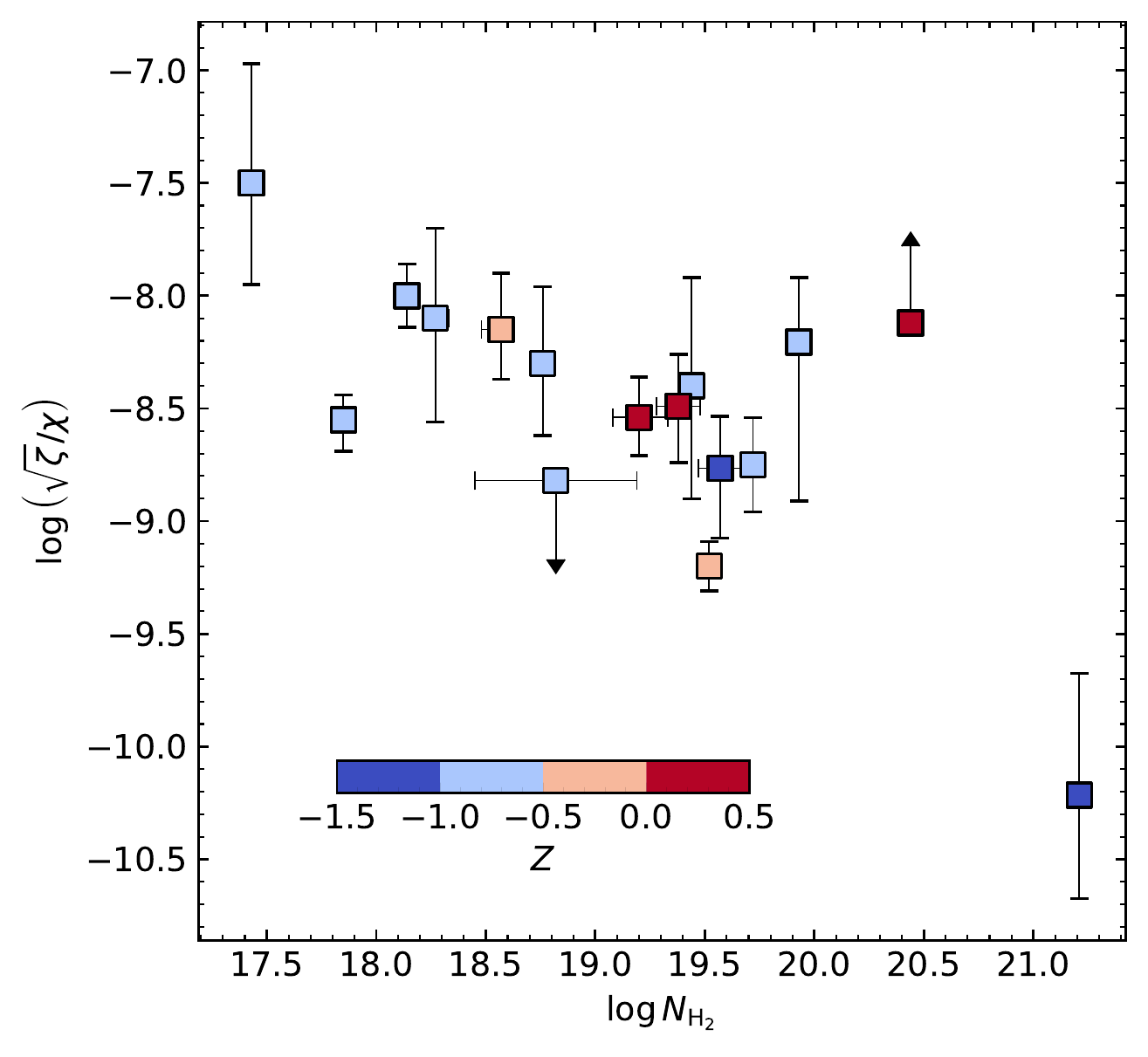}}
    \caption{$\sqrt{\zeta}/\chi$ as a function of $\rm H_2$ column density.
    The points are color-coded by metallicity using the color bar shown at the bottom.
    \label{fig:sqrt_cr_uv_H2}
    }
\end{figure}

Assuming that $\zeta \propto \chi^2$ dependence is real, we plot the $\sqrt{\zeta}/\chi$ as a function of $\log N(\rm H_2)$ in Fig.~\ref{fig:sqrt_cr_uv_H2}. One can see that indeed for the main bulk of the systems with $\log N(\rm H_2)$ in the range 18 -- 20 the dispersion is significantly reduced and only within 1~dex, in comparison with the 3-dex dispersion of $\zeta$ in Figure~\ref{fig:cr_H2_data} (at the same time, we have not found a significant correlation of $\sqrt{\zeta}/\chi$ with neither $Z$ nor $n$).
As already discussed, the single outlier value of $\sqrt{\zeta}/\chi$ with the highest $\log N(\rm H_2)$ (corresponding to J\,0843$+$0221) can be related to cosmic-ray propagation effects at high column densities, or its exceptionally-low metallicity. In turn, at the lower column-density end, absorption systems probe very diffuse gas, where the application of the $\rm H/H_2$ and $\rm D/\rm HD$ transition model that we used should be taken with caution. The most important issue, in our opinion, is that the $\rm HD/H_2$ ratio primarily depends on the hydrogen ionization fraction of the medium, while the CRIR was derived assuming that this ionization state of the diffuse ISM is mainly determined by CRIR and recombination on dust grains \citep{Balashev2020}. In the case of a very diffuse medium, one can expect that the hydrogen ionization fraction is higher due to mixing with ionized and/or warm neutral media, the latter being mostly atomic and hence having higher ionization fractions. All this can effectively mimic the increase of the $\sqrt{\zeta}/\chi$ ratio at lower $\log N(\rm H_2)$ values, that one can notice in Fig.~\ref{fig:sqrt_cr_uv_H2}. Additionally, the recombination rate coefficient can have a non-linear dependence on the metallicity (in our model, we assume it is linear), since it strongly depends on the properties of the dust and we have no strict constraints on it as a function of the metallicity. Indeed a possible correlation of $\zeta$ with $Z$ (see right panel of Fig.~\ref{fig:cr_uv_n}; excluding J\,0843$+$0221 as a possible outlier, we obtain a correlation coefficient of 0.67 with p-value 0.012)
can be caused by this non-linear behaviour. Therefore, we caution to use a simple homogeneous model to estimate CRIR at low $\rm H_2$ column densities. Otherwise, one can propose a physical explanation of the correlation of $\zeta$ with $Z$, as higher metallicity systems probe  more massive galaxies \citep[from well-known mass-metallicity relations (see, e.g.,][]{Sanders2015}, where star formation on average is expected to be higher than in low-mass galaxies and the cosmic-ray flux (and therefore CRIR) is expected to be enhanced.

\section{Conclusion}
\label{sec:conclusion}

We have presented new measurements of HD molecules in high-$z$ absorption systems found in quasar spectra. We looked for HD in all known strong H$_2$-bearing systems and we detected HD molecules in four DLAs and placed upper limits on the HD column density for another twelve DLAs. 
%
So with this study, we significantly increased the sample of HD-bearing DLAs. We find that HD/H$_2$ relative abundances show large dispersion around the D/H isotopic ratio.
This together with previously-known inputs from the modelling of the ISM chemistry indicate that HD/H$_2$ ratios cannot be used to constrain the primordial D/H value.
%
In turn, observed HD/H$_2$ ratios can be used to estimate the gas physical conditions, in particular the cosmic-ray ionization rate (CRIR; \citealt{Balashev2020}). We find that the CRIR varies from a few $10^{-18}\,{\rm s^{-1}}$ to a few $10^{-15}\,{\rm s^{-1}}$ in our sample of high-redshift absorbers, that likely reflects the 
wide range of environments and physical conditions probed by DLAs. These ranges and dispersion are also in line with previous measurements using various methods in the Galaxy as well as other galaxies. 

We find that the CRIR is highly correlated with the UV field intensity in our sample, while it does not correlate with number density and only slightly with metallicity.
These correlations suggest a physical connection between the sources of cosmic rays and those of UV radiation. 
Moreover, we find a quadratic dependence of $\zeta$ on $\chi$ in our sample, which is probably due to transport effects of low-energy cosmic rays. 
We caution however that these correlations may be artificial because of the dependence of $N(\rm HD)/N{(\rm H_2)}$ on combination of $\zeta$, $\chi$, and $Z$.

Additionally, most of the methods currently used to determine CRIR involve a detailed chemical modeling of the
regions related to the diffuse and translucent ISM. Being the dominant species in this region that determine the chemical network results, H$_2$ can be subject to strong systematics concerning the time-dependent chemistry, since the formation timescale of H$_2$ can be relatively high in comparison with cloud lifetimes and steady-state models may not be appropriate \citep[e.g.,][]{Balashev2010}.

\section*{Acknowledgements}
This work was supported by RSF grant 18-12-00301.
We acknowledge support from the French {\sl Centre National de la Recherche Scientifique} through the Russian-French collaborative research program 
``The diffuse interstellar medium of high redshift galaxies'' and from the 
French {\sl Agence Nationale de la Recherche} under ANR grant 17-CE31-0011-01, project ``HIH2'' (PI: Noterdaeme).

\section*{Data availability} 
The results of this paper are based on open data retrieved from the ESO and KECK telescope archives. These data can be shared on reasonable requests to the authors.

\bibliographystyle{mnras}
\bibliography{references.bib}


\appendix

\section{Profile fitting results}

\subsection{HD line profiles}

\begin{figure*}
    \centering
    \includegraphics[width=\textwidth,trim=0.6cm 0cm 0cm 0cm,clip]{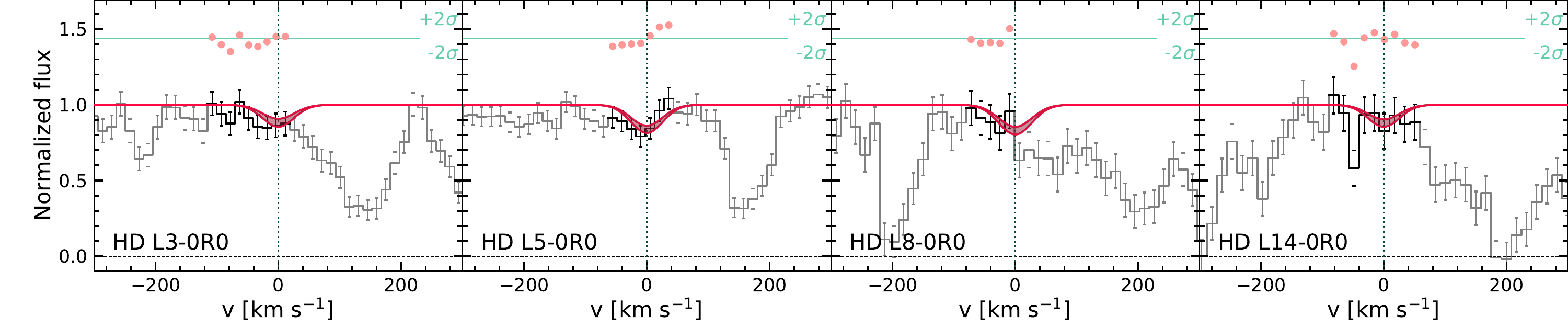}
    \caption{Fit to HD absorption lines in DLA at z=2.77943 in J\,0136$+$0440. The gray and black lines show whole spectra and only pixels used to derive column density, respectively. The red lines indicate the fit profile to HD absorption lines. The region between lines enclose to 0.68 central percentile interval for the fit models, which are sampled from the obtained posterior probability distribution of the parameters. The red points at the top of each panel show the residuals, for the median profile. The vertical dashed lines indicate the central position of the velocity components.}
    \label{fig:J0136}
\end{figure*}

\begin{figure*}
    \centering
    \includegraphics[width=\textwidth,trim=0.6cm 0cm 0cm 0cm,clip]{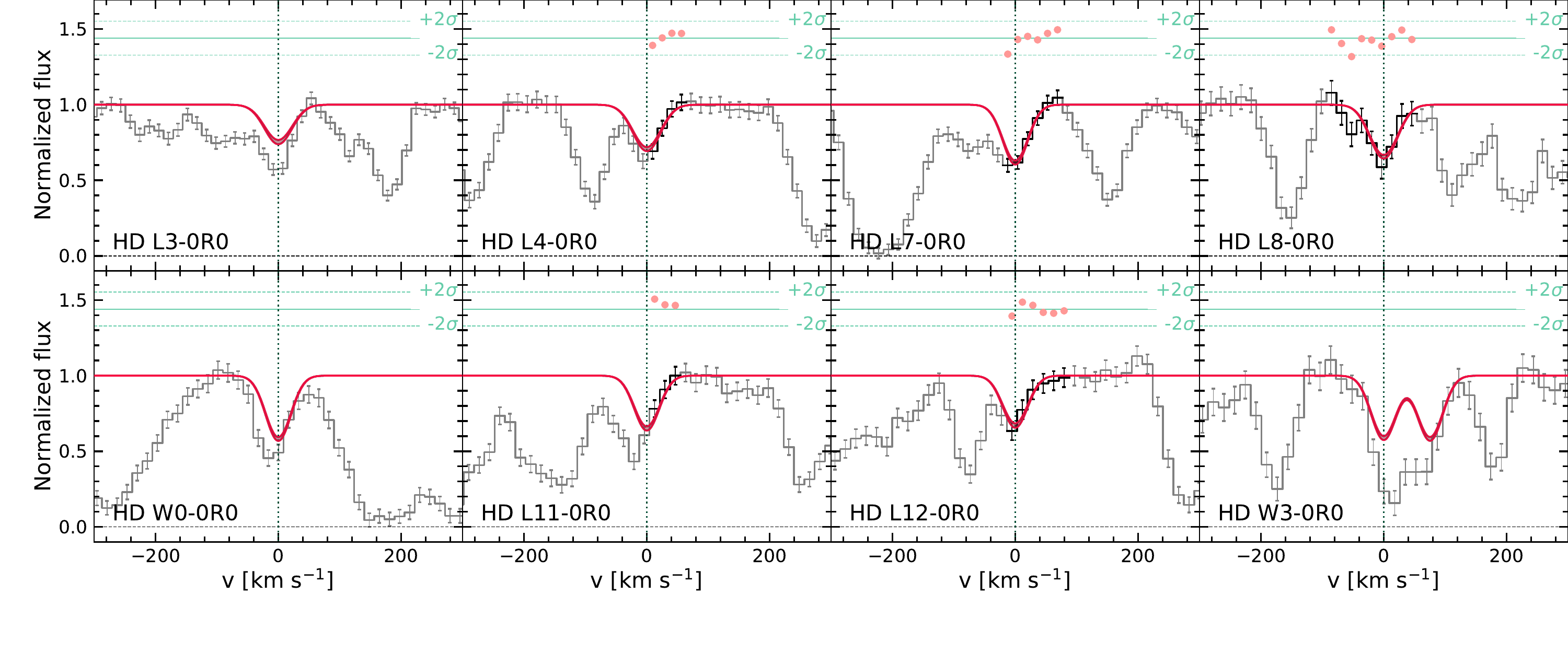}
    \caption{Fit to HD absorption lines in DLA at z=2.62524 in J\,0858$+$1749. The lines are the same as in Fig.~\ref{fig:J0136}.}
    \label{fig:J0858}
\end{figure*}

\begin{figure*}
    \centering
    \includegraphics[width=\textwidth,trim=0.6cm 0cm 0cm 0cm,clip]{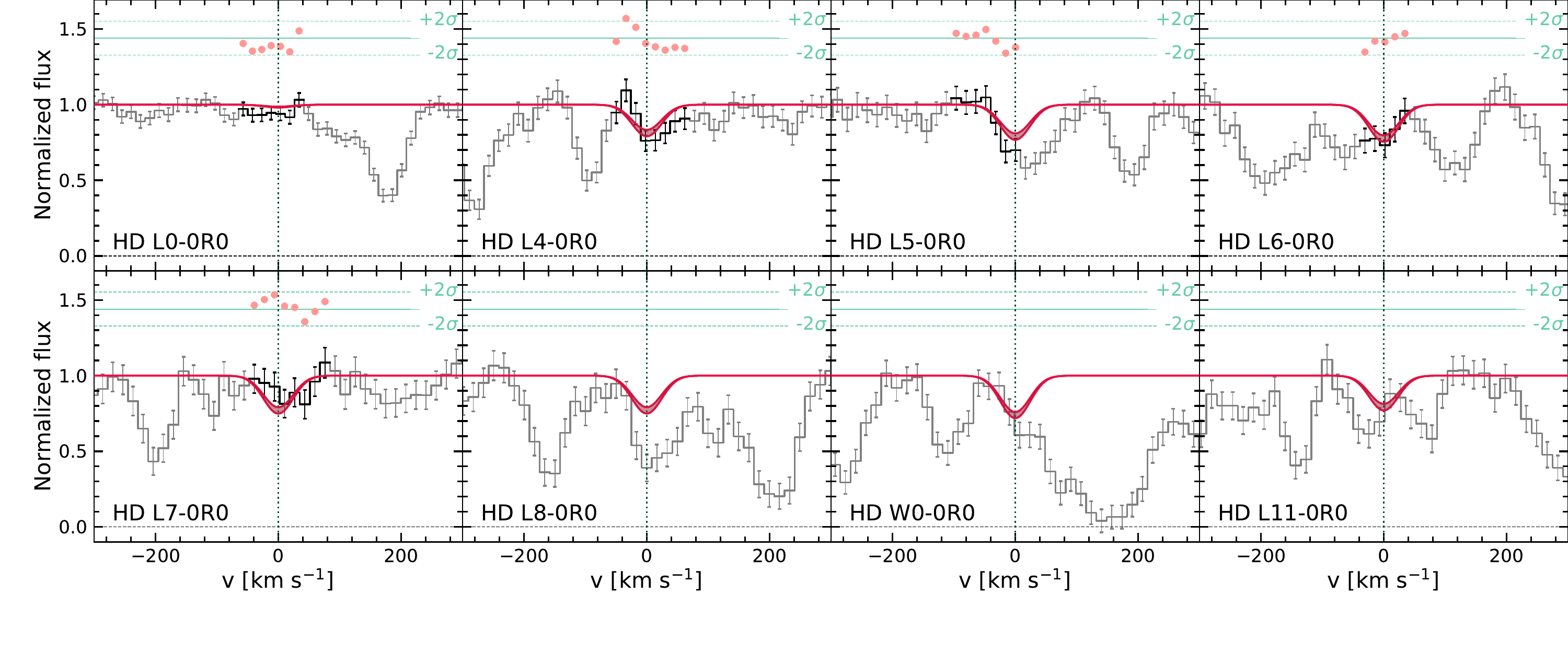}
    \caption{Fit to HD absorption lines in DLA at z=2.56918 in J\,0906$+$0548. The lines are the same as in Fig.~\ref{fig:J0136}.}
    \label{fig:J0906}
\end{figure*}

\begin{figure*}
    \centering
    \includegraphics[width=\textwidth,trim=0.0cm 0cm 0cm 0cm,clip]{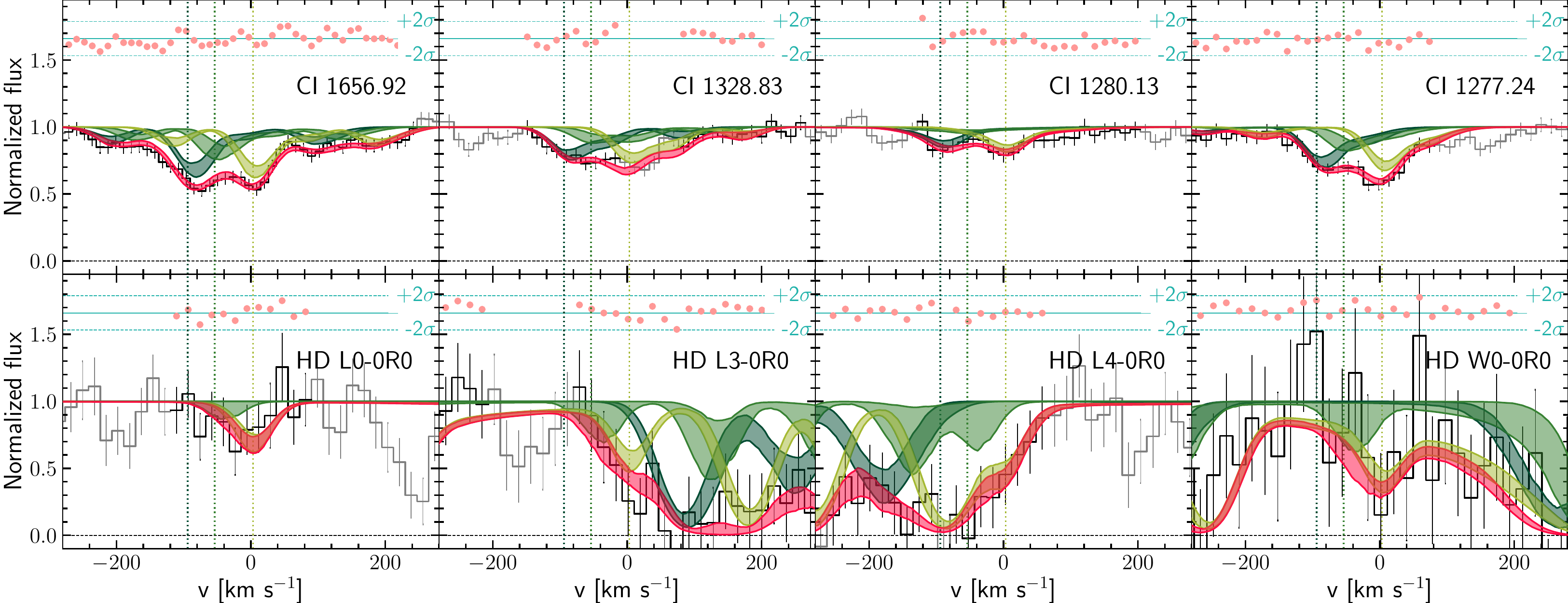}
    \caption{Fit to \ion{C}{I} and HD absorption lines in DLA in J\,0917$-$0154. The lines are the same as in Fig.~\ref{fig:J0136}. The green lines and regions correspond to the contribution of each velocity component. In that particular system all HD lines, except L0-0R(0), are significantly blended with $\rm H_2$ absorption lines, for which the fit in shown in Fig.~\ref{fig:J0917_H2}.}
    \label{fig:J0917}
\end{figure*}

\begin{figure*}
    \centering
    \includegraphics[width=\textwidth,trim=0.6cm 0cm 0cm 0cm,clip]{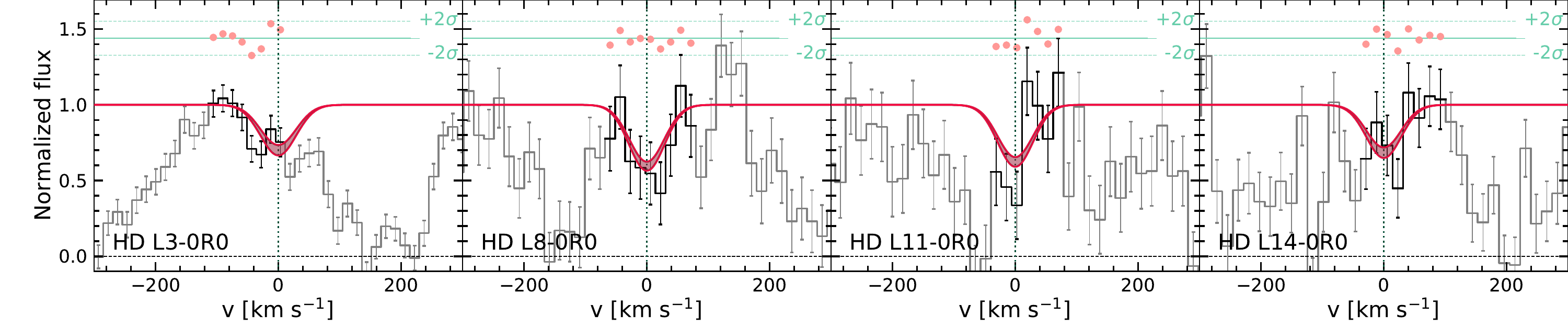}
    \caption{Fit to HD absorption lines in DLA at z=2.606406 in J\,0946$+$1216. The lines are the same as in Fig.~\ref{fig:J0136}.}
    \label{fig:J0946}
\end{figure*}

\begin{figure*}
    \centering
    \includegraphics[width=\textwidth,trim=0.6cm 0cm 0cm 0cm,clip]{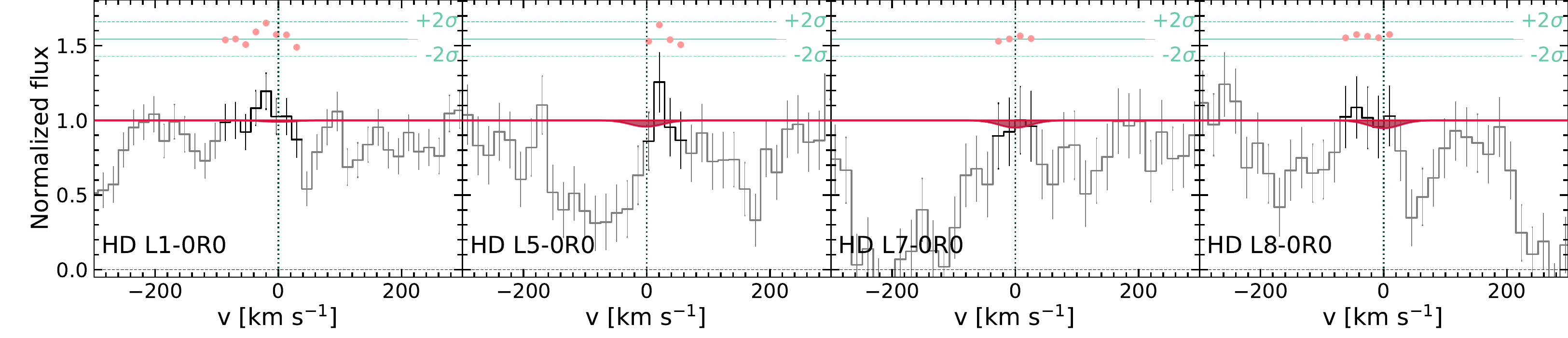}
    \caption{Fit to HD absorption lines in DLA at z=2.3228054 in J\,1143$+$1420. The lines are the same as in Fig.~\ref{fig:J0136}.}
    \label{fig:J1143}
\end{figure*}

\begin{figure*}
    \centering
    \includegraphics[width=\textwidth,trim=0.6cm 0cm 0cm 0cm,clip]{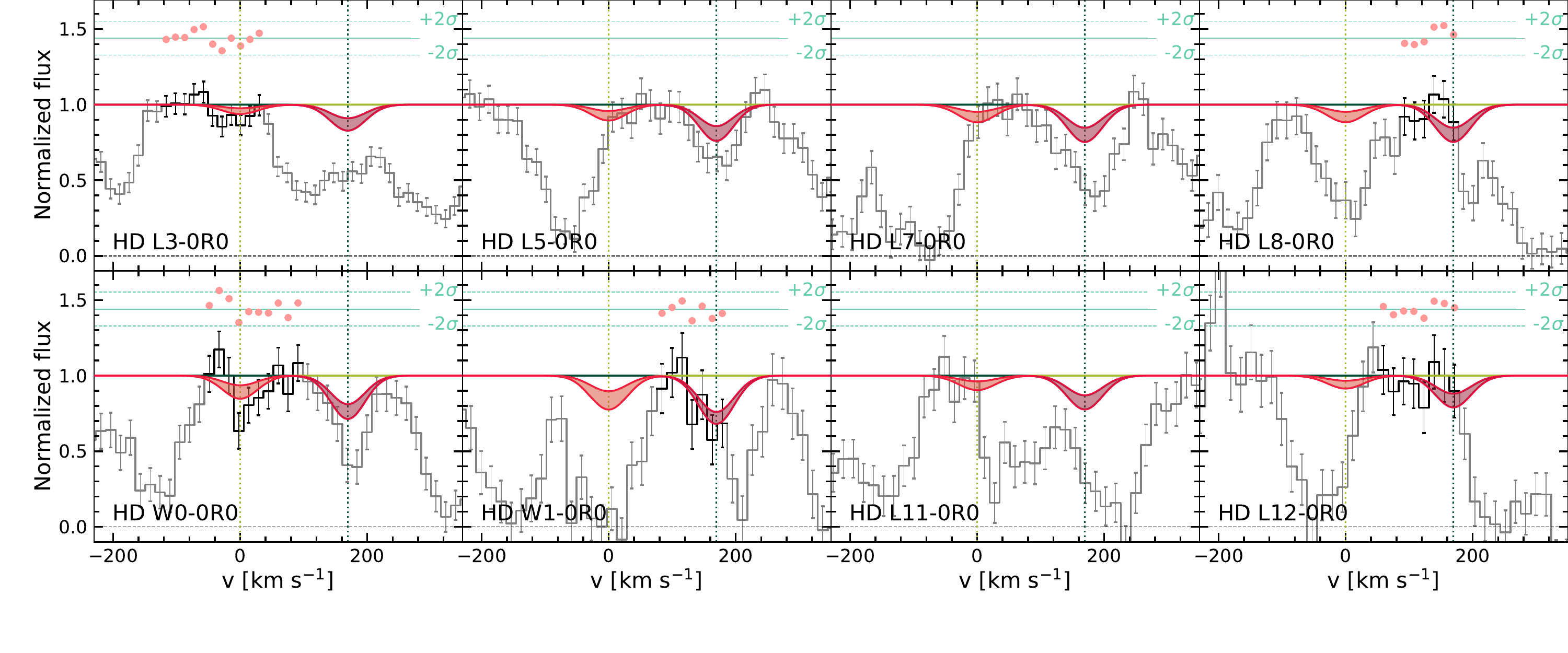}
    \caption{Speсtrum of J\,1146$+$0743 at the position of expected HD absorption lines from DLA at z$\approx$2.840. The lines are the same as in Fig.~\ref{fig:J0917}, except the red lines show the profiles of HD lines with column densities correspond to the upper limits reported in Table~\ref{tab:fit_results}.}
    \label{fig:J1146}
\end{figure*}

\begin{figure*}
    \centering
    \includegraphics[width=\textwidth,trim=0.6cm 0cm 0cm 0cm,clip]{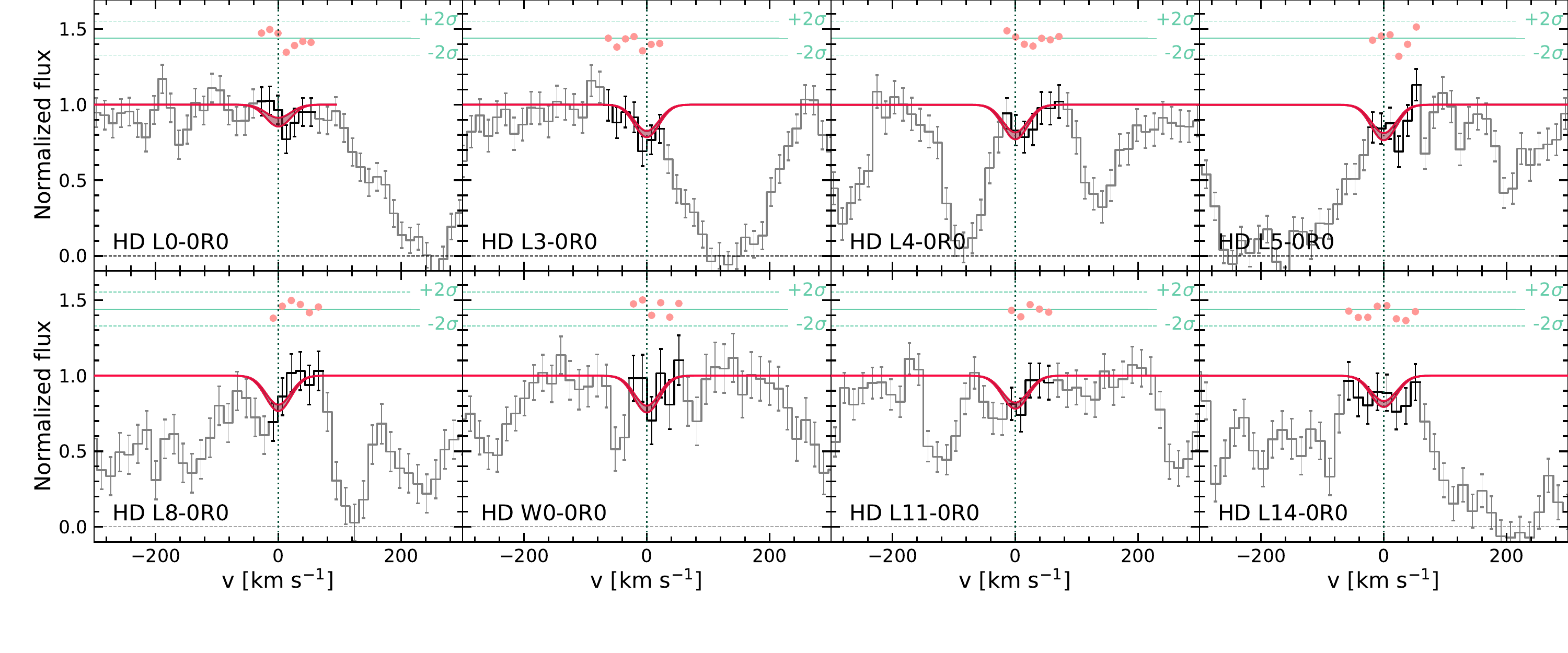}
    \caption{Fit to HD absorption lines in DLA at z=3.03292 in J\,1236$+$0010. The lines are the same as in Fig.~\ref{fig:J0136}.}
    \label{fig:J1236}
\end{figure*}

\begin{figure*}
    \centering
    \includegraphics[width=\textwidth,trim=0.6cm 0cm 0cm 0cm,clip]{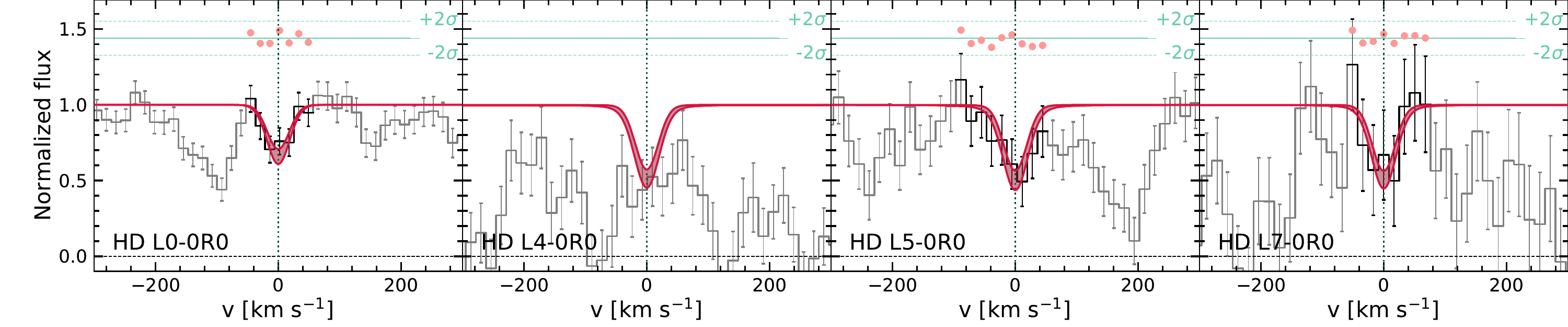}
    \caption{Fit to HD absorption lines in DLA at z=2.58796 in J\,1513$+$0352. The lines are the same as in Fig.~\ref{fig:J0136}. 
    }
    \label{fig:J1513}
\end{figure*}

\begin{figure*}
    \centering
    \includegraphics[width=\textwidth,trim=0.8cm 0cm 0cm 0cm,clip]{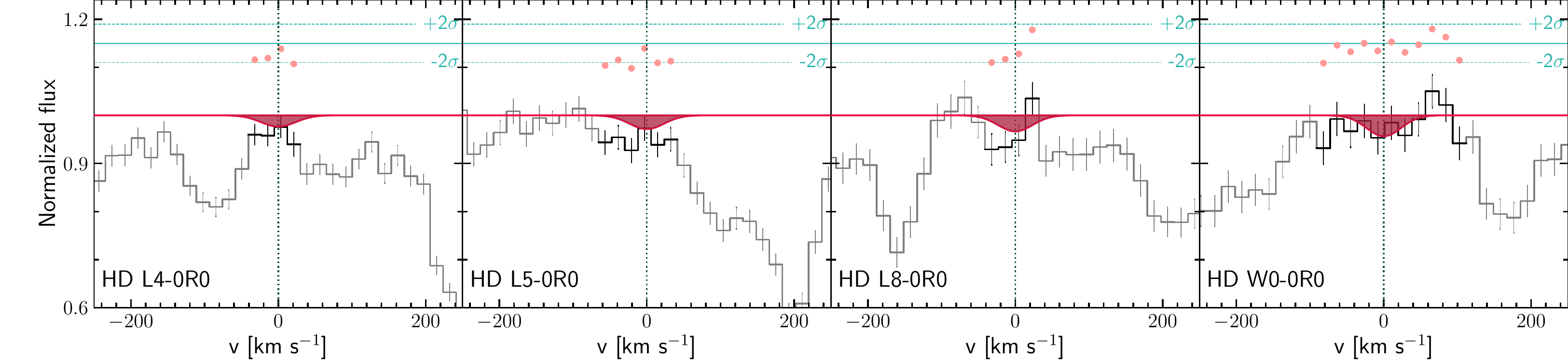}
    \caption{Fit to HD absorption lines in DLA at z=2.2279378 in J\,2232$+$1242. The lines are the same as in Fig.~\ref{fig:J0136}.}
    \label{fig:J2232}
\end{figure*}

\begin{figure*}
    \centering
    \includegraphics[width=\textwidth,trim=0.6cm 0cm 0cm 0cm,clip]{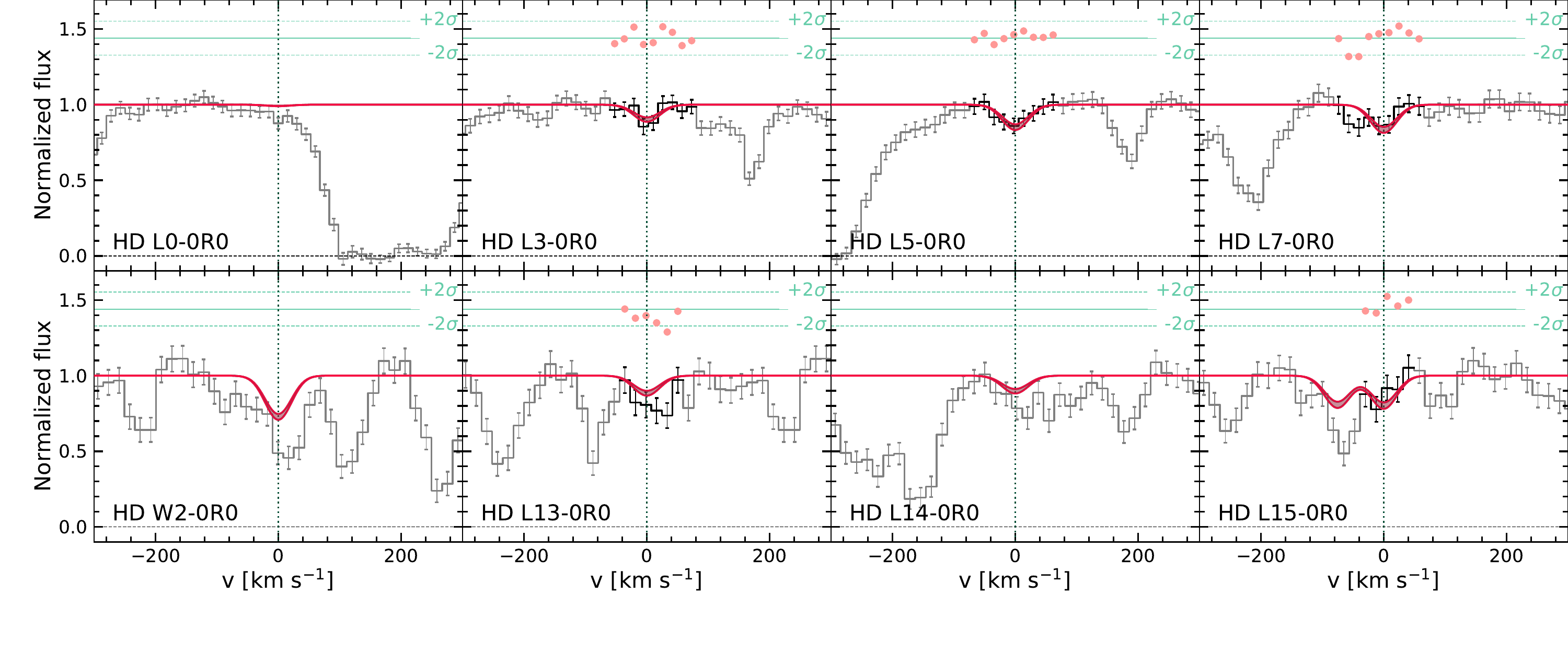}
    \caption{Fit to HD absorption lines in DLA at z=2.58797 in J\,2347$+$0051. The lines are the same as in Fig.~\ref{fig:J0136}.}
    \label{fig:J2347}
\end{figure*}

\begin{figure*}
    \centering
    \includegraphics[width=\textwidth,trim=0.6cm 0cm 0cm 0cm,clip]{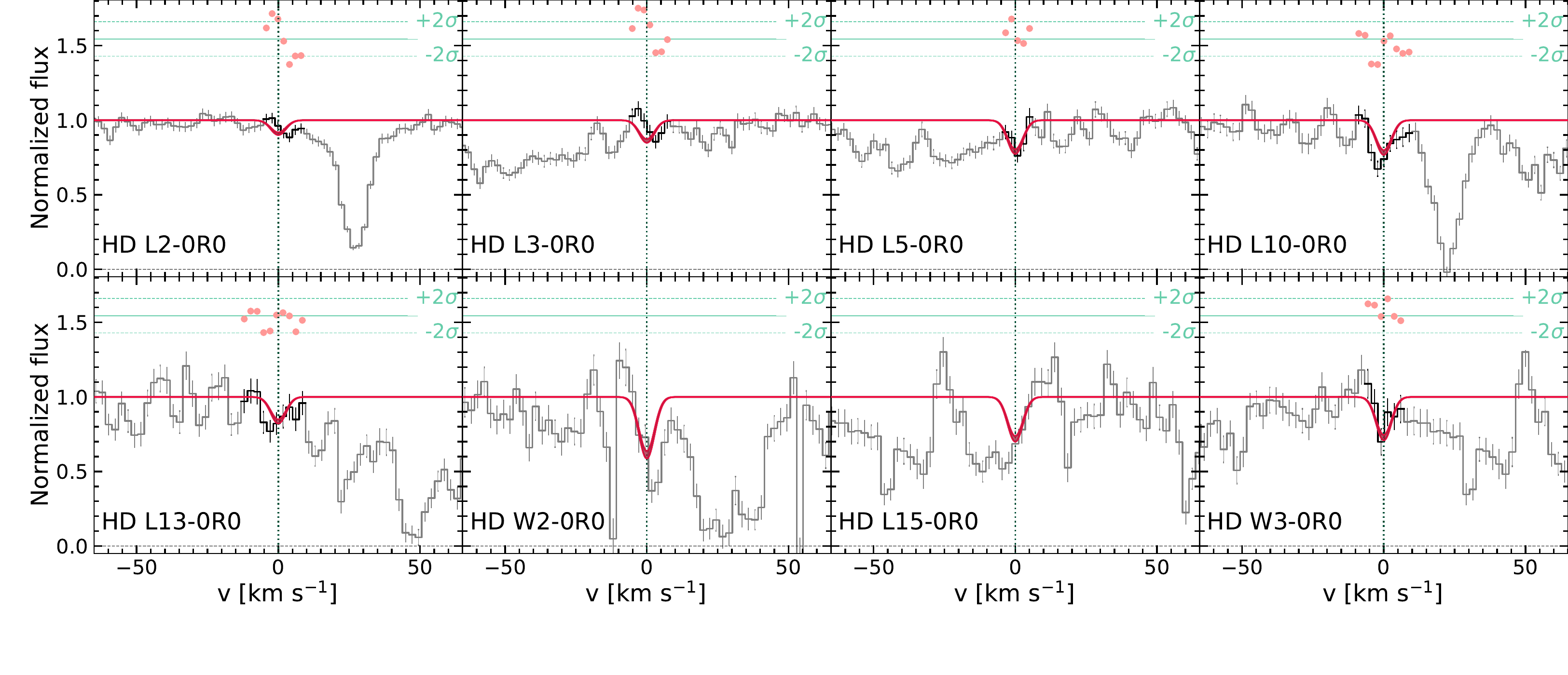}
    \caption{Fit to HD absorption lines in DLA towards HE\,0027+1836. The lines are the same as in Fig.~\ref{fig:J0136}.}
    \label{fig:HE0027_HD}
\end{figure*}

\begin{figure*}
    \centering
    \includegraphics[width=\textwidth]{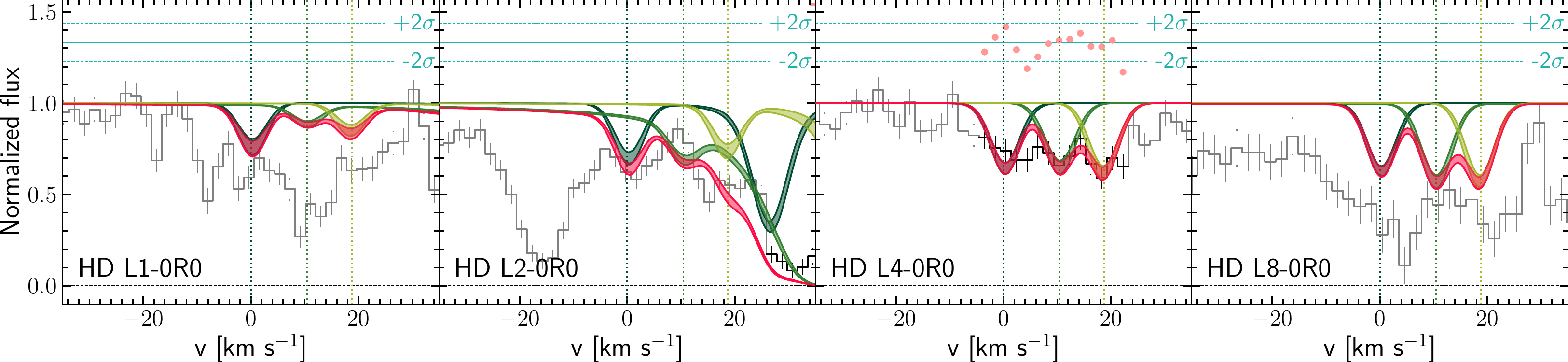}
    \caption{Fit to HD absorption lines in DLA towards J\,0816+1446. The lines are the same as in Fig.~\ref{fig:J0917}.}
    \label{fig:J0816_HD}
\end{figure*}

\begin{figure*}
    \centering
    \includegraphics[width=\textwidth]{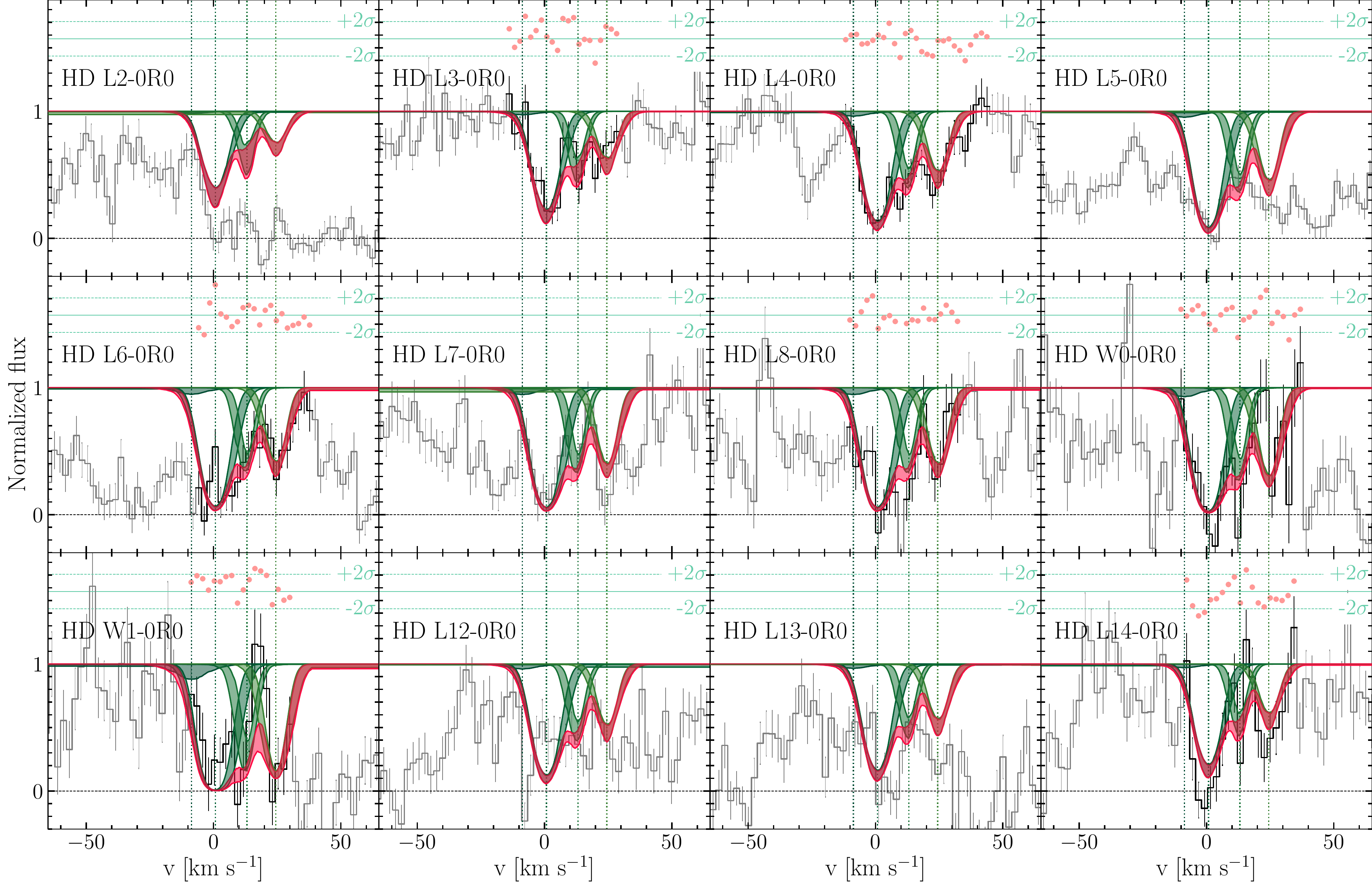}
    \caption{Fit to HD absorption lines in DLA towards J\,1311+2225. The lines are the same as in Fig.~\ref{fig:J0917}.}
    \label{fig:J1311_HD}
\end{figure*}

\begin{figure*}
    \centering
    \includegraphics[width=\textwidth,trim=0.8cm 0cm 0cm 0cm,clip]{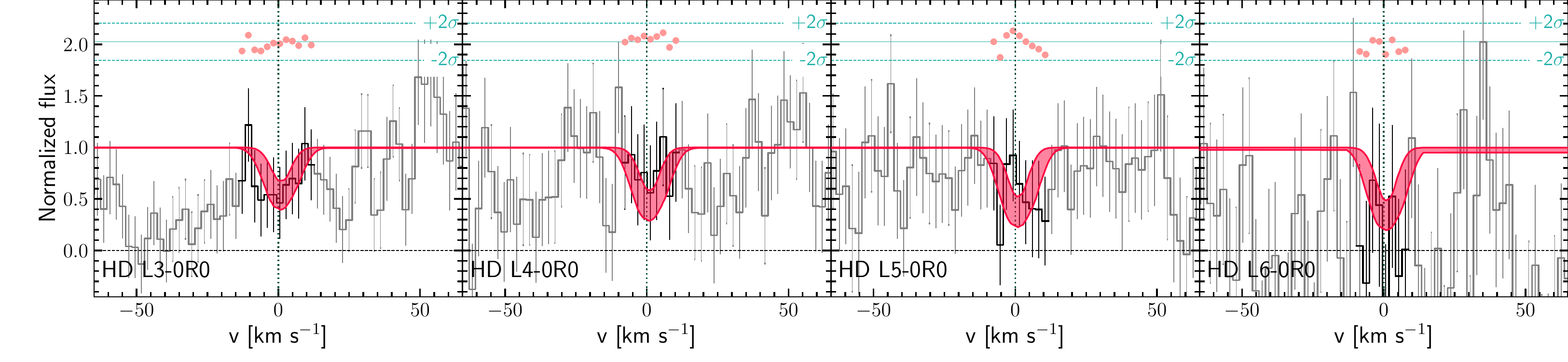}
    \caption{Fit to HD absorption lines in DLA towards J\,2140$-$1840. The lines are the same as in Fig.~\ref{fig:J0136}.}
    \label{fig:J2140_HD}
\end{figure*}

\begin{figure*}
    \centering
    \includegraphics[width=\textwidth,trim=0.6cm 0cm 0cm 0cm,clip]{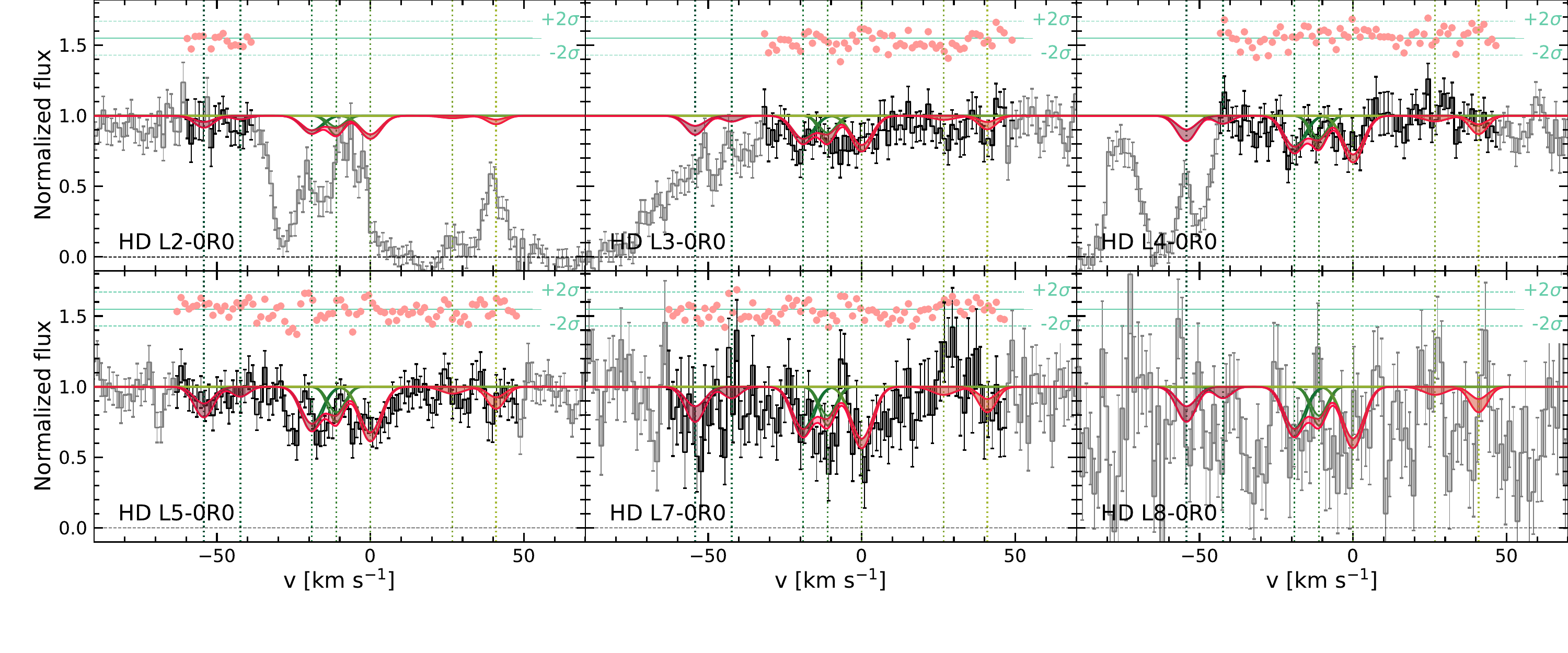}
    \caption{Fit to HD absorption lines in DLA at $z = 2.055$ in J\,2340$-$0053. The lines are the same as in Fig.~\ref{fig:J0917}.}
    \label{fig:J2340_HD}
\end{figure*}

Figures~\ref{fig:J0136}-\ref{fig:J2340_HD} show the line profiles of HD lines in the DLAs studied in this paper and listed in Table~\ref{tab:fit_results}.

\subsection{Fit results of absorption lines towards J\,0917$+$0154}
\label{appendix:J0917}

\begin{figure*}
    \centering
    \includegraphics[width=\textwidth]{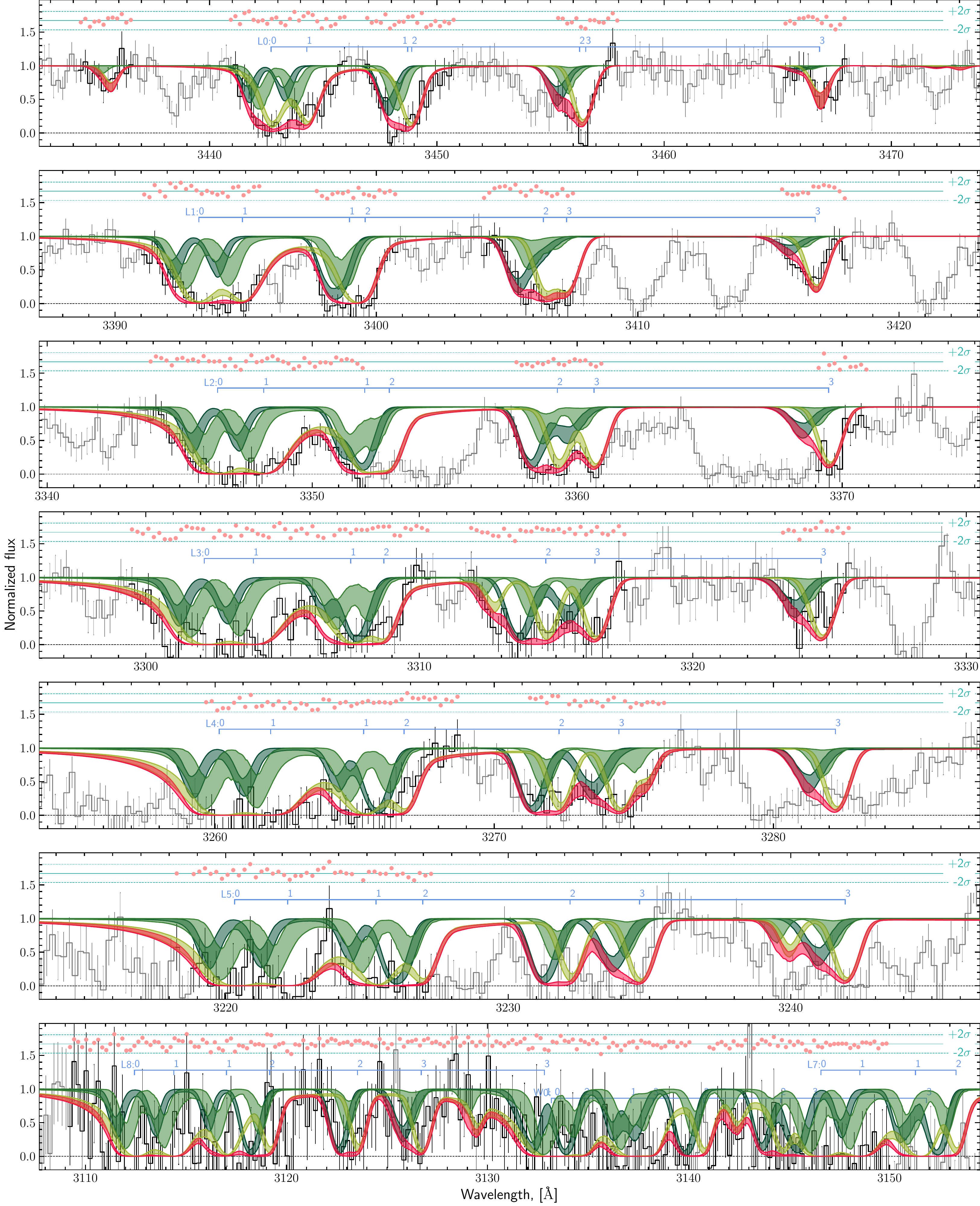}
    \caption{Fit to H$_2$ and HD absorption lines in DLA at $z = 2.106$ towards J\,091721.37$+$015448.1. The lines are the same as in Fig.~\ref{fig:J0917}. The blue ticks marks a position of H$_2$ lines.}
    \label{fig:J0917_H2}
\end{figure*}

In Table~\ref{tab:J0917} we present results of the fitting of \CI, H$_2$ and HD lines in the DLA at $z=2.106$ towards J\,0917$+$0154 using 3-component model. Figure~\ref{fig:J0917_H2} shows the profiles of H$_2$ absorption lines, while in Figure~\ref{fig:J0917} we present the fit to \CI\ and HD lines.

\begin{table*}
\centering
\caption{Fit results of \CI, H$_2$, and HD lines $z = 2.106$ towards J\,0917$-$0154.}
\label{tab:J0917}
\begin{tabular}{cccccc}
\hline
\hline
species & comp & 1 & 2 & 3 & $\log N_{\rm tot}$ \\
& $z$ & $2.10586(^{+6}_{-3})$ & $2.10624(^{+13}_{-7})$ & $2.106812(^{+49}_{-11})$ &  \\
& $\Delta$v, km/s & -95.8 & -46.7 & 0.0 &  \\
\hline
& $b$, km/s & $5.2^{+1.1}_{-1.8}$ & $6.4^{+1.5}_{-2.4}$ & $4.7^{+1.1}_{-1.3}$ &  \\
CI & $\log N$ & $14.15^{+0.16}_{-0.18}$ & $13.46^{+0.12}_{-0.26}$ & $14.35^{+0.20}_{-0.18}$ & $14.59^{+0.18}_{-0.10}$ \\
CI* & $\log N$ & $13.55^{+0.12}_{-0.20}$ & $13.41^{+0.12}_{-0.19}$ & $13.66^{+0.12}_{-0.15}$ & $13.99^{+0.06}_{-0.08}$ \\
CI** & $\log N$ & $13.30^{+0.17}_{-0.28}$ & $13.02^{+0.21}_{-0.42}$ & $13.29^{+0.18}_{-0.29}$ & $13.67^{+0.09}_{-0.18}$ \\
\hline
H$_2$ $J=0$ &  $\log N$ & $18.11^{+0.30}_{-0.83}$ & $18.48^{+0.67}_{-0.96}$ & $19.73^{+0.14}_{-0.13}$ & $19.80^{+0.07}_{-0.16}$ \\
H$_2$ $J=1$ & $\log N$ & $17.77^{+0.78}_{-0.35}$ & $17.98^{+1.13}_{-1.13}$ & $19.81^{+0.09}_{-0.07}$ & $19.87^{+0.05}_{-0.09}$ \\
H$_2$ $J=2$ & $\log N$ & $16.30^{+0.26}_{-0.22}$ & $15.95^{+0.69}_{-1.09}$ & $18.14^{+0.26}_{-0.64}$ & $18.18^{+0.13}_{-0.73}$ \\
& $b$, km/s & $26.9^{+5.5}_{-7.2}$ & $9.2^{+9.6}_{-2.0}$ & $12.1^{+5.3}_{-1.5}$ &\\
H$_2$ $J=3$ & $\log N$ & $15.53^{+0.16}_{-0.16}$ & $14.86^{+0.66}_{-0.70}$ & $16.43^{+0.24}_{-0.17}$ & $16.46^{+0.24}_{-0.12}$ \\
& $b$, km/s & $29.5^{+9.9}_{-9.6}$ & $15.3^{+0.4}_{-0.5}$ & $21.7^{+3.4}_{-9.1}$ & \\
\hline
& $\log N_{\rm tot}$ & $17.96^{+0.82}_{-0.16}$ & $18.4^{+1.0}_{-0.3}$ &  $20.09^{+0.07}_{-0.08}$ \\
\hline
HD $J=0$ & $\log N$ & $<12$ & $<15.9$ & $<18.1$ & $<18.1$ \\
\hline
\end{tabular}
\begin{tablenotes}
\item The Doppler parameters of \CI\ fine-structure levels, H$_2$ J=0,1 rotational levels, and HD were tied together, while H$_2$ J=2,3 were varied independently, but taking into account penalty function for the Doppler parameters as described in Sect.~\ref{sect:high}.
\end{tablenotes}
\end{table*}

\subsection{DLA at J\,0812$+$3208}

\begin{figure*}
\centering
\includegraphics[width=\textwidth,trim=0.6cm 0cm 0cm 0cm,clip]{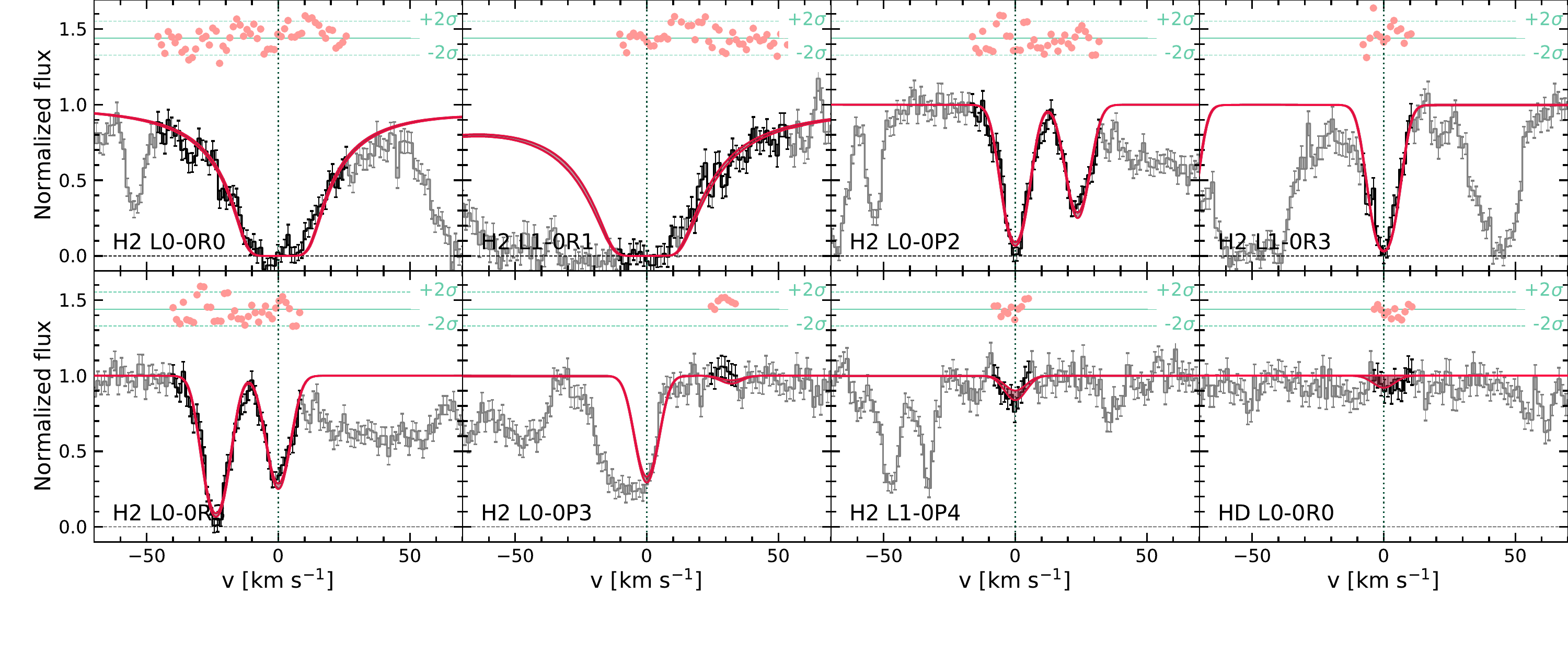}
\caption{Fit to H$_2$ and HD absorption lines in DLA at z=2.06678 in J\,0812$+$0032. The lines are the same as in Fig.~\ref{fig:J0136}.}
\label{fig:J0812}
\end{figure*}

In this section we present results of the fitting of H$_2$ and HD lines in the DLA at $z = 2.067$ towards J\,0812$+$3208. Table~\ref{table:J0812} provides detailed result of the fitting and Figure~\ref{fig:J0812} shows line profiles of H$_2$ and HD absorption lines.

\begin{center}
\begin{table}
\centering
\caption{Results from H$_2$ and HD analysis at  $z = 2.066780(1)$ towards J\,0812$+$3208.}
\label{table:J0812}
\begin{tabular}{lcccccccl}
 \hline
 & $J$ & $b$ & $\log N$\\
\hline
H$_2$ & 0 & $4.4^{+0.1}_{-0.1}$ & $19.03^{+0.02}_{-0.02}$ \\
          & 1 & -''- & $18.88^{+0.03}_{-0.03}$ \\
          & 2 & -''- & $16.19^{+0.04}_{-0.03}$ \\
          & 3 & -''- & $15.76^{+0.03}_{-0.02}$ \\
          & 4 & -''- & $14.28^{+0.12}_{-0.12}$ \\
          & total &  & $19.26^{+0.02}_{-0.01}$ \\
\hline
HD    & 0 & -''- & $<14.43$ \\
\hline
HD/2H$_2$ & & & $< 7.39\times10^{-6}$ \\
\hline
\end{tabular}
\end{table}
\end{center}

\subsection{Fit results of H$_2$/HD towards J\,0816+1446}

In this section we present results of HD$_2$ and HD fitting in the DLA at $z =3.287$ using 3-component model. Table~\ref{tab:J0816} provides HD and H$_2$ fit results and HD line profiles are shown in Figure~\ref{fig:J0816_HD}. 

\begin{table*}
\caption{Fit results of H$_2$, and HD lines $z = 3.287$ towards J\,0816$+$1446.}
\label{tab:J0816}
\begin{tabular}{cccccc}
\hline
\hline
species & comp & 1 & 2 & 3 & $\log N_{\rm tot}$ \\
& z & $3.287252(^{+3}_{-2})$ & $3.287399(^{+2}_{-3})$ & $3.287515(^{+2}_{-3})$ &  \\
& $\Delta$v, km/s & -10.0 & 0 & 8.1 &  \\
\hline
& b, km/s & $0.62^{+0.03}_{-0.04}$ & $1.49^{+0.05}_{-0.09}$ & $1.07^{+0.11}_{-0.12}$ &  \\
H$_2$ J=0 & $\log N$ & $16.25^{+0.10}_{-0.12}$ & $17.61^{+0.06}_{-0.03}$ & $17.13^{+0.09}_{-0.06}$ & $17.76^{+0.03}_{-0.02}$ \\
H$_2$ J=1 & $\log N$ & $16.83^{+0.09}_{-0.11}$ & $18.30^{+0.03}_{-0.03}$ & $17.39^{+0.09}_{-0.15}$ & $18.36^{+0.02}_{-0.02}$ \\
H$_2$ J=2 & $\log N$ & $15.88^{+0.12}_{-0.10}$ & $17.43^{+0.05}_{-0.06}$ & $16.42^{+0.13}_{-0.17}$ & $17.47^{+0.04}_{-0.04}$ \\
& b, km/s & $0.76^{+0.06}_{-0.05}$ & $1.86^{+0.06}_{-0.07}$ & $1.27^{+0.07}_{-0.06}$ &  \\
H$_2$ J=3 & $\log N$ & $14.91^{+0.09}_{-0.11}$ & $15.55^{+0.03}_{-0.02}$ & $15.81^{+0.20}_{-0.20}$ & $16.00^{+0.15}_{-0.08}$ \\
& b, km/s & $0.77^{+0.06}_{-0.08}$ & $3.42^{+0.13}_{-0.17}$ & $1.24^{+0.07}_{-0.08}$ &  \\
\hline
 & $\log N_{\rm tot}$ & $16.97^{+0.09}_{-0.10}$ &  $18.43^{+0.04}_{-0.03}$ &   $17.60^{+0.10}_{-0.10}$ \\
\hline
HD & $\log N$ & $<14.9$ & $<14$ & $<14.2$ & $<15.0$ \\
\hline
\end{tabular}
\begin{tablenotes}
\item The Doppler parameters of H$_2$ J=0,1 rotational levels, and HD were tied together, while H$_2$ J=2,3 were varied independently, but taking into account penalty function for the Doppler parameters as described in Sect.~\ref{sect:high}.
\end{tablenotes}
\end{table*}

\subsection{Fit results of H$_2$/CI towards J\,1311$+$2225}
\label{appendix:J1311}

\begin{figure*}
    \centering
    \includegraphics[width=\textwidth,trim=0.2cm 0cm 0cm 0cm]{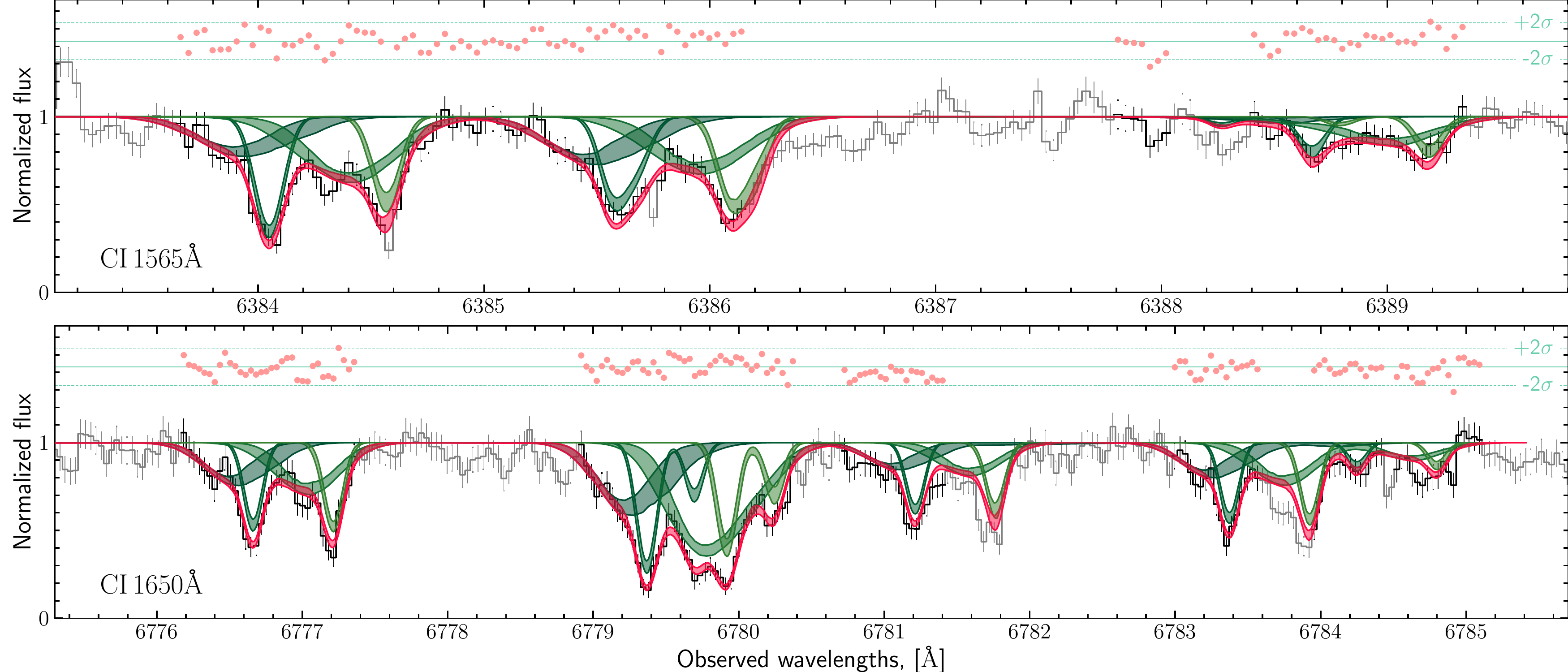}
    \caption{Fit to \ion{C}{I} absorption lines in DLA towards J\,1311+2225. The lines are the same as in Fig.~\ref{fig:J0917}.}
    \label{fig:J1311_CI}
\end{figure*}

\begin{figure*}
    \centering
    \includegraphics[width=\textwidth]{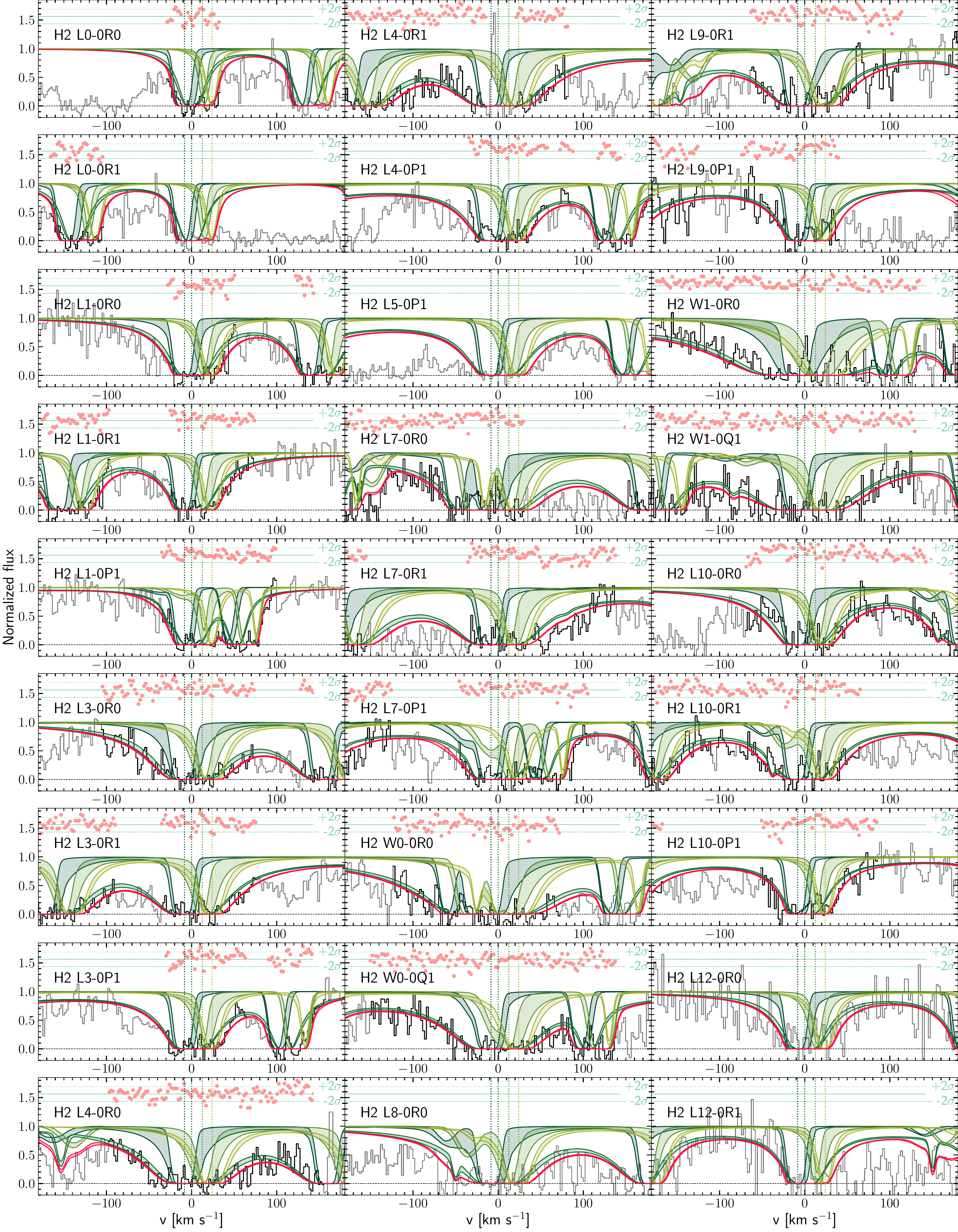}
    \caption{Fit to J=0, J=1 H$_2$ absorption lines in DLA towards J\,1311+2225. The lines are the same as in Fig.~\ref{fig:J0917}.}
    \label{fig:J1311_H2_low}
\end{figure*}

\begin{figure*}
    \centering
    \includegraphics[width=\textwidth]{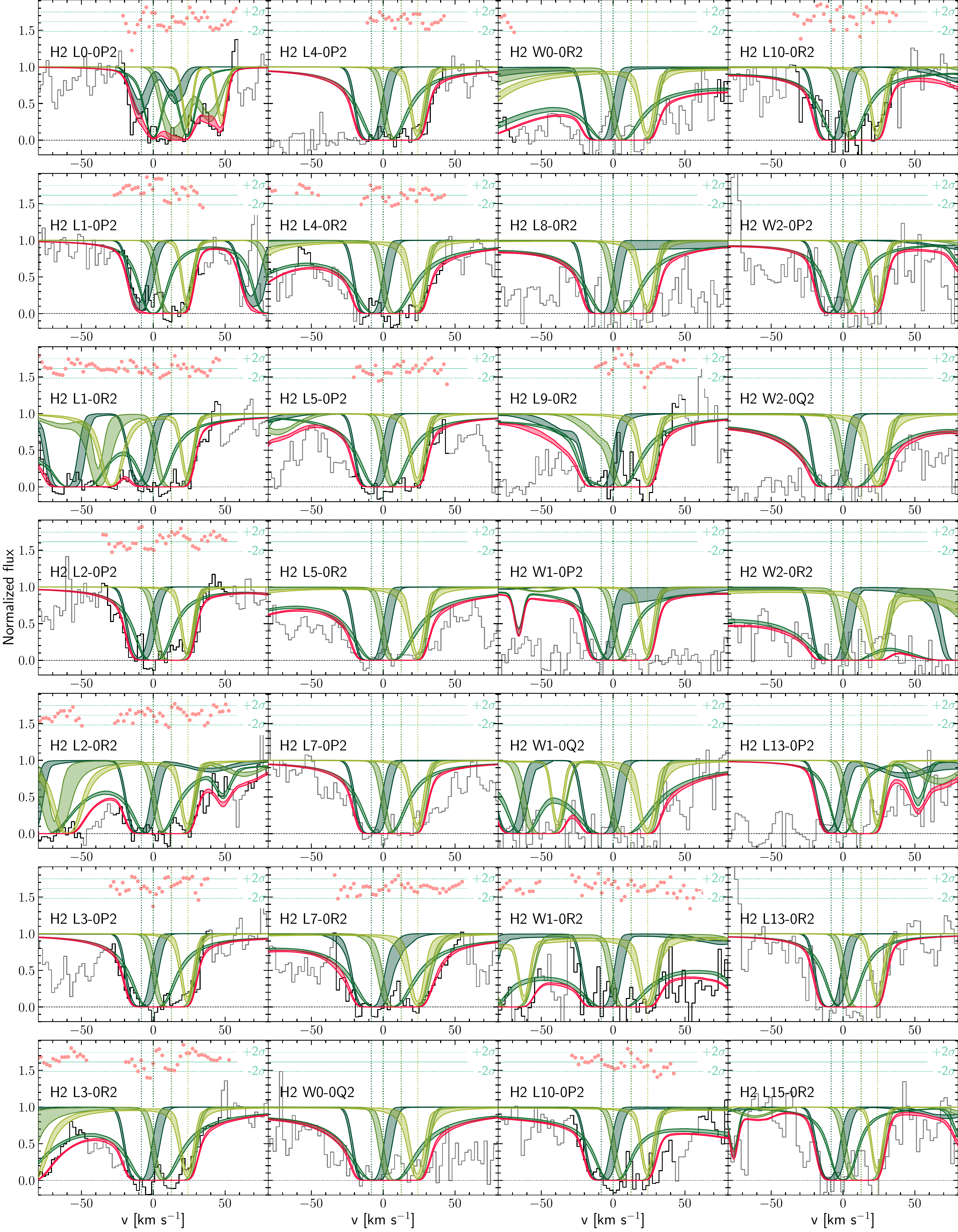}
    \caption{Fit to J=2 H$_2$ absorption lines in DLA towards J\,1311+2225. The lines are the same as in Fig.~\ref{fig:J0917}.}
    \label{fig:J1311_H2_j2}
\end{figure*}

\begin{figure*}
    \centering
    \includegraphics[width=\textwidth]{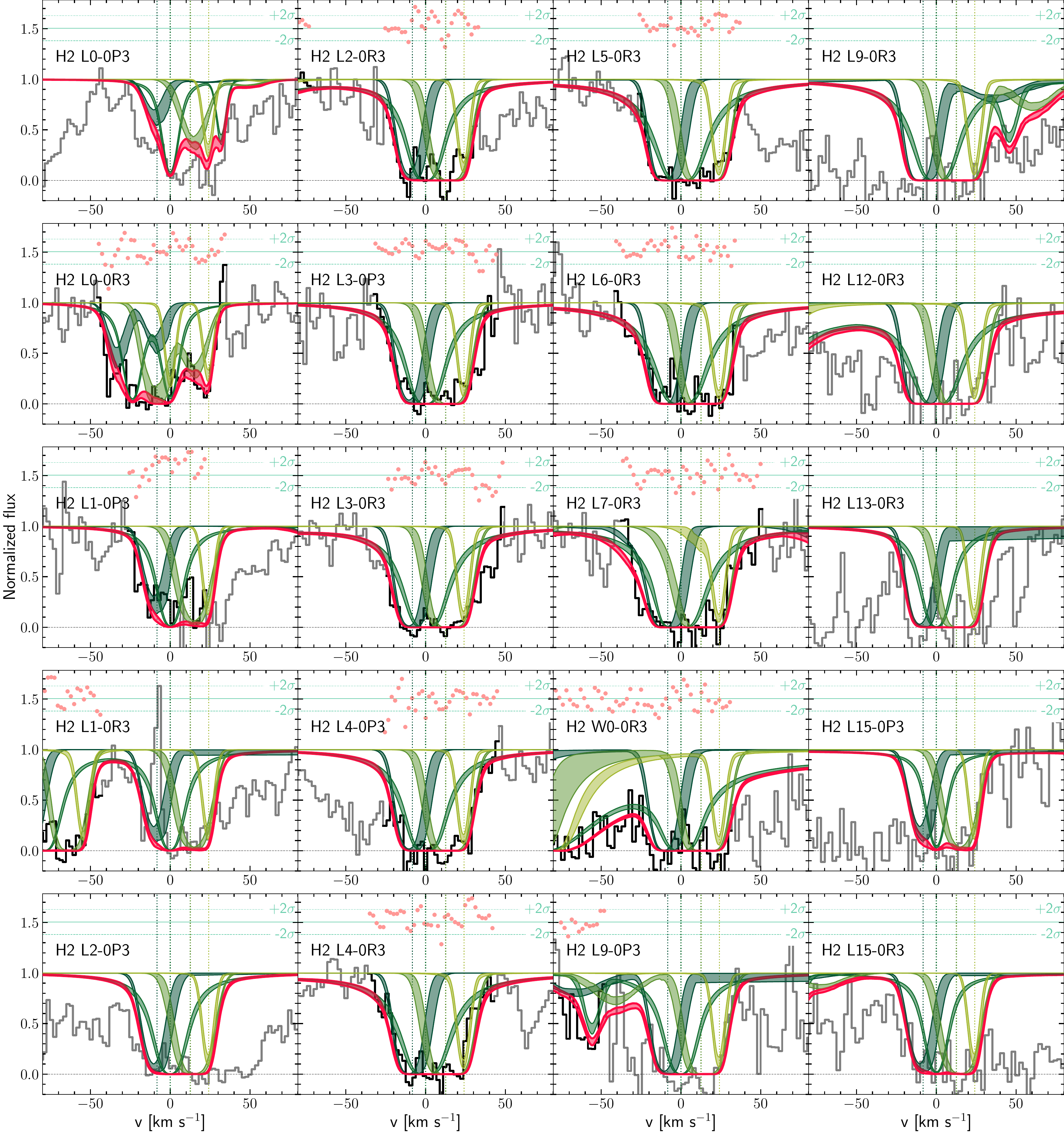}
    \caption{Fit to J=3 H$_2$ absorption lines in DLA towards J\,1311+2225. The lines are the same as in Fig.~\ref{fig:J0917}.}
    \label{fig:J1311_H2_j3}
\end{figure*}

\begin{figure*}
    \centering
    \includegraphics[width=\textwidth]{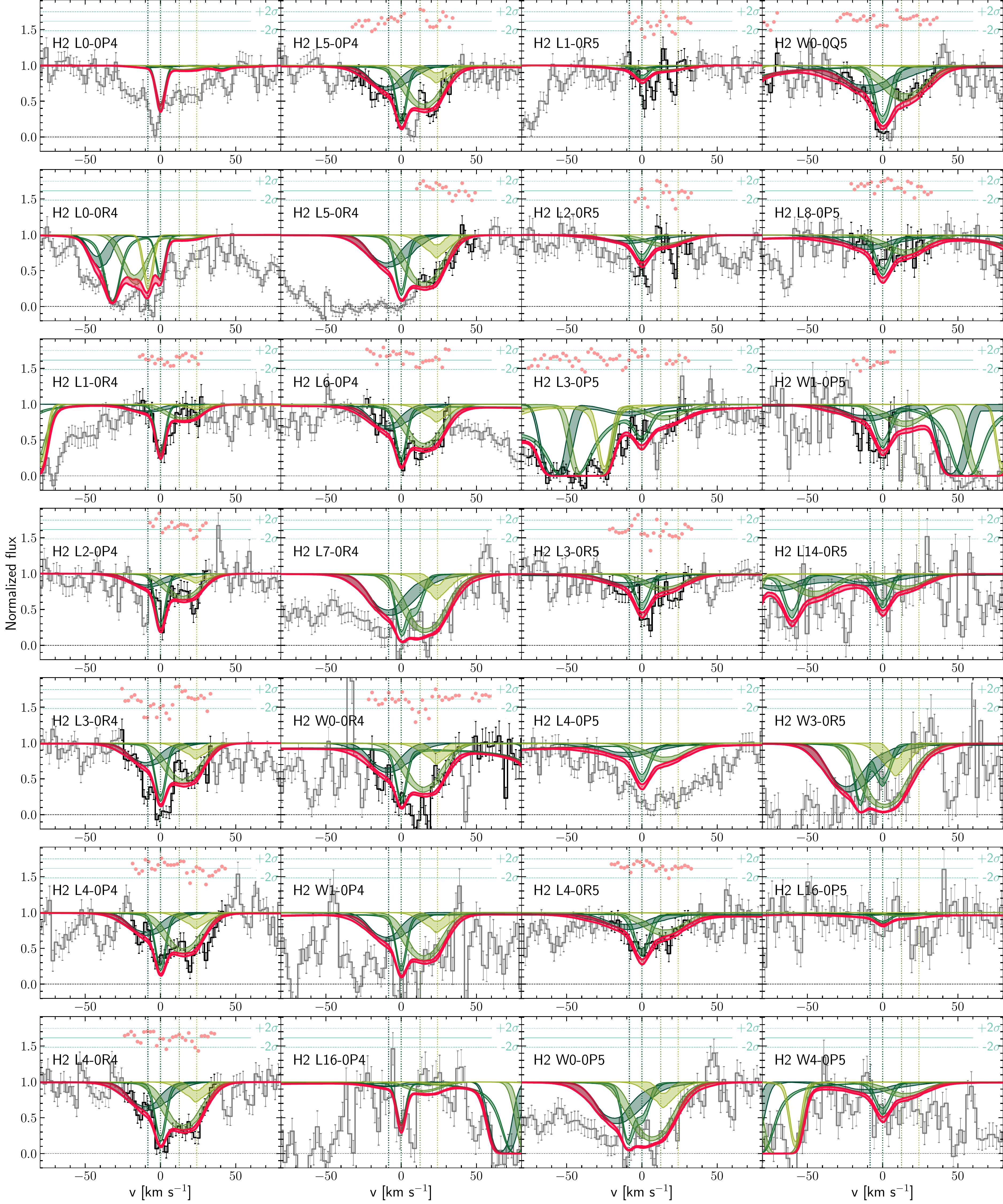}
    \caption{Fit to J=4 and J=5 H$_2$ absorption lines in DLA towards J\,1311+2225. The lines are the same as in Fig.~\ref{fig:J0917}.}
    \label{fig:J1311_H2_j45}
\end{figure*}

In this section we present detailed results of the fitting of $\rm H_2$ and \CI\ in the DLA at z=3.09 towards J\,1311$+$2225 using 4-component model. Table~\ref{tab:J1311_results} provided the fitted values, while Figures~\ref{fig:J1311_CI} and \ref{fig:J1311_H2_low}-\ref{fig:J1311_H2_j45} show line profiles of \ion{C}{I} and H$_2$ absorption lines, respectively.

\begin{table*}
\caption{Fit results of \ion{C}{I}, H$_2$, and HD in DLA at $z = 3.09$ towards J\,131129.11$+$222552.6. \label{tab:J1311_results}}
\begin{tabular}{ccccccc}
\hline
\hline
& comp & 1 & 2 & 3 & 4 & $\log N_{\rm tot}$ \\
\hline
 & $z$ & $3.091410(^{+21}_{-14})$ & $3.0915350(^{+23}_{-21})$ & $3.091735(^{+12}_{-10})$ & $3.0918577(^{+25}_{-25})$ &  \\
& $\Delta$v, km/s & -10.0 & 0.0 & 14.5 & 23.7 &  \\
\hline
& $b$, km/s & $11^{+1}_{-3}$ & $1.2^{+0.4}_{-0.2}$ & $12^{+1}_{-2}$ & $1.1^{+0.7}_{-0.1}$ & \\
\ion{C}{I} & $\log N$ & $13.17^{+0.11}_{-0.11}$ & $13.81^{+0.73}_{-0.27}$ & $13.48^{+0.05}_{-0.05}$ & $13.32^{+0.31}_{-0.20}$ & $14.42^{+0.51}_{-0.13}$ \\
\ion{C}{I$^{*}$} & $\log N$ & $13.28^{+0.10}_{-0.11}$ & $13.45^{+0.11}_{-0.08}$ & $13.42^{+0.07}_{-0.08}$ & $13.40^{+0.16}_{-0.09}$ & $13.99^{+0.08}_{-0.04}$ \\
\ion{C}{I$^{**}$} & $\log N$ & $12.00^{+0.5}_{-0.4}$ & $12.86^{+0.07}_{-0.11}$ & $13.14^{+0.04}_{-0.09}$ & $12.81^{+0.10}_{-0.11}$ & $13.43^{+0.04}_{-0.02}$ \\
& $\log N_{\rm tot}$ & $13.56^{+0.10}_{-0.12}$ & $13.95^{+0.61}_{-0.16}$ & $13.85^{+0.04}_{-0.06}$ & $13.71^{+0.21}_{-0.11}$ & $14.42^{+0.34}_{-0.10}$ \\
\hline
H$_2$ $J=0$ & $\log N$ & $17.57^{+0.56}_{-0.33}$ & $19.07^{+0.03}_{-0.05}$ & $17.70^{+0.40}_{-0.30}$ & $18.21^{+0.08}_{-0.11}$ & $19.15^{+0.01}_{-0.02}$ \\
& $b$, km/s & $5.9^{+0.4}_{-0.6}$ & $0.7^{+0.3}_{-0.1}$ & $1.3^{+0.1}_{-0.5}$ & $1.4^{+0.2}_{-0.3}$ &  \\
H$_2$ $J=1$ & $\log N$ & $17.34^{+0.42}_{-0.31}$ & $19.24^{+0.03}_{-0.01}$ & $17.89^{+0.41}_{-0.33}$ & $18.29^{+0.09}_{-0.08}$ & $19.31^{+0.01}_{-0.01}$ \\
& $b$, km/s & $\dagger$ & $\dagger$ & $\dagger$ & $\dagger$ & \\
H$_2$ $J=2$ & $\log N$ & $15.75^{+0.15}_{-0.09}$ & $18.45^{+0.04}_{-0.04}$ & $16.33^{+0.32}_{-0.22}$ & $17.10^{+0.18}_{-0.42}$ & $18.47^{+0.04}_{-0.04}$ \\
& $b$, km/s & $5.9^{+1.1}_{-0.3}$ & $1.1^{+0.1}_{-0.1}$ & $7.3^{+1.0}_{-0.9}$ & $1.5^{+0.2}_{-0.1}$ & \\
H$_2$ $J=3$ & $\log N$ & $15.56^{+0.10}_{-0.13}$ & $18.23^{+0.06}_{-0.10}$ & $15.90^{+0.08}_{-0.05}$ & $15.95^{+0.38}_{-0.23}$ & $18.24^{+0.05}_{-0.10}$ \\
& $b$, km/s & $7.7^{+0.8}_{-1.1}$ & $1.0^{+0.2}_{-0.1}$ & $9.2^{+0.9}_{-0.5}$ & $1.6^{+0.4}_{-0.2}$ & \\
H$_2$ $J=4$ & $\log N$ & $14.57^{+0.09}_{-0.06}$ & $16.32^{+0.24}_{-0.21}$ & $14.89^{+0.04}_{-0.05}$ & $13.93^{+0.25}_{-0.19}$ & $16.34^{+0.23}_{-0.19}$ \\
& $b$, km/s & $16.8^{+2.3}_{-3.3}$ & $1.1^{+0.2}_{-0.1}$ & $15.0^{+1.1}_{-1.3}$ & $4.3^{+3.9}_{-2.4}$ & \\
H$_2$ $J=5$ & $\log N$ & $14.39^{+0.09}_{-0.07}$ & $14.35^{+0.08}_{-0.06}$ & $14.52^{+0.05}_{-0.09}$ & $12.98^{+0.44}_{-0.65}$ & $14.91^{+0.02}_{-0.02}$ \\
& $b$, km/s & $19.1^{+4.3}_{-3.3}$ & $4.5^{+1.3}_{-0.7}$ & $18.2^{+2.2}_{-3.3}$ & $8.0^{+7.1}_{-3.1}$ & \\
\hline
& $\log N_{\rm tot}$  & $17.87^{+0.37}_{-0.33}$ & $19.52^{+0.02}_{-0.02}$ & $18.25^{+0.22}_{-0.39}$ & $18.57^{+0.05}_{-0.09}$ & $19.59^{+0.01}_{-0.01}$ \\
\hline
HD $J=0$  & $z$ & $3.091410(^{+21}_{-14})$ & $3.0915397(^{+66}_{-77})$ & $3.091714(^{+28}_{-48})$ & $3.091871(^{+11}_{-26})$ \\ 
        &$\log N$ & $\lesssim 12.81$ & $14.82^{+0.08}_{-0.08}$ & $14.30^{+0.37}_{-0.31}$ & $ 14.27^{+0.10}_{-0.13}$ & $15.02^{+0.11}_{-0.07}$ \\
      & $b$, km/s & $8.0^{+4.6}_{-5.4}$ & $5.4^{+0.8}_{-0.8}$ & $\lesssim 2.8$ & $4.0^{+1.6}_{-1.2}$ &  \\
\hline
HD/2H$_2$ &      & $\lesssim 4.4\times 10^{-6}$ & $\left(1.0^{+0.3}_{-0.2}\right)\times 10^{-5}$ & $\left(5.6^{+13.7}_{-3.2}\right)\times 10^{-5}$ & $\left(2.5^{+0.9}_{-0.7}\right)\times 10^{-6}$ & $\left(1.34^{+0.39}_{-0.21}\right)\times 10^{-5}$ \\
\hline
\hline
$\log n$ &  & $2.1^{+0.5}_{-0.3}$ & $1.7^{+0.2}_{-0.2}$ & $1.9^{+0.1}_{-0.1}$ & $2.1^{+0.2}_{-0.3}$ & \\
$\log \chi$ &  & $0.9^{+0.2}_{-0.2}$ & $1.1^{+0.1}_{-0.1}$ & $0.9^{+0.1}_{-0.1}$ & $0.6^{+0.2}_{-0.2}$\\
\hline
\hline
\end{tabular}
\begin{tablenotes}
\item $\dagger$ The Doppler parameters of H$_2$ $J=1$ was tied to H$_2$ $J=0$.
\end{tablenotes}
\end{table*}

\begin{figure*}
    \centering
    \includegraphics[width=0.6\textwidth]{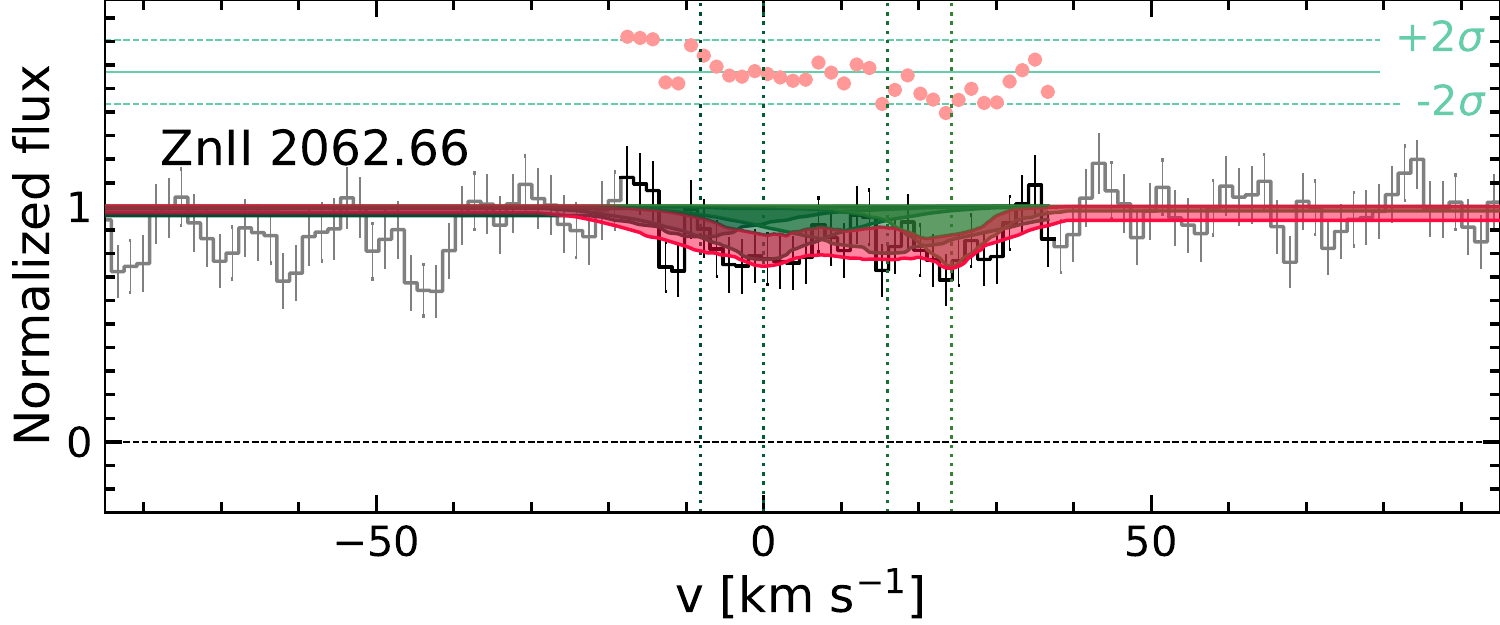}
    \caption{Fit to \ion{Zn}{II} absorption lines in DLA towards J\,1311+2225. The lines are the same as in Fig.~\ref{fig:J0917}.}
    \label{fig:J1311_ZnII}
\end{figure*}

\subsection{Fit results of H$_2$/CI towards J\,2340$-$0053}
\label{appendix:J2340}

\begin{figure*}
    \centering
    \includegraphics[width=\textwidth]{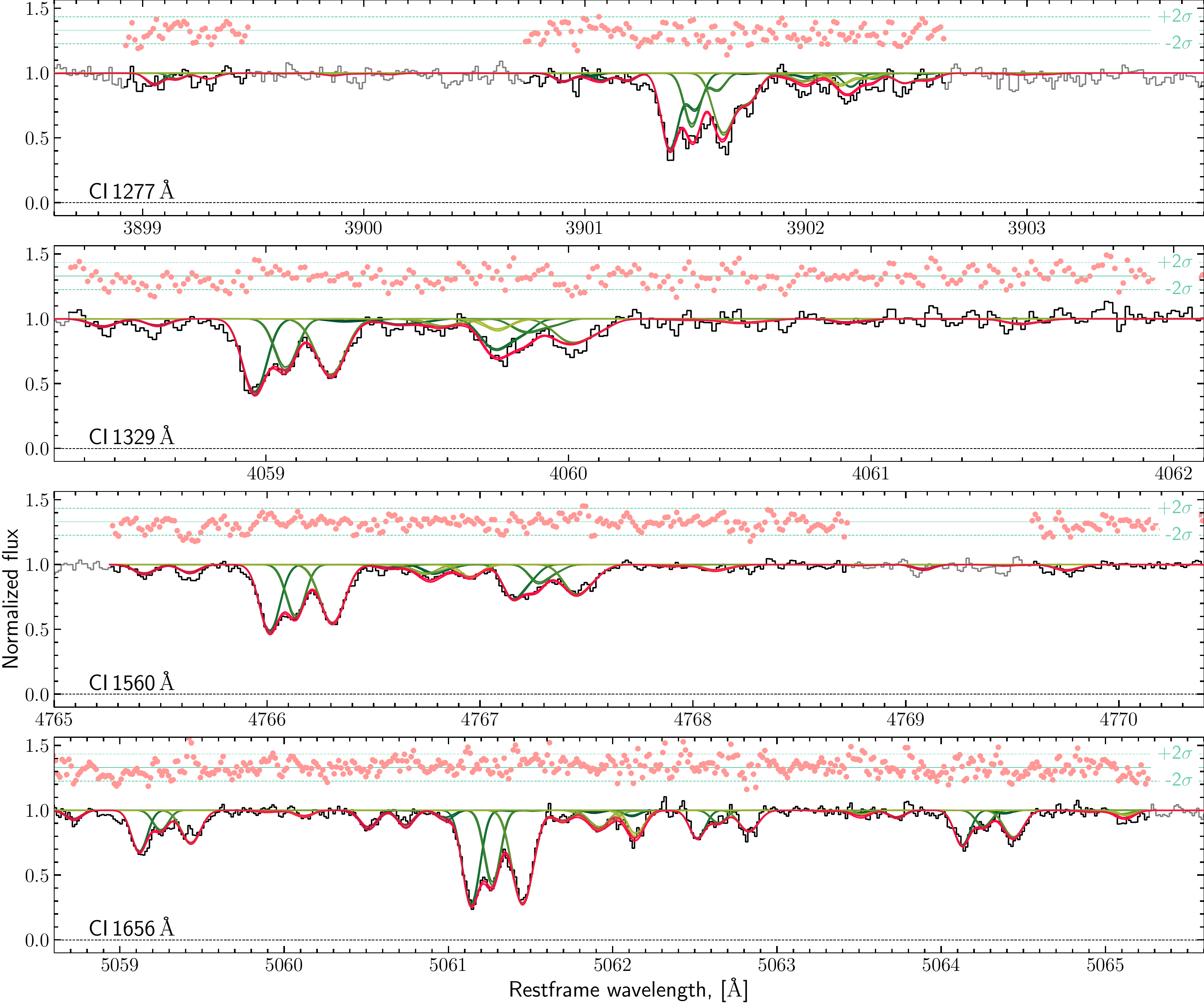}
    \caption{Fit to \ion{C}{I} absorption lines in DLA at 2.05 towards J\,2340$-$0053. The lines are the same as in Fig.~\ref{fig:J0136}.}
    \label{fig:J2340_CI}
\end{figure*}

\begin{figure*}
    \centering
    \includegraphics[width=\textwidth]{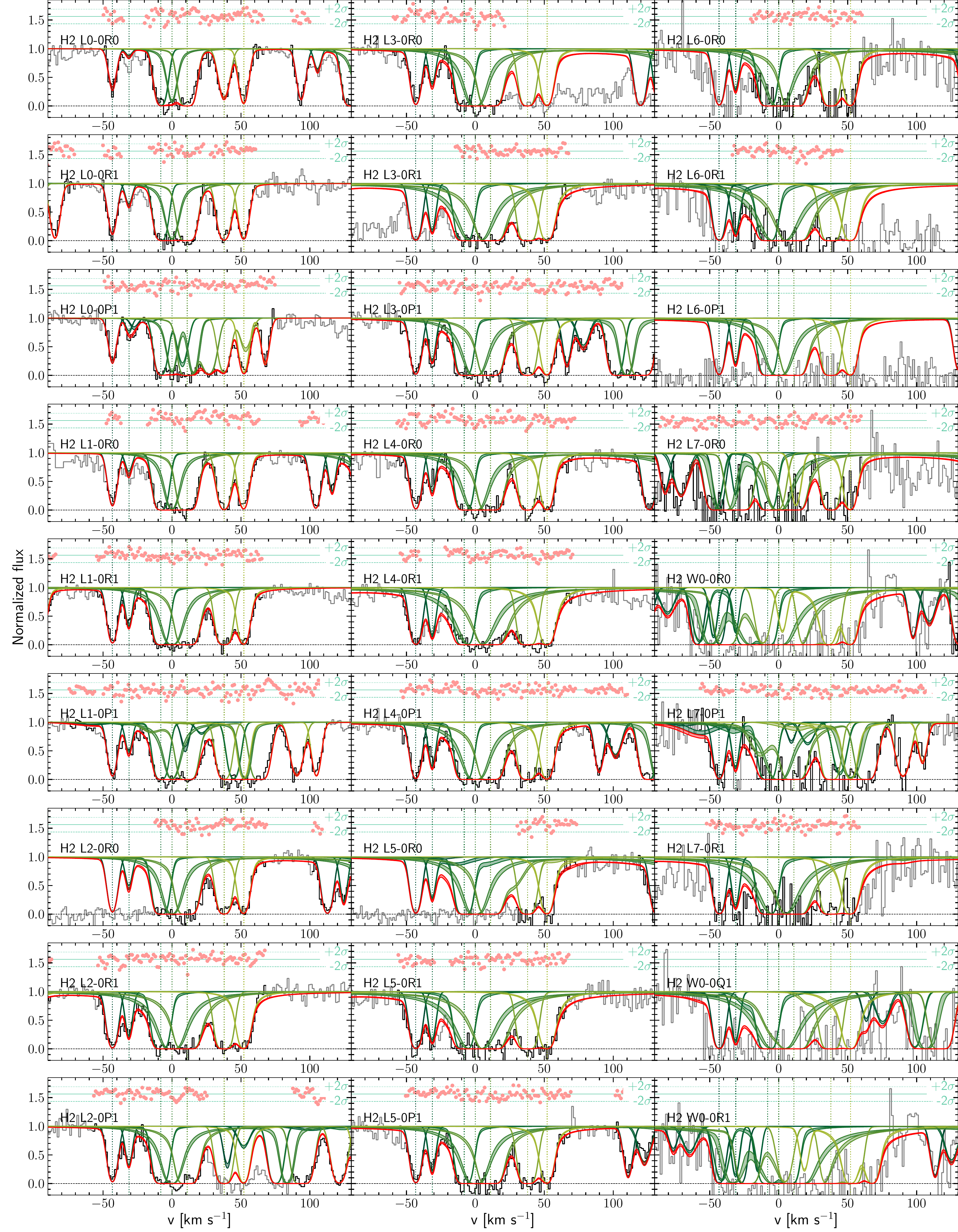}
    \caption{Fit to $\rm H_2$ absorption lines of transitions from $J=0$ and $J=1$ rotational levels in DLA at 2.05 towards J\,2340$-$0053. The lines are the same as in Fig.~\ref{fig:J0136}.}
    \label{fig:J2340_H2_J01}
\end{figure*}

\begin{figure*}
    \centering
    \includegraphics[width=\textwidth]{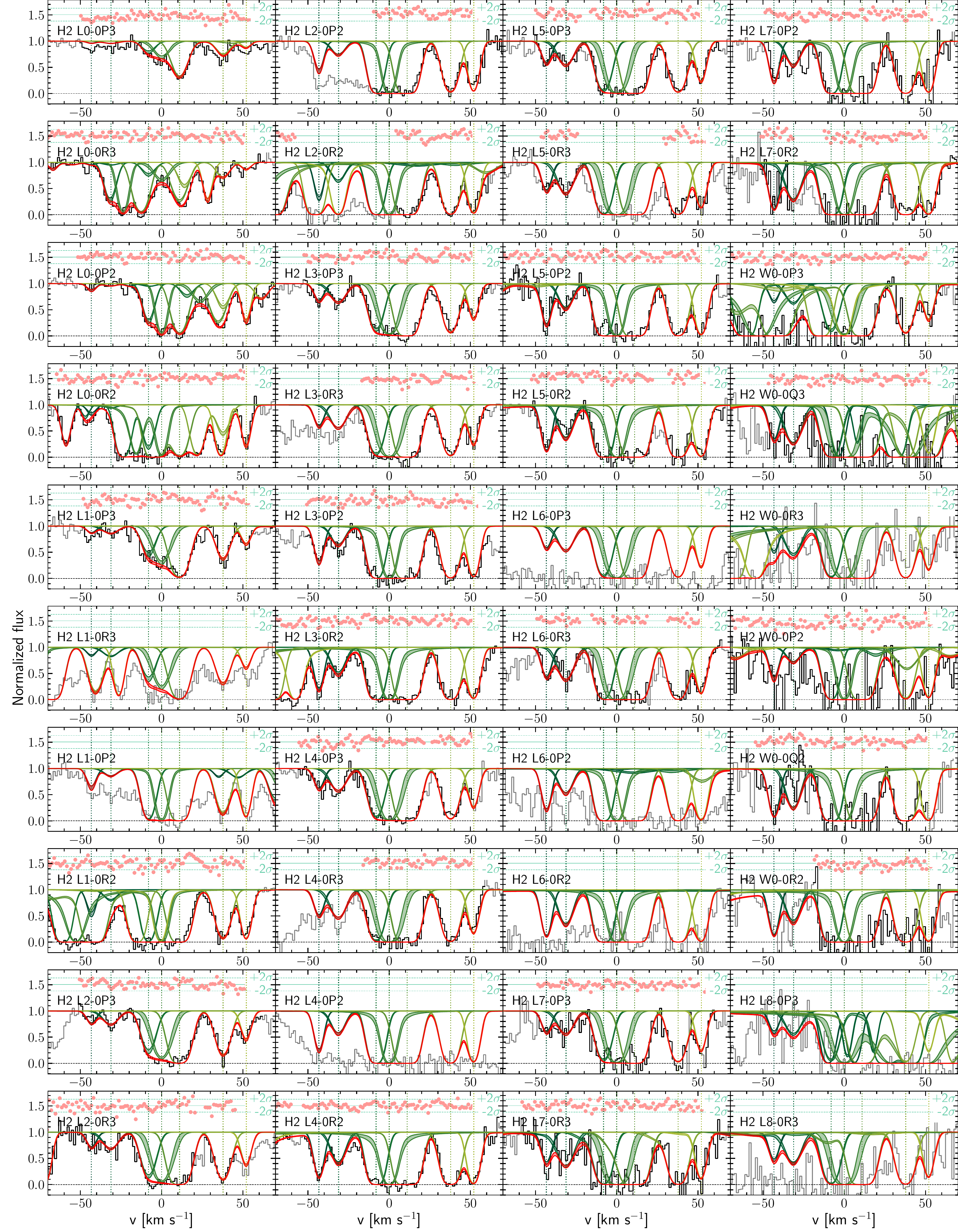}
    \caption{Fit to $\rm H_2$ absorption lines of transitions from $J=2$ and $J=3$ rotational levels in DLA at 2.05 towards J\,2340$-$0053. The lines are the same as in Fig.~\ref{fig:J0136}.}
    \label{fig:J2340_H2_J23}
\end{figure*}

\begin{figure*}
    \centering
    \includegraphics[width=\textwidth]{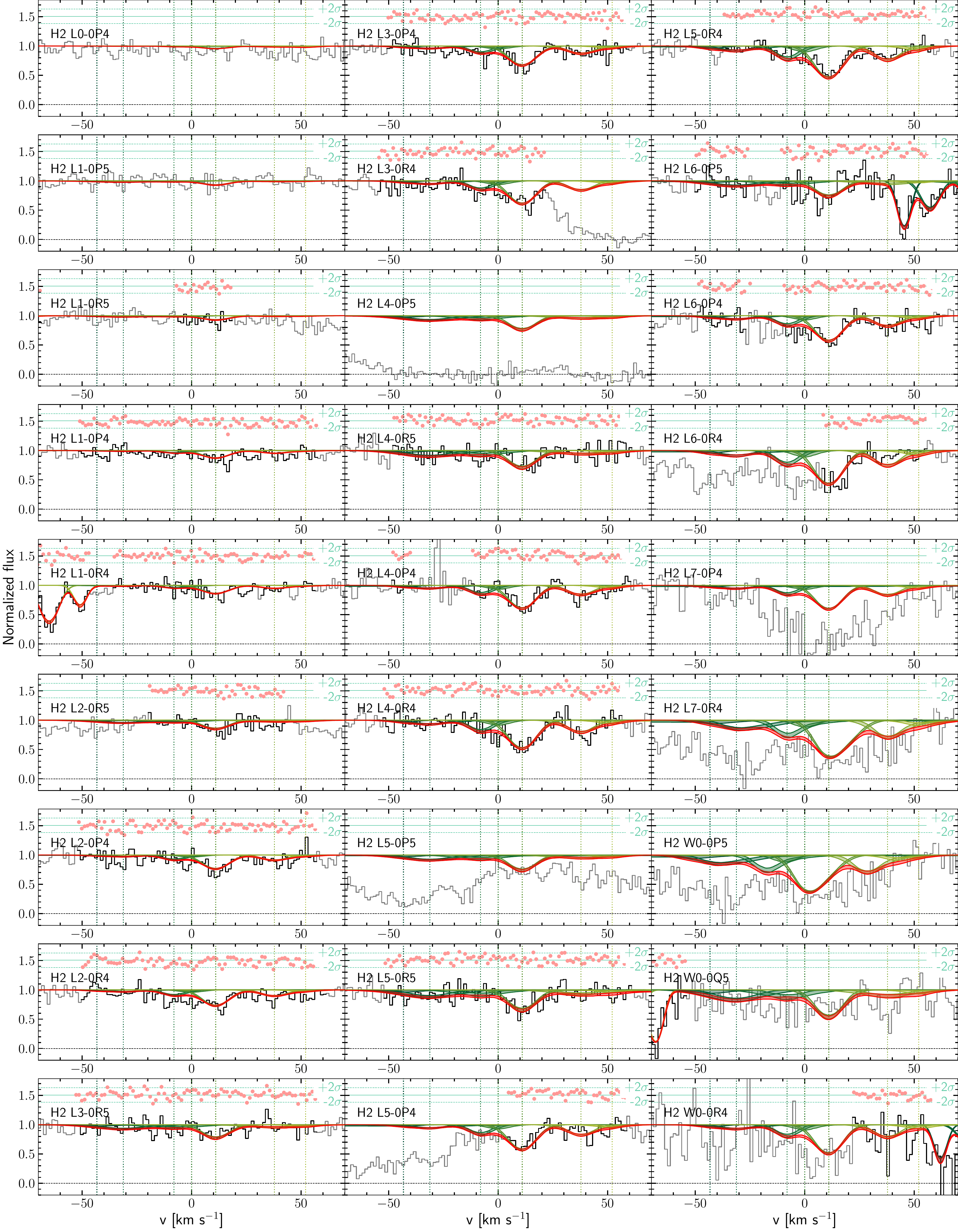}
    \caption{Fit to $\rm H_2$ absorption lines of transitions from $J=4$ and $J=5$ rotational levels in DLA at 2.05 towards J\,2340$-$0053. The lines are the same as in Fig.~\ref{fig:J0136}.}
    \label{fig:J2340_H2_J45}
\end{figure*}

In this section we present detailed results of the fitting of $\rm H_2$ and \CI\ in the DLA at z=2.05 towards J\,2340$-$0053 using 7-component model. Table~\ref{tab:J2340_results} provided the fitted values, while Figures~\ref{fig:J2340_CI} and \ref{fig:J2340_H2_J01}-\ref{fig:J2340_H2_J45} show \ion{C}{I} and H$_2$ absorption lines, respectively.

\begin{landscape}
\begin{table}
\caption{Fit results of H$_2$ in DLAs at $z = 2.055$ towards J\,2340$-$0053 with seven components.}
\label{tab:J2340_results}
\begin{tabular}{cccccccccc}
\hline
\hline
& comp & 1 & 2 & 3 & 4 & 5 & 6 & 7 & $\log N_{\rm tot}$ \\
\hline
& $z$ & $2.0541703(^{+6}_{-4})$ & $2.0542913(^{+4}_{-9})$ & $2.054528(^{+3}_{-3})$ & $2.054610(^{+1}_{-1})$ & $2.054723(^{+3}_{-3})$ & $2.0549952(^{+5}_{-4})$ & $2.0551398(^{+6}_{-4})$ &  \\
& $\Delta$v, km/s & -54.2 & -42.3 & -19.1 & -11.0 & 0.0 & 26.7 & 40.9 &  \\
\hline
& $b$, km/s & $2.4^{+0.7}_{-0.6}$ & $1.6^{+0.6}_{-0.1}$ & $1.9^{+0.1}_{-0.1}$ & $1.5^{+0.2}_{-0.1}$ & $3.3^{+0.2}_{-0.1}$ & $3.9^{+1.3}_{-0.9}$ & $3.0^{+0.9}_{-0.8}$ &  \\
\CI & $\log N$ & $12.27^{+0.03}_{-0.05}$ & $12.17^{+0.04}_{-0.04}$ & $13.51^{+0.04}_{-0.03}$ & $13.20^{+0.05}_{-0.04}$ & $13.29^{+0.01}_{-0.01}$ & $12.34^{+0.05}_{-0.05}$ & $12.42^{+0.05}_{-0.02}$ & $13.89^{+0.01}_{-0.02}$ \\
\CI$^*$ & $\log N$ & $11.97^{+0.08}_{-0.13}$ & $12.29^{+0.05}_{-0.05}$ & $13.06^{+0.01}_{-0.01}$ & $12.67^{+0.02}_{-0.03}$ & $12.97^{+0.01}_{-0.01}$ & $11.72^{+0.15}_{-0.22}$ & $12.18^{+0.06}_{-0.07}$ & $13.48^{+0.01}_{-0.01}$ \\
\CI$^{**}$ & $\log N$ & $<10.7$ & $12.06^{+0.07}_{-0.09}$ & $<11.0$ & $<11.0$ & $12.19^{+0.06}_{-0.05}$ & $<11.0$ & $<10.6$ & $12.47^{+0.05}_{-0.04}$ \\
& $\log N_{\rm tot}$ & $12.46^{+0.02}_{-0.06}$ & $12.66^{+0.03}_{-0.03}$ & $13.66^{+0.02}_{-0.03}$ & $13.32^{+0.04}_{-0.04}$ & $13.48^{+0.01}_{-0.01}$ & $12.43^{+0.06}_{-0.06}$ & $12.61^{+0.04}_{-0.03}$ & $14.04^{+0.01}_{-0.01}$ \\
cf$\dagger$ & & & & $0.86^{+0.02}_{-0.02}$ & $0.94^{+0.04}_{-0.02}$ & $0.92^{+0.03}_{-0.02}$ & & & \\
\hline
H$_2$ $J=0$ & $\log N$ & $15.30^{+0.03}_{-0.07}$ & $14.22^{+0.09}_{-0.05}$ & $16.45^{+0.19}_{-0.14}$ & $17.58^{+0.09}_{-0.08}$ & $17.50^{+0.05}_{-0.07}$ & $15.59^{+0.04}_{-0.05}$ & $16.78^{+0.05}_{-0.07}$ & $17.90^{+0.03}_{-0.03}$ \\
& $b$, km/s & $2.5^{+0.1}_{-0.1}$ & $1.7^{+0.1}_{-0.2}$ & $3.0^{+0.1}_{-0.2}$ & $1.0^{+0.1}_{-0.3}$ & $3.1^{+0.1}_{-0.1}$ & $3.8^{+0.1}_{-0.1}$ & $1.8^{+0.1}_{-0.1}$ &  \\
H$_2$ $J=1$ & $\log N$ & $15.85^{+0.05}_{-0.04}$ & $14.99^{+0.06}_{-0.04}$ & $16.90^{+0.18}_{-0.09}$ & $18.13^{+0.06}_{-0.06}$ & $18.03^{+0.04}_{-0.04}$ & $16.28^{+0.04}_{-0.03}$ & $17.31^{+0.03}_{-0.05}$ & $18.43^{+0.02}_{-0.02}$ \\
& $b$, km/s & $\ddagger$ & $\ddagger$ & $\ddagger$ & $\ddagger$ & $\ddagger$ & $\ddagger$ & $\ddagger$ &  \\
H$_2$ $J=2$ &  $\log N$ & $14.67^{+0.06}_{-0.03}$ & $14.36^{+0.02}_{-0.03}$ & $15.74^{+0.06}_{-0.07}$ & $17.15^{+0.14}_{-0.21}$ & $16.20^{+0.04}_{-0.06}$ & $15.45^{+0.03}_{-0.03}$ & $15.72^{+0.06}_{-0.06}$ & $17.20^{+0.14}_{-0.14}$ \\
& $b$, km/s & $2.4^{+0.2}_{-0.1}$ & $4.8^{+0.6}_{-0.4}$ & $3.6^{+0.2}_{-0.1}$ & $1.9^{+0.2}_{-0.3}$ & $4.8^{+0.1}_{-0.1}$ & $4.4^{+0.2}_{-0.1}$ & $2.3^{+0.1}_{-0.1}$ &  \\
H$_2$ $J=3$ & $\log N$ & $14.15^{+0.03}_{-0.04}$ & $14.38^{+0.02}_{-0.02}$ & $15.03^{+0.04}_{-0.06}$ & $15.28^{+0.04}_{-0.08}$ & $15.76^{+0.02}_{-0.02}$ & $15.14^{+0.01}_{-0.03}$ & $14.61^{+0.04}_{-0.02}$ & $16.03^{+0.02}_{-0.01}$ \\
& $b$, km/s & $2.8^{+0.3}_{-0.3}$ & $5.6^{+0.5}_{-0.5}$ & $4.6^{+0.3}_{-0.3}$ & $5.5^{+1.2}_{-0.9}$ & $5.4^{+0.1}_{-0.1}$ & $4.7^{+0.1}_{-0.2}$ & $2.8^{+0.2}_{-0.2}$ &  \\
H$_2$ $J=4$ & $\log N$ & $12.62^{+0.65}_{-0.46}$ & $13.68^{+0.06}_{-0.13}$ & $13.88^{+0.06}_{-0.08}$ & $13.41^{+0.17}_{-0.34}$ & $14.43^{+0.02}_{-0.03}$ & $14.00^{+0.04}_{-0.06}$ & $13.33^{+0.18}_{-0.19}$ & $14.73^{+0.01}_{-0.02}$ \\
& $b$, km/s & $6.2^{+4.0}_{-2.2}$ & $12.0^{+2.7}_{-5.3}$ & $6.4^{+1.2}_{-1.1}$ & $5.8^{+2.3}_{-1.1}$ & $7.7^{+0.5}_{-0.5}$ & $7.6^{+1.6}_{-0.9}$ & $7.3^{+2.2}_{-3.4}$ &  \\
H$_2$ $J=5$ & $\log N$ & $12.35^{+0.94}_{-0.58}$ & $13.93^{+0.12}_{-0.12}$ & $13.60^{+0.17}_{-0.13}$ & $12.31^{+0.43}_{-0.83}$ & $14.15^{+0.05}_{-0.03}$ & $13.66^{+0.12}_{-0.15}$ & $13.07^{+0.31}_{-0.35}$ & $14.54^{+0.02}_{-0.04}$ \\
& $b$, km/s & $12.9^{+3.4}_{-5.2}$ & $17.6^{+1.8}_{-4.2}$ & $7.9^{+5.9}_{-1.9}$ & $10.0^{+2.1}_{-3.8}$ & $7.5^{+0.8}_{-0.5}$ & $15.8^{+3.2}_{-3.3}$ & $10.4^{+1.3}_{-5.2}$ &  \\
\hline
& $\log N_{\rm tot}$ & $15.99^{+0.04}_{-0.04}$ & $15.24^{+0.04}_{-0.03}$ & $17.11^{+0.12}_{-0.14}$ & $18.27^{+0.06}_{-0.06}$ & $18.14^{+0.04}_{-0.04}$ & $16.43^{+0.03}_{-0.03}$ & $17.43^{+0.04}_{-0.05}$ & $18.57^{+0.02}_{-0.02}$ \\
\hline
HD $J=0$ & $\log N$ & $\lesssim 13.51$ & $\lesssim 12.69$ & $\lesssim 13.76$ & $13.60^{+0.15}_{-0.14}$ & $13.84^{+0.05}_{-0.05}$ & $\lesssim 12.58$ & $13.29^{+0.15}_{-0.21}$ & $14.11^{+0.06}_{-0.06}$ \\
 & $b$, km/s & $2.5^{+0.1}_{-0.1}$ & $1.7^{+0.1}_{-0.2}$ & $3.0^{+0.1}_{-0.2}$ & $1.0^{+0.1}_{-0.3}$ & $3.1^{+0.1}_{-0.1}$ & $3.8^{+0.1}_{-0.1}$ & $1.8^{+0.1}_{-0.1}$ &  \\
 \hline
 HD/2H$_2$ & & $\lesssim 1.7\times 10^{-3}$ & $\lesssim 1.4\times 10^{-3}$ & $\lesssim 2.2\times 10^{-4}$ & $\left(1.1^{+0.5}_{-0.3}\right)\times 10^{-5}$ & $\left(2.5^{+0.4}_{-0.3}\right)\times 10^{-5}$ & $\lesssim 7.1\times 10^{-5}$ & $\left(3.6^{+1.6}_{-0.3}\right)\times 10^{-5}$ & $\left(1.7^{+0.3}_{-0.2}\right)\times 10^{-5}$ \\
\hline
\hline
$\log n$ & & & & $0.8^{+0.3}_{-0.3}$ & $0.6^{+0.3}_{-0.4}$ & $1.2^{+0.1}_{-0.1}$ & & $0.8^{+0.3}_{-0.5}$ \\
$\log \chi$ & & & & $0.1^{+0.2}_{-0.2}$ & $-0.2^{+0.2}_{-0.2}$ & $0.5^{+0.1}_{-0.1}$ & & $-0.2^{+0.2}_{-0.2}$ & \\
\hline
\hline
\end{tabular}
\begin{tablenotes}
\item $\dagger$ covering factors used for \ion{C}{i} $\sim1560$\,\AA\ lines. 
\item $\ddagger$ The Doppler parameters of H$_2$ $J=1$ was tied to H$_2$ $J=0$.
\end{tablenotes}
\end{table}
\end{landscape}

\section{1d and 2d Posteriors on physical conditions}

\begin{center}
\begin{figure*}
\begin{minipage}[h]{0.47\linewidth}
\center{\includegraphics[width=1\linewidth]{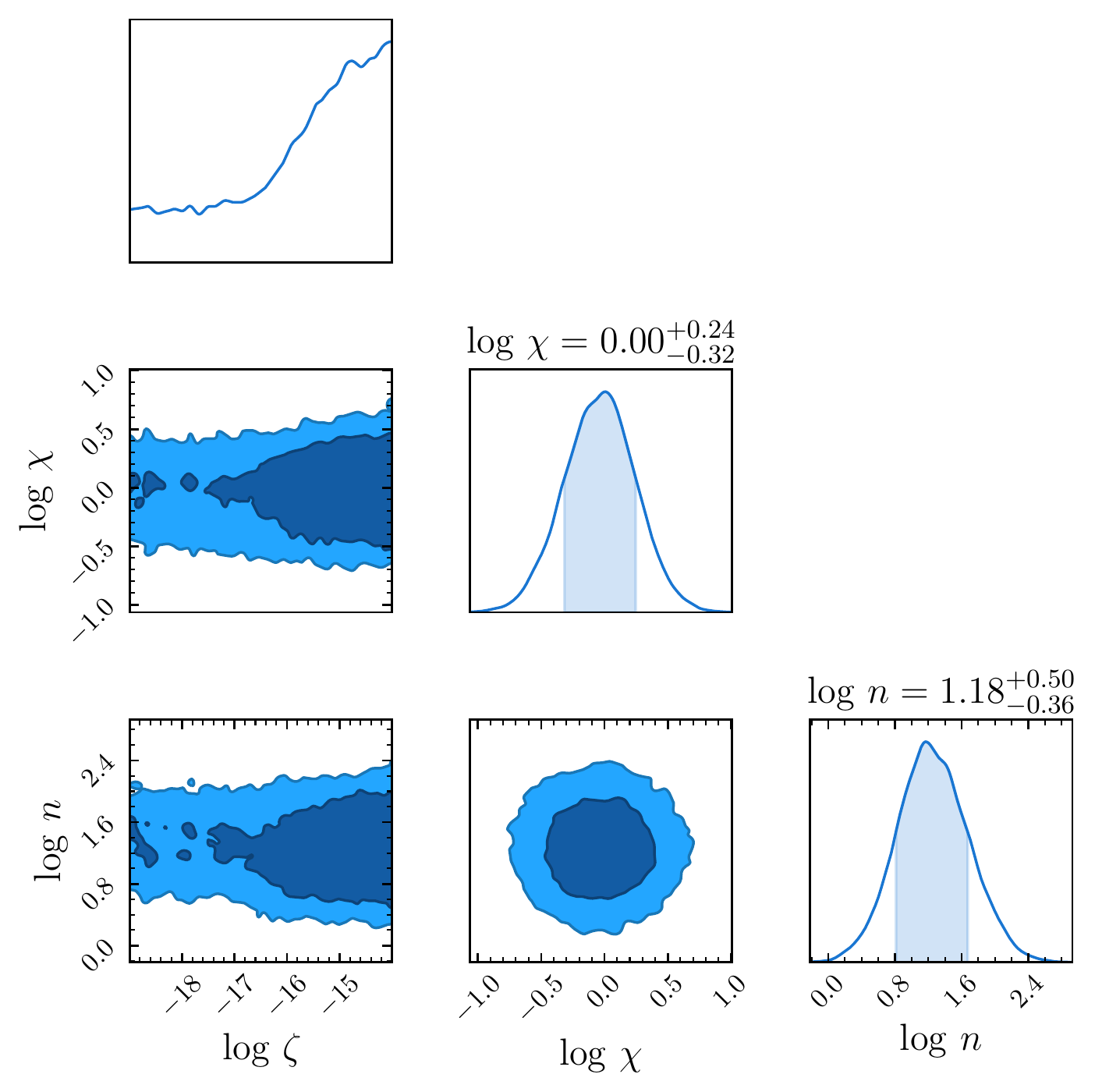}}  
{DLA at $z=2.526$ towards J\,0000+0048}
\end{minipage}
\hfill
\begin{minipage}[h]{0.47\linewidth}
\center{\includegraphics[width=1\linewidth]{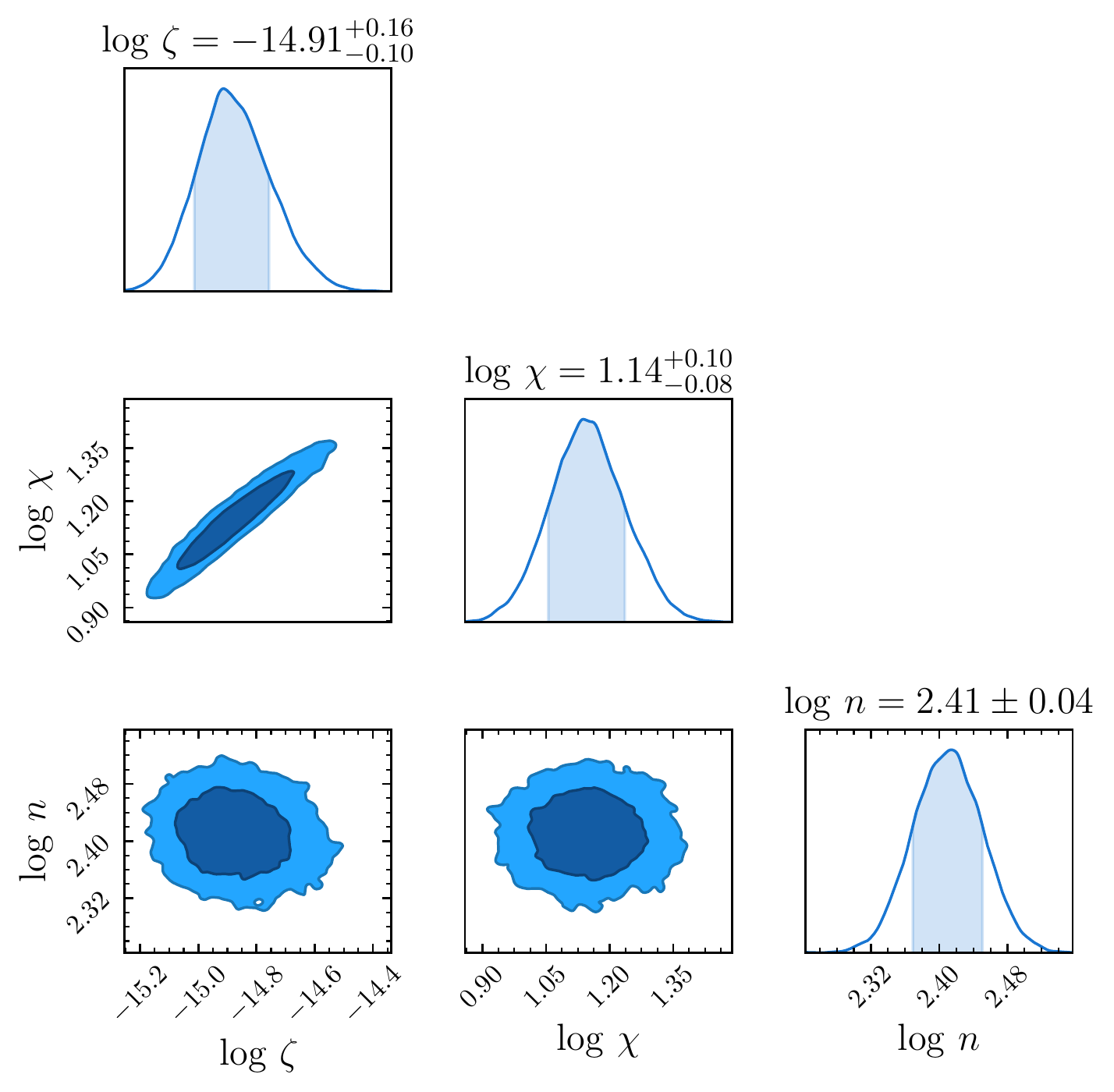}} 
{DLA at $z=2.811$ towards J\,0528$-$2505}
\end{minipage}
\vfill
\vspace{0.5cm}
\begin{minipage}[h]{0.47\linewidth}
\center{\includegraphics[width=1\linewidth]{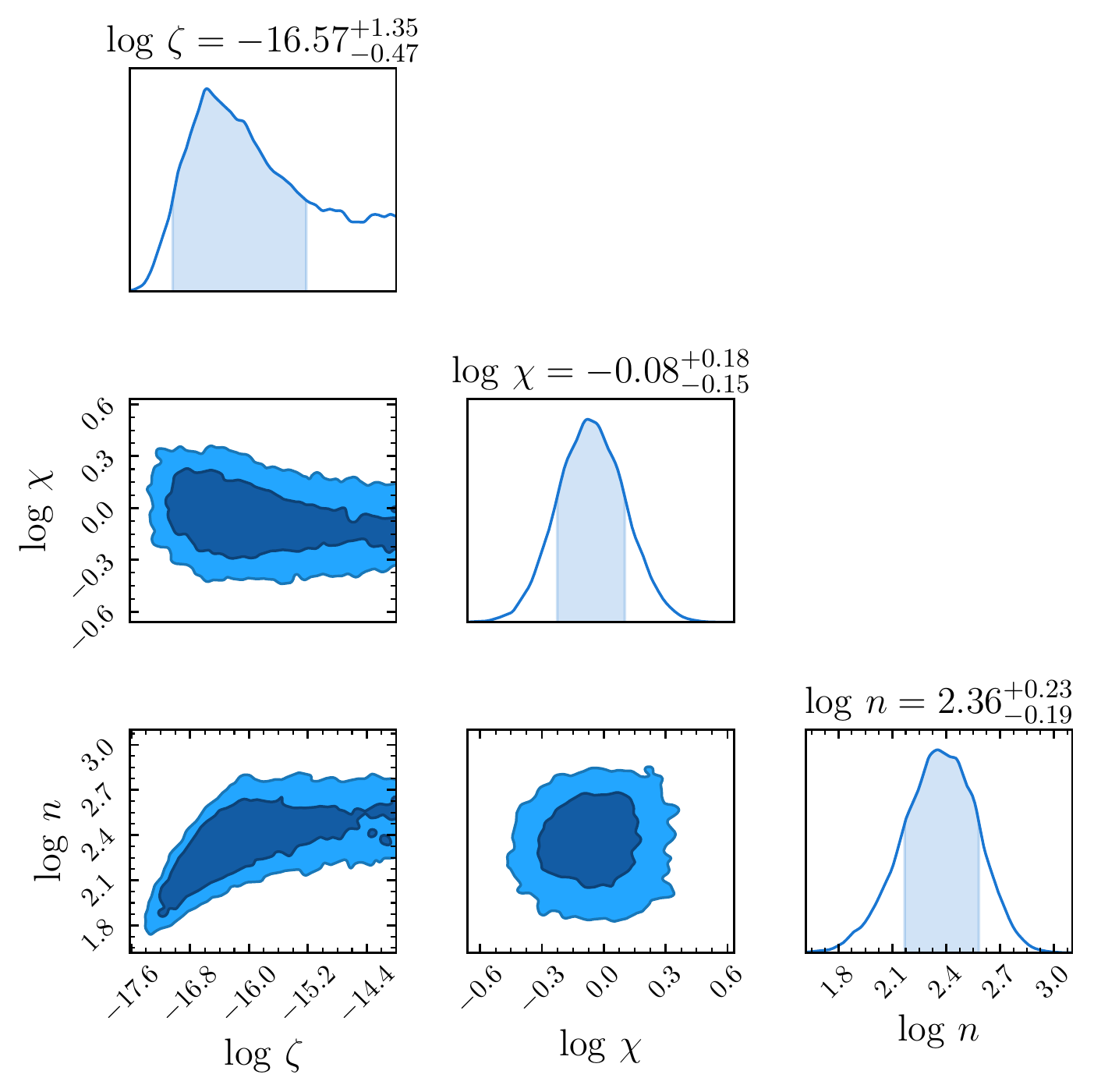}} 
{DLA at $z=2.6264$ towards J\,0812+3208 (comp 1)}
\end{minipage}
\hfill
\begin{minipage}[h]{0.47\linewidth}
\center{\includegraphics[width=1\linewidth]{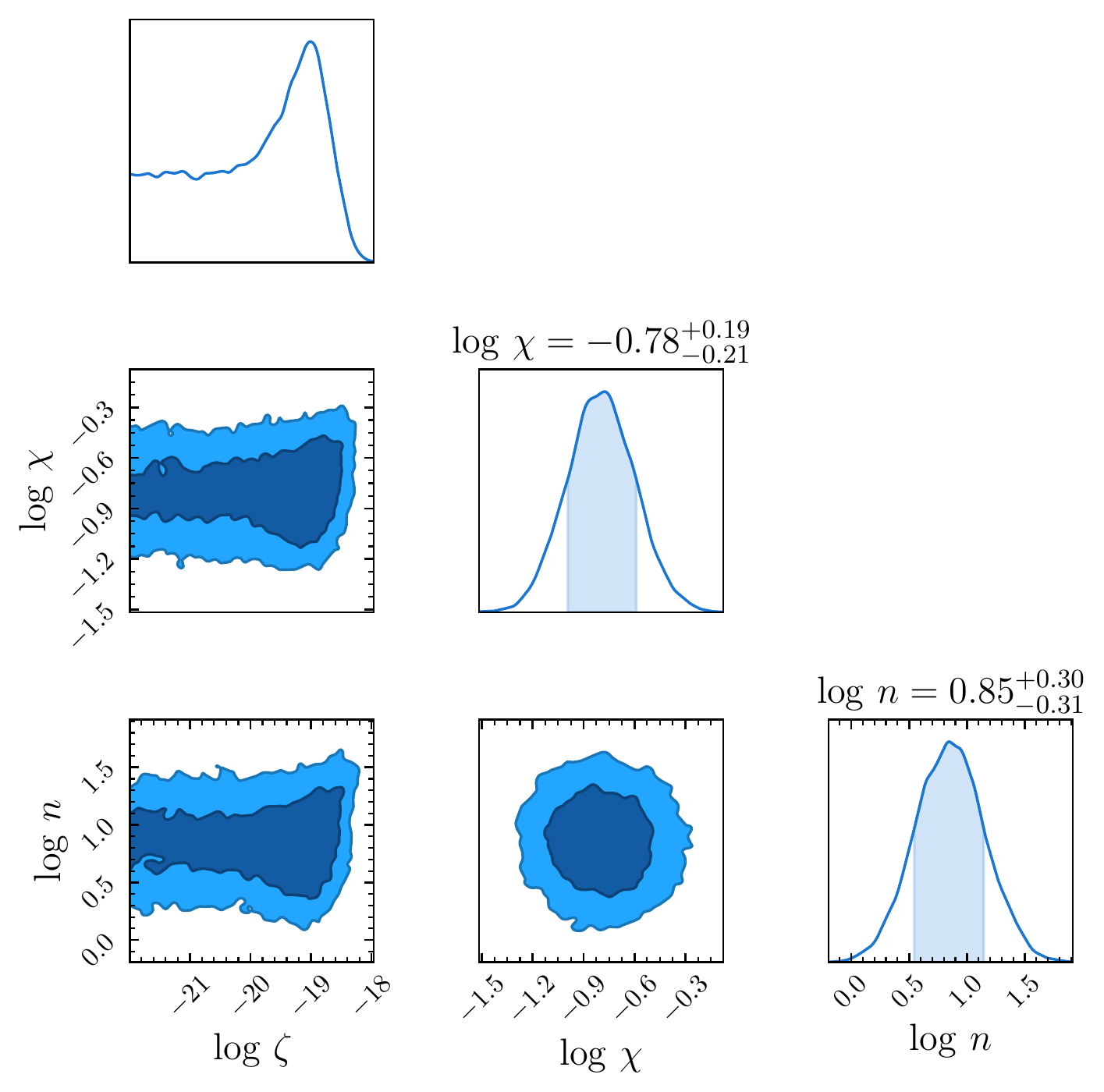}}
{DLA at $z=2.6263$ towards J\,0812+3208 (comp 2)}
\end{minipage}

\caption{The marginalized posterior probability functions for CRIR, UV field strength, and number density, obtained from fitting measured $N_{\rm HD}/N_{\rm H_2}$ in the systems at high redshift. The diagonal panels on each subplot indicate 1d posterior function, where the blue shade area corresponds to 0.68 confidence level. The non-diagonal panels show the 2d posterior function, where dark and light blue regions correspond to 0.68 and 0.95 confidence levels. }
\label{fig: MCMC_results1}
\end{figure*}
\end{center}

\begin{center}
\begin{figure*}
\begin{minipage}[h]{0.47\linewidth}
\center{\includegraphics[width=1\linewidth]{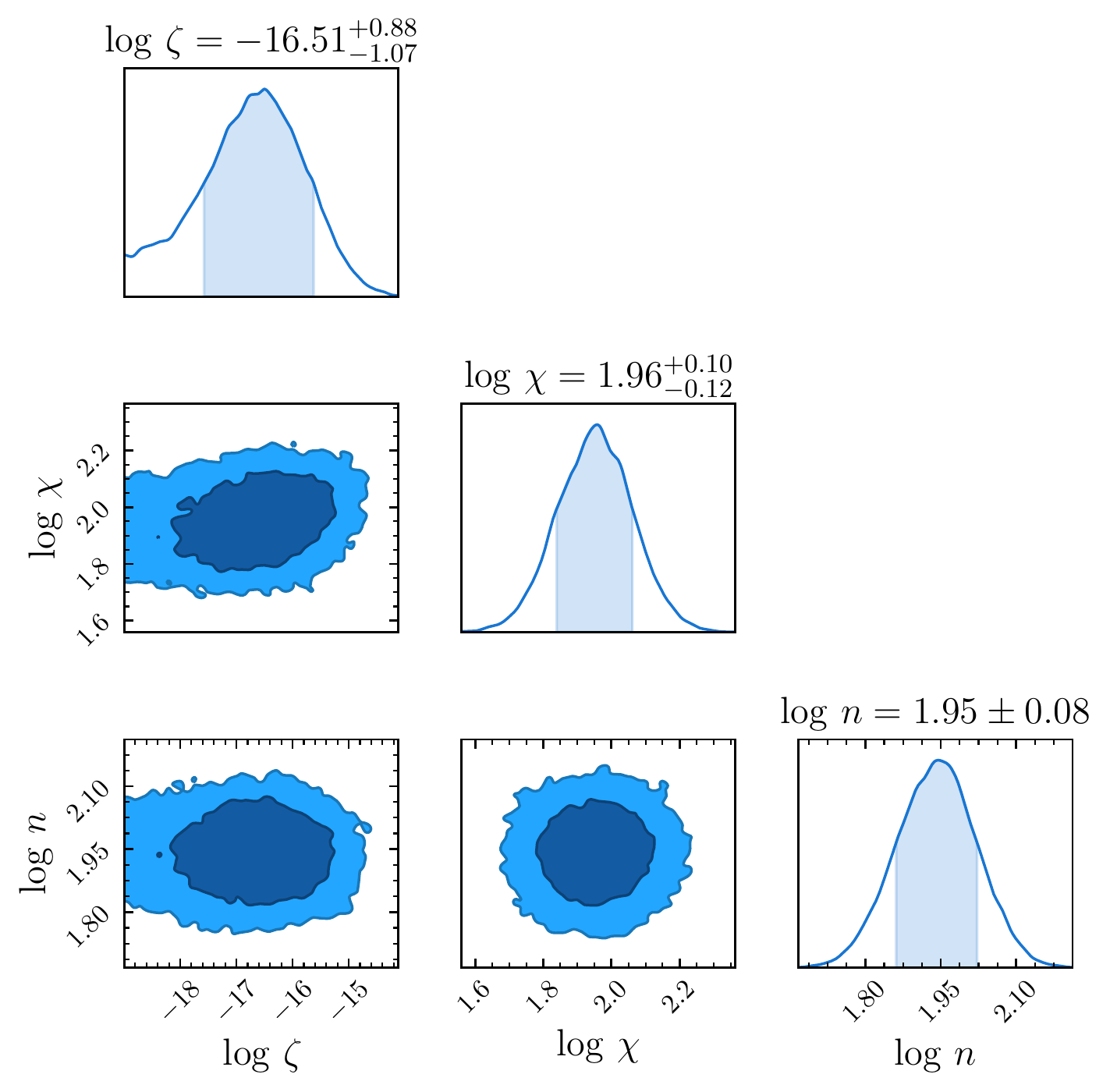}}  
{DLA at $z=2.786$ towards J\,0843+0221}
\end{minipage}
\hfill
\begin{minipage}[h]{0.47\linewidth}
\center{\includegraphics[width=1\linewidth]{figures/MCMC/J0858_MCMC2.pdf}} 
{DLA at $z=2.625$ towards J\,0858+1749}
\end{minipage}
\vfill
\vspace{0.5cm}
\begin{minipage}[h]{0.47\linewidth}
\center{\includegraphics[width=1\linewidth]{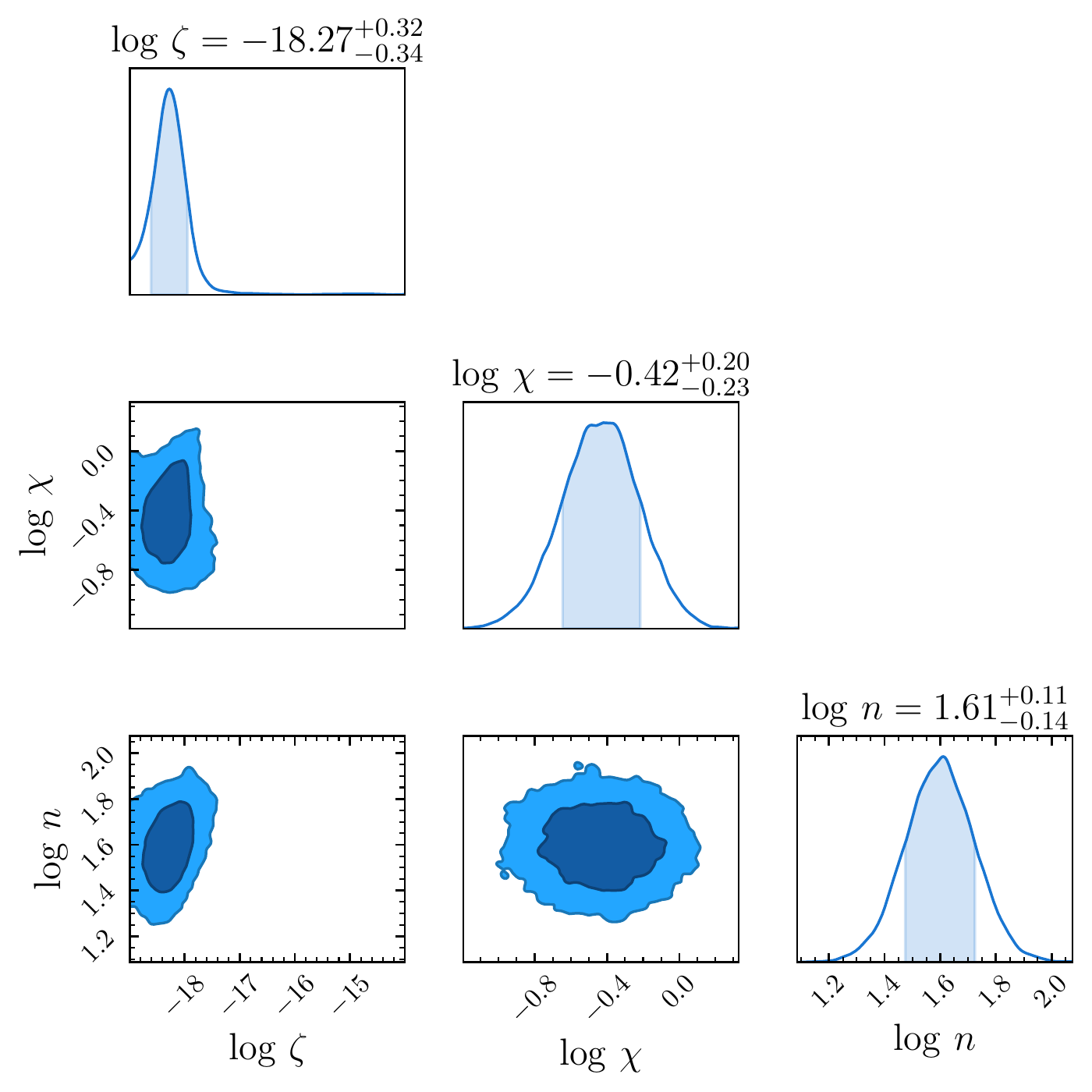}} 
{DLA at $z=2.338$ towards J\,1232+0815}
\end{minipage}
\hfill
\begin{minipage}[h]{0.47\linewidth}
\center{\includegraphics[width=1\linewidth]{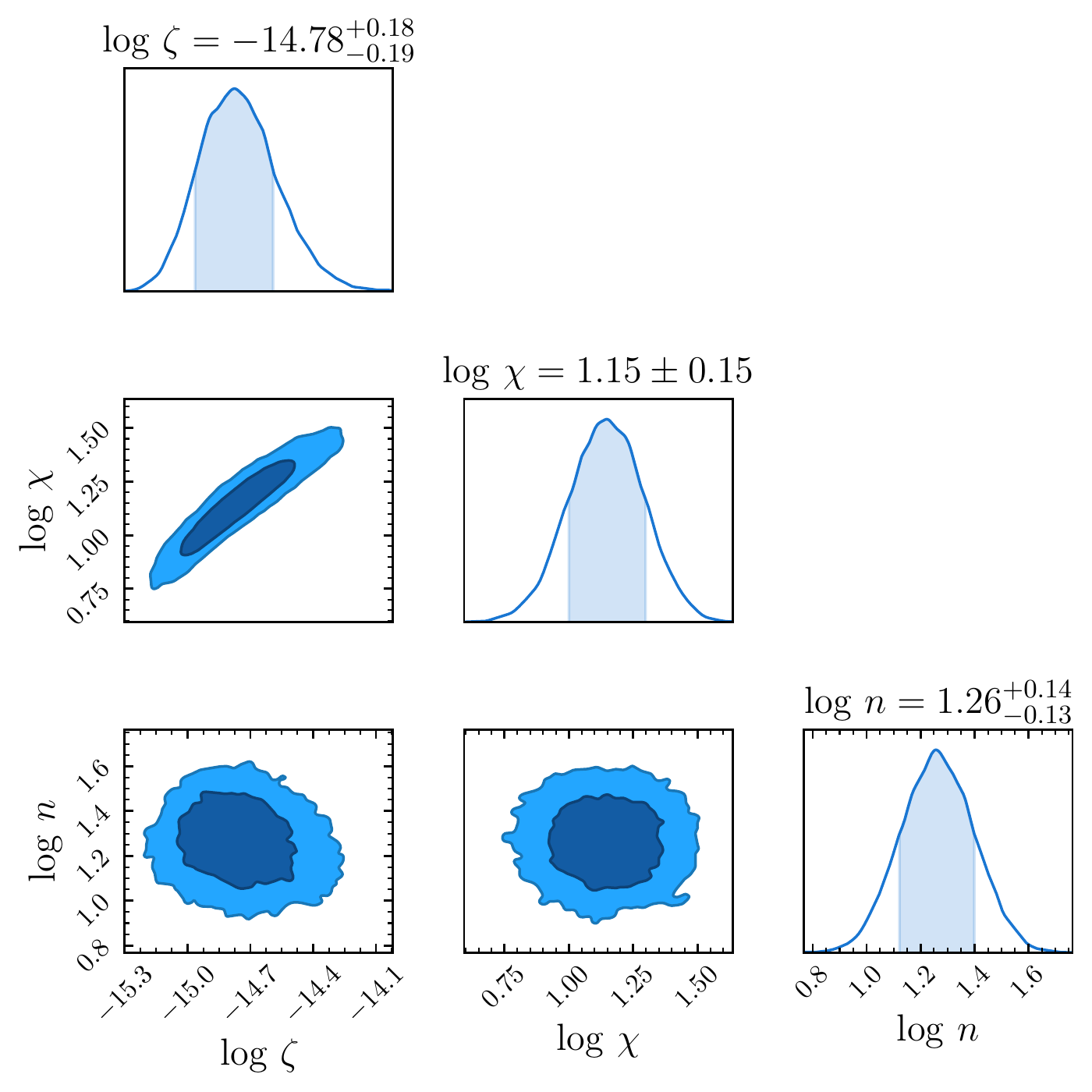}}  
{DLA at $z=2.6896$ towards J\,1237+0647}
\end{minipage}

\caption{The marginalized posterior probability functions for CRIR, UV field strength, and number density (continued).
}
\label{fig: MCMC_results2}
\end{figure*}
\end{center}

\begin{center}
\begin{figure*}
\begin{minipage}[h]{0.47\linewidth}
\center{\includegraphics[width=1\linewidth]{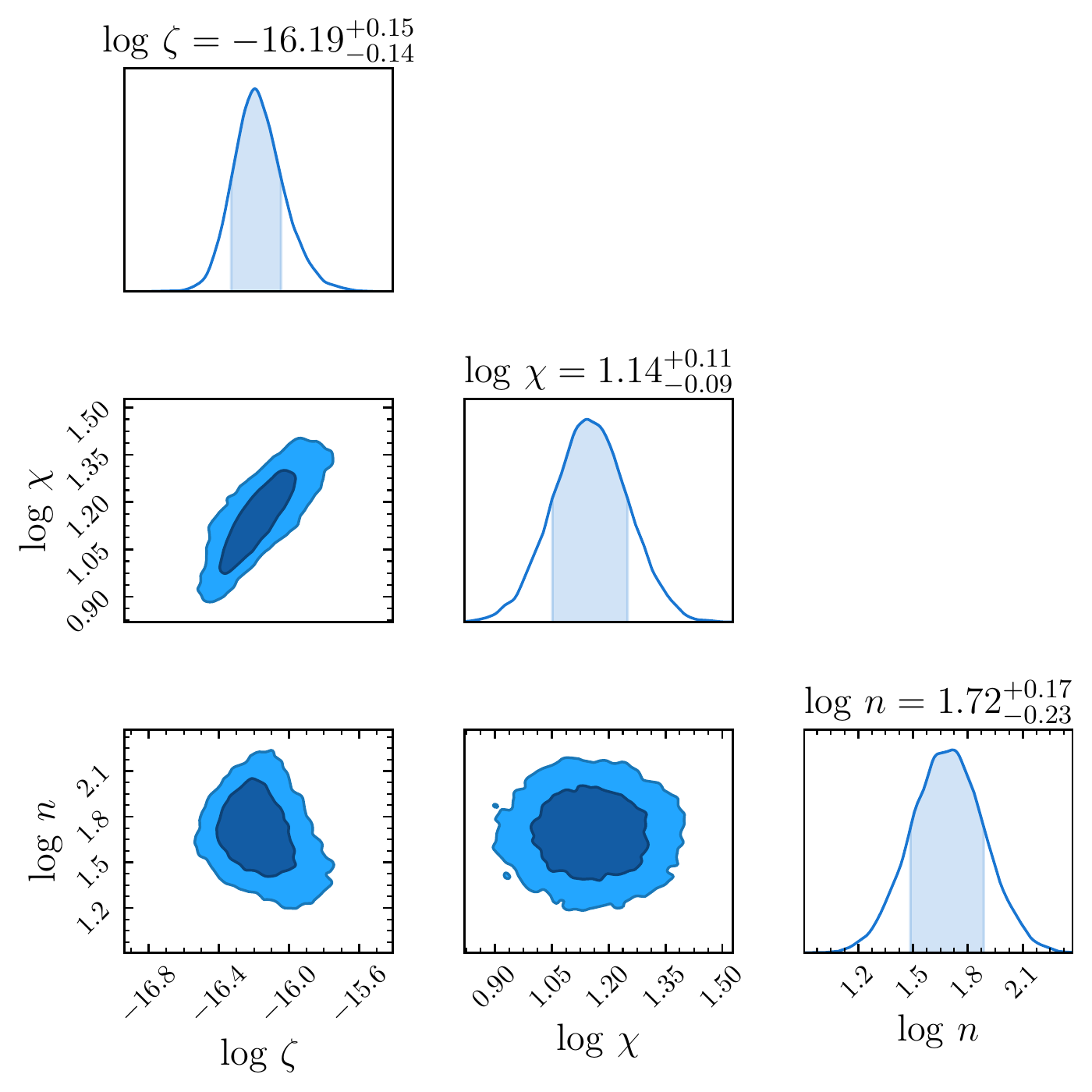}} 
{DLA at $z=3.0915$ towards J\,1311+2225 (comp 2)}
\end{minipage}
\hfill
\begin{minipage}[h]{0.47\linewidth}
\center{\includegraphics[width=1\linewidth]{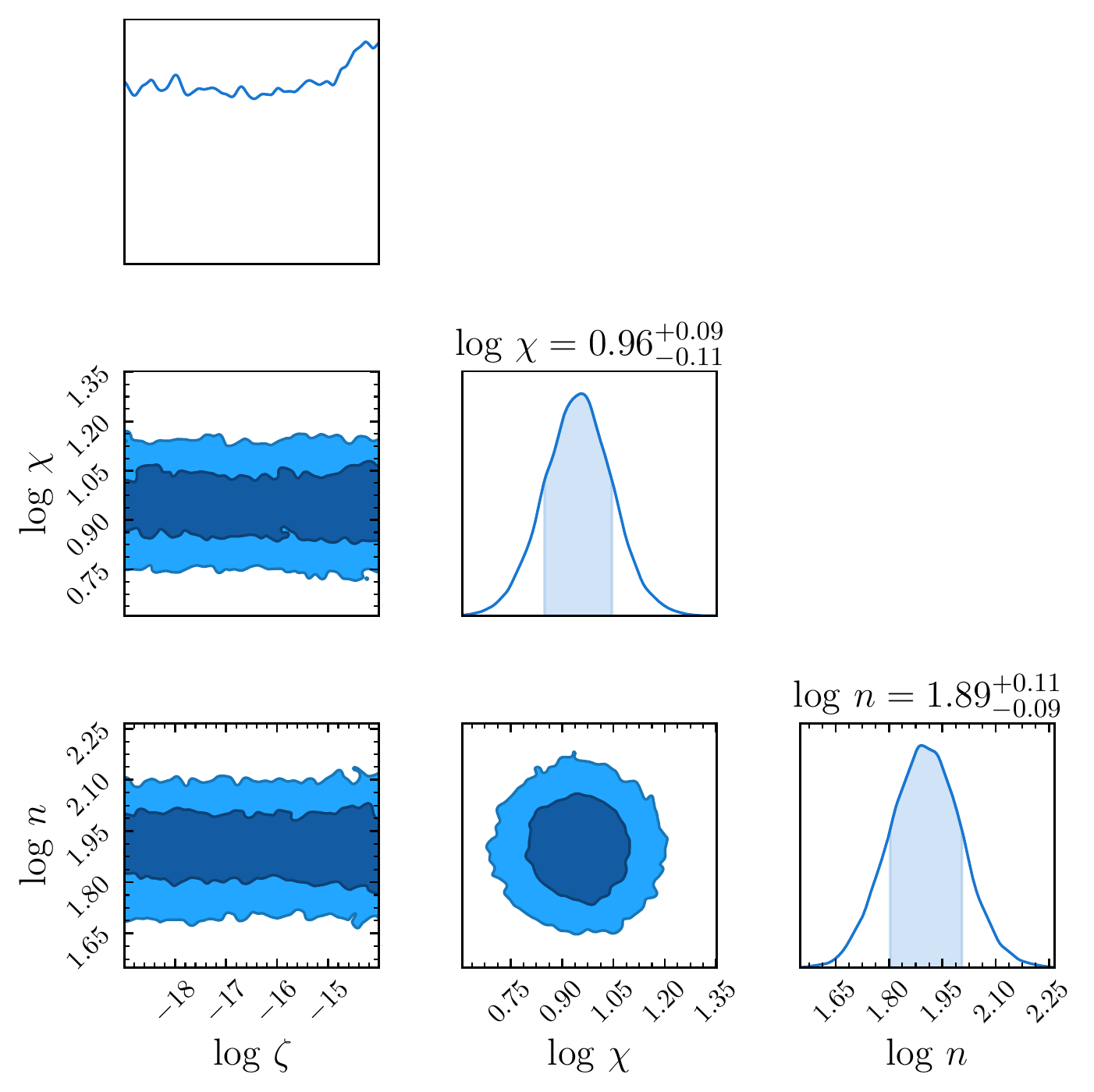}} 
{DLA at $z=3.0917$ towards J\,1311+2225 (comp 3)}
\end{minipage}
\vfill
\vspace{0.5cm}
\begin{minipage}[h]{0.47\linewidth}
\center{\includegraphics[width=1\linewidth]{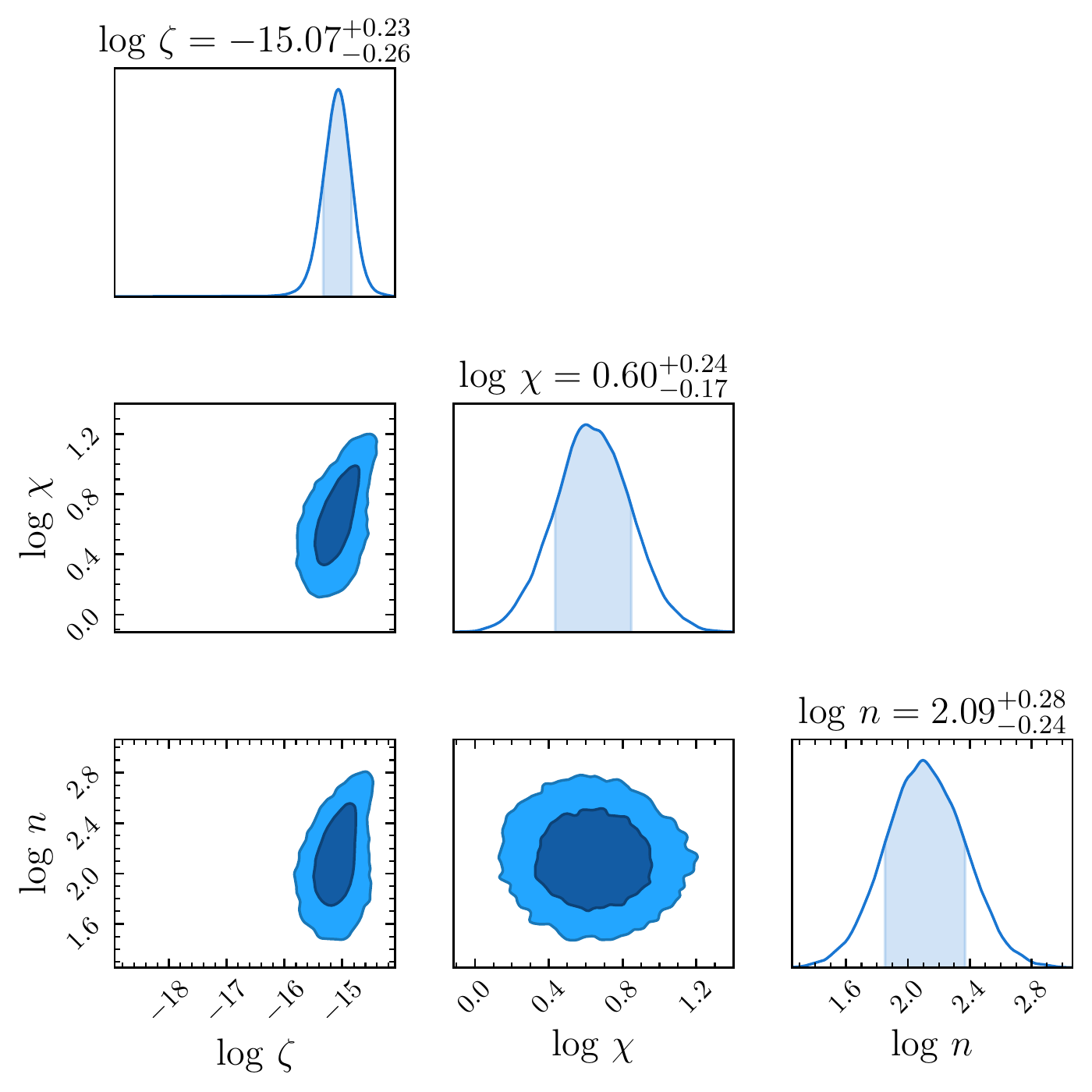}}
{DLA at $z=3.0919$ towards J\,1311+2225 (comp 4)}
\end{minipage}
\hfill
\begin{minipage}[h]{0.47\linewidth}
\center{\includegraphics[width=1\linewidth]{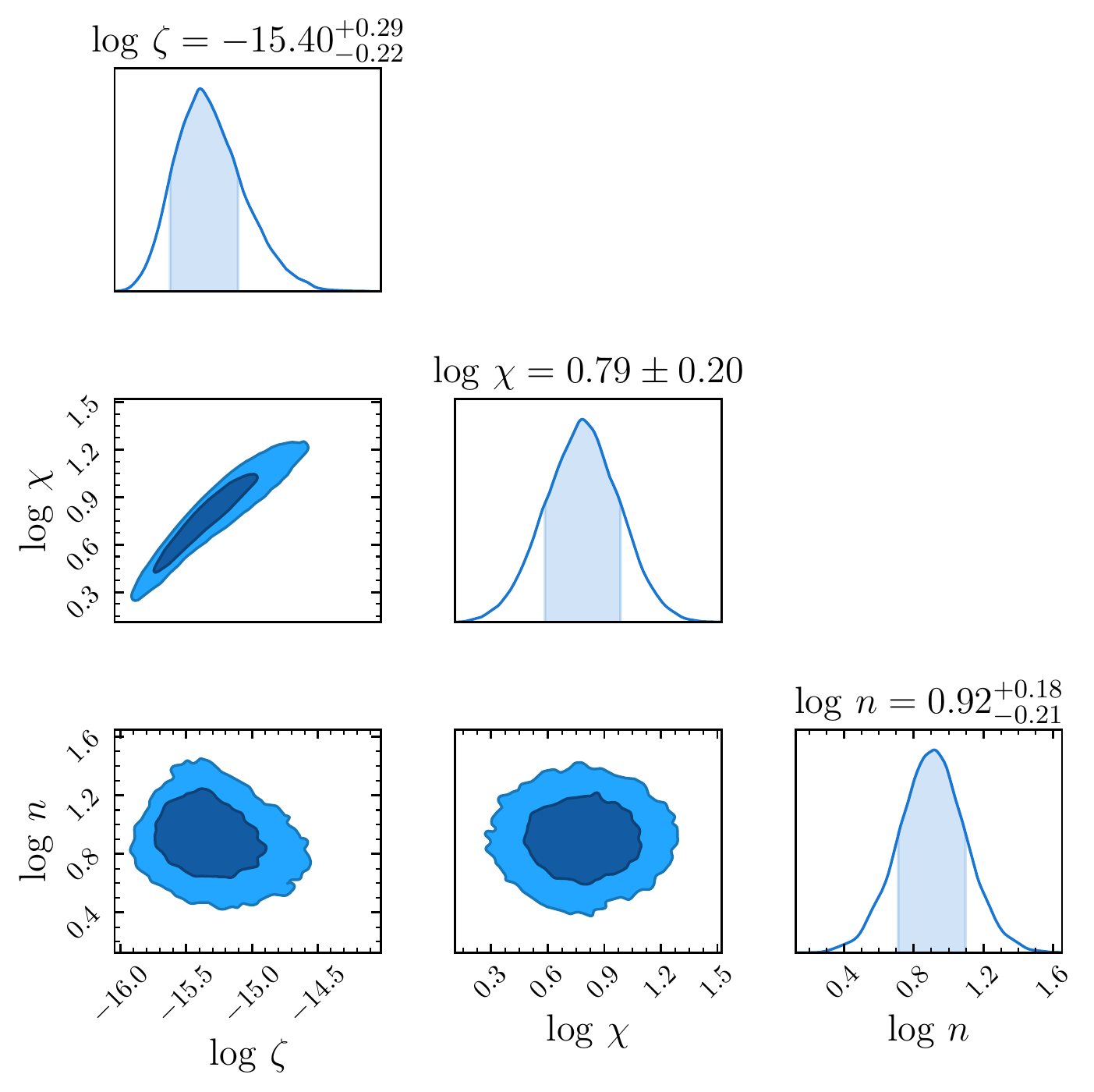}}  
{DLA at $z=2.418$ towards J\,1439+1117}
\end{minipage}

\caption{The marginalized posterior probability functions for CRIR, UV field strength, and number density (continued).
}
\label{fig: MCMC_results3}
\end{figure*}
\end{center}

\begin{center}
\begin{figure*}
\begin{minipage}[h]{0.47\linewidth}
\center{\includegraphics[width=1\linewidth]{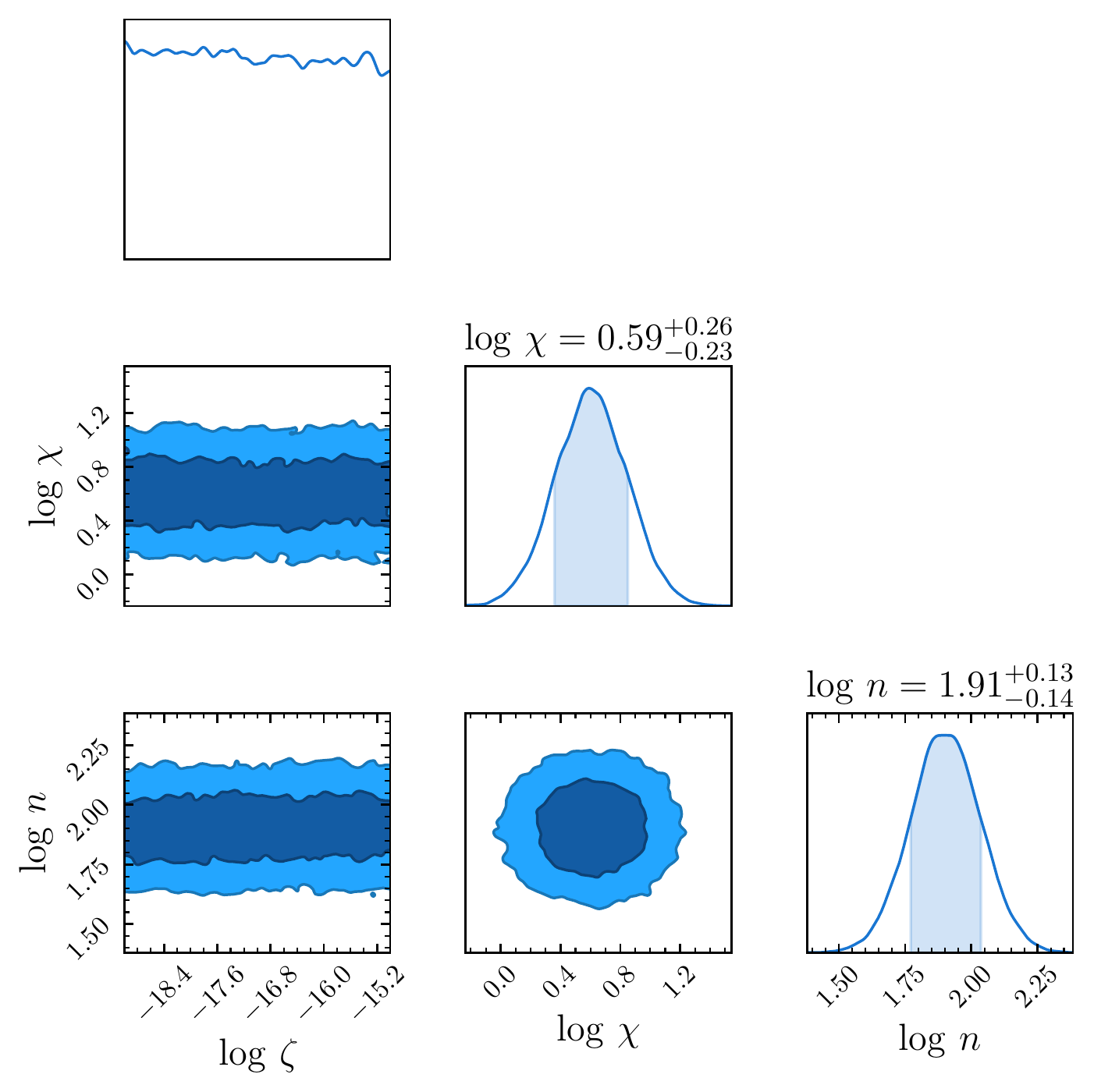}} 
{DLA at $z=2.464$ towards J\,1513+0352}
\end{minipage}
\hfill
\begin{minipage}[h]{0.47\linewidth}
\center{\includegraphics[width=1\linewidth]{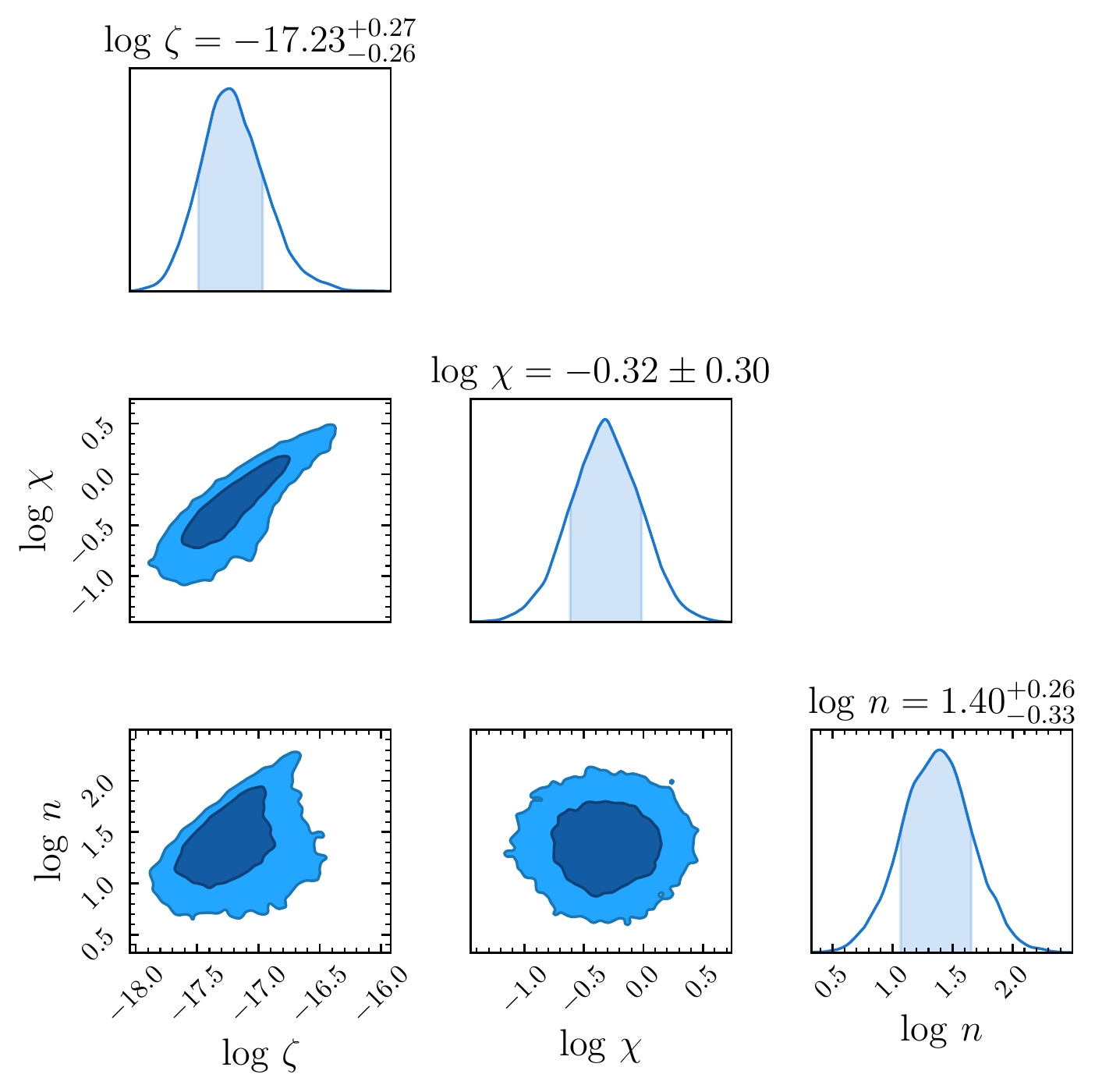}} 
{DLA at $z=3.091$ towards J\,2100+0641}
\end{minipage}
\vfill
\vspace{0.5cm}
\begin{minipage}[h]{0.47\linewidth}
\center{\includegraphics[width=1\linewidth]{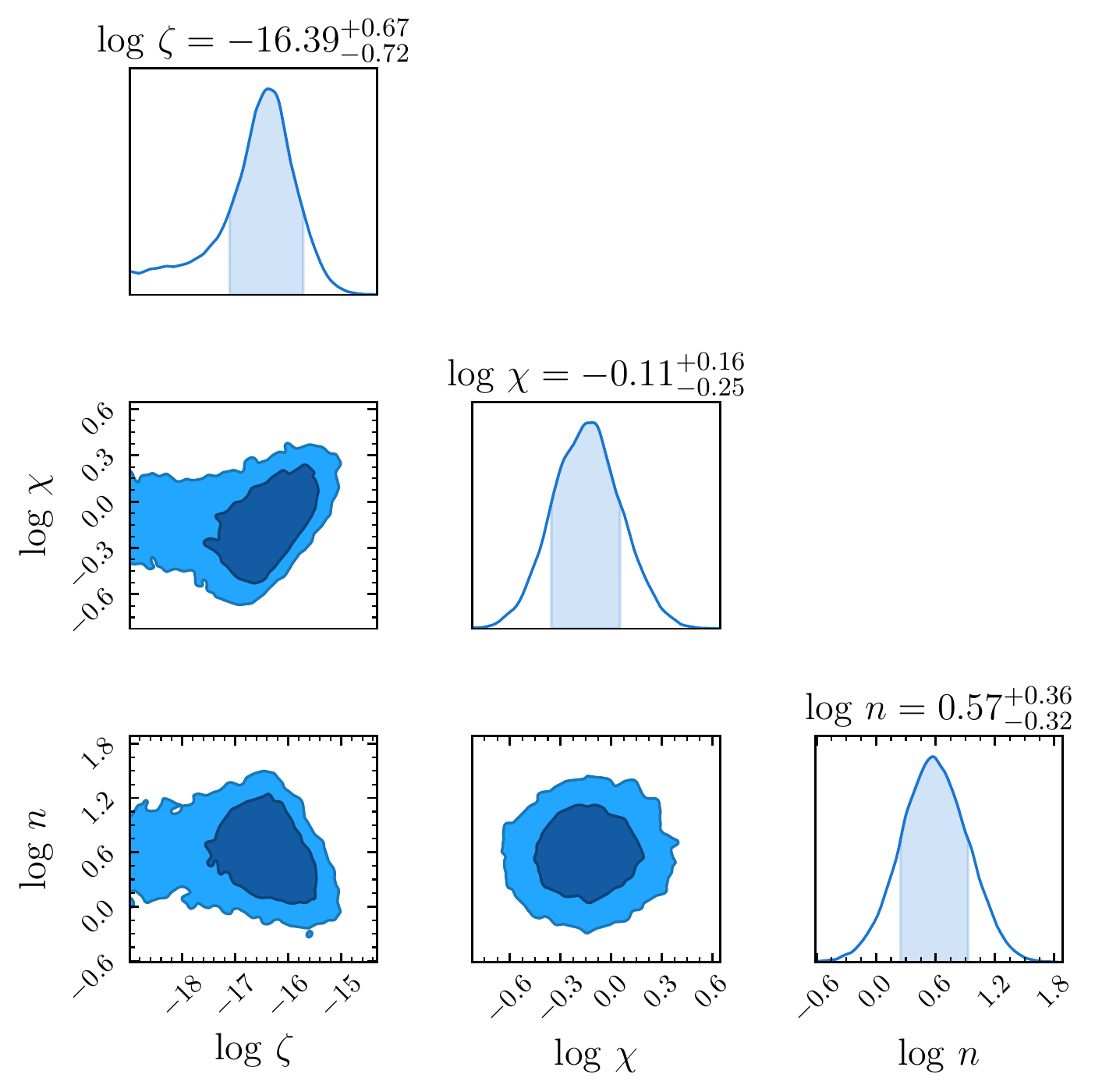}}
{DLA at $z=2.0546$ towards J\,2340$-$0053 (comp 4)}
\end{minipage}
\hfill
\begin{minipage}[h]{0.47\linewidth}
\center{\includegraphics[width=1\linewidth]{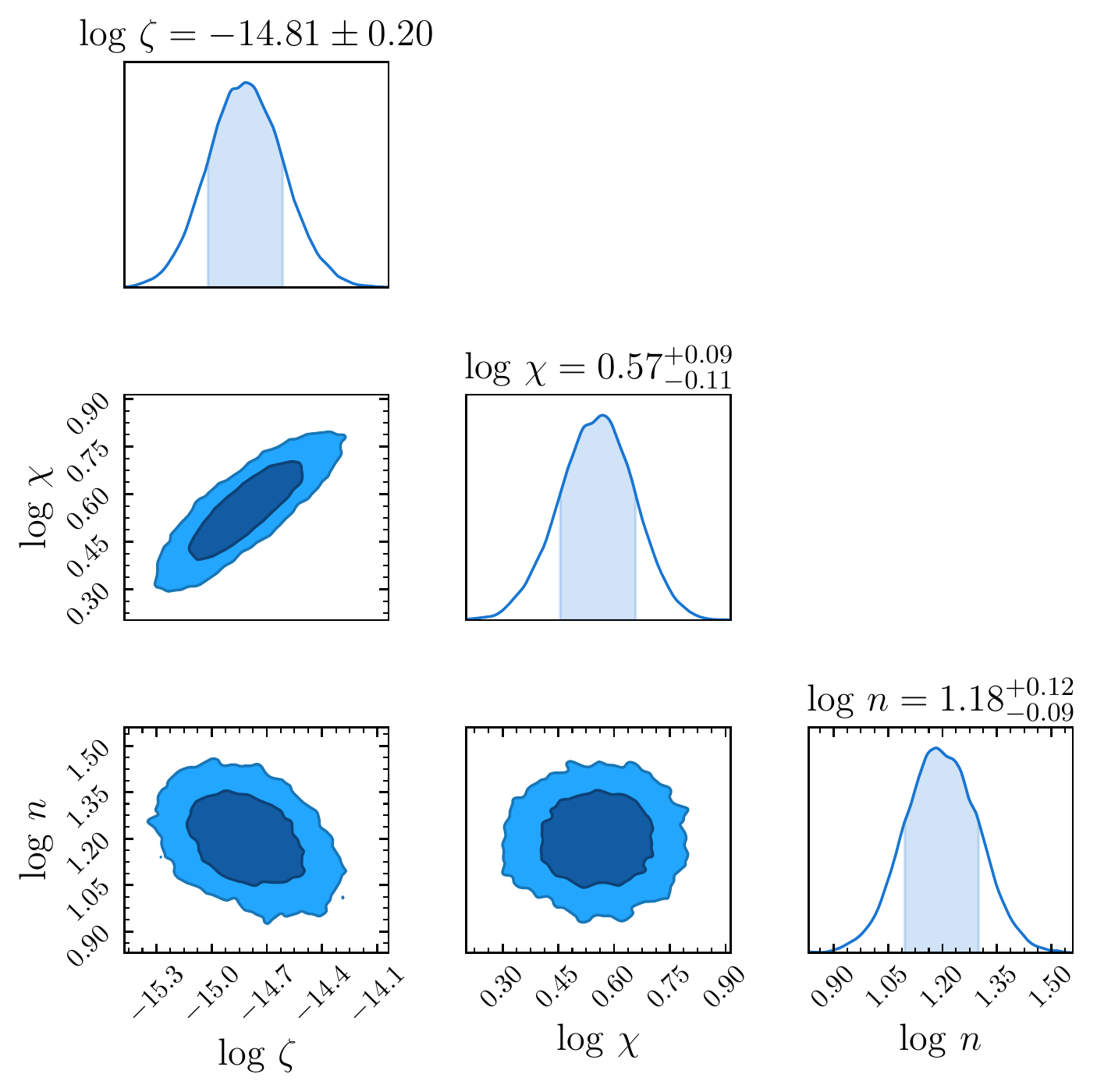}}  
{DLA at $z=2.0547$ towards J\,2340$-$0053 (comp 5)}
\end{minipage}

\caption{The marginalized posterior probability functions for CRIR, UV field strength, and number density (continued).
}
\label{fig: MCMC_results4}
\end{figure*}
\end{center}

\begin{center}
\begin{figure*}
\begin{minipage}[h]{0.47\linewidth}
\center{\includegraphics[width=1\linewidth]{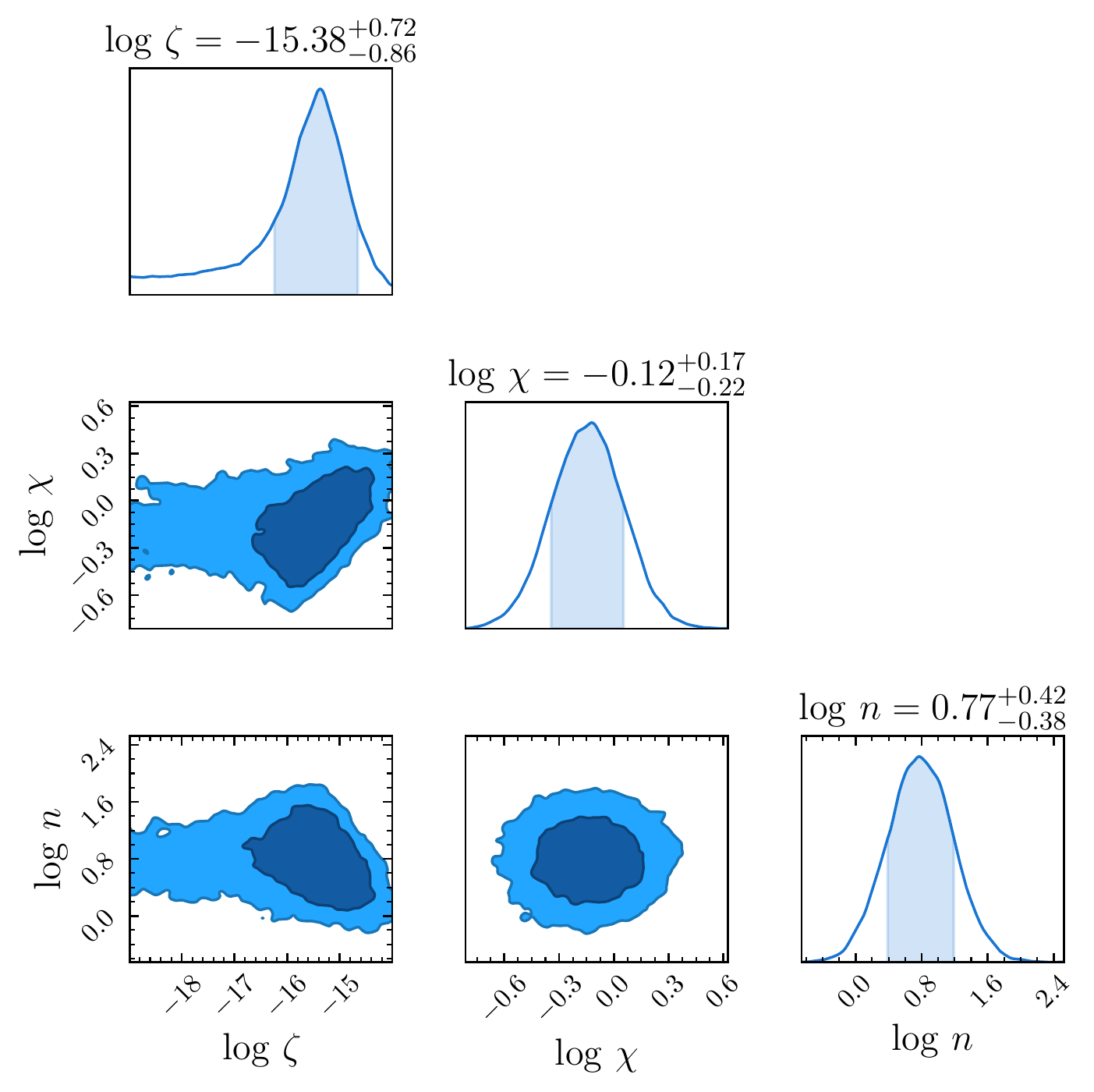}} 
{DLA at $z=2.0551$ towards J\,2340$-$0053 (comp 7)}
\end{minipage}
\hfill
\begin{minipage}[h]{0.47\linewidth}
\center{\includegraphics[width=1\linewidth]{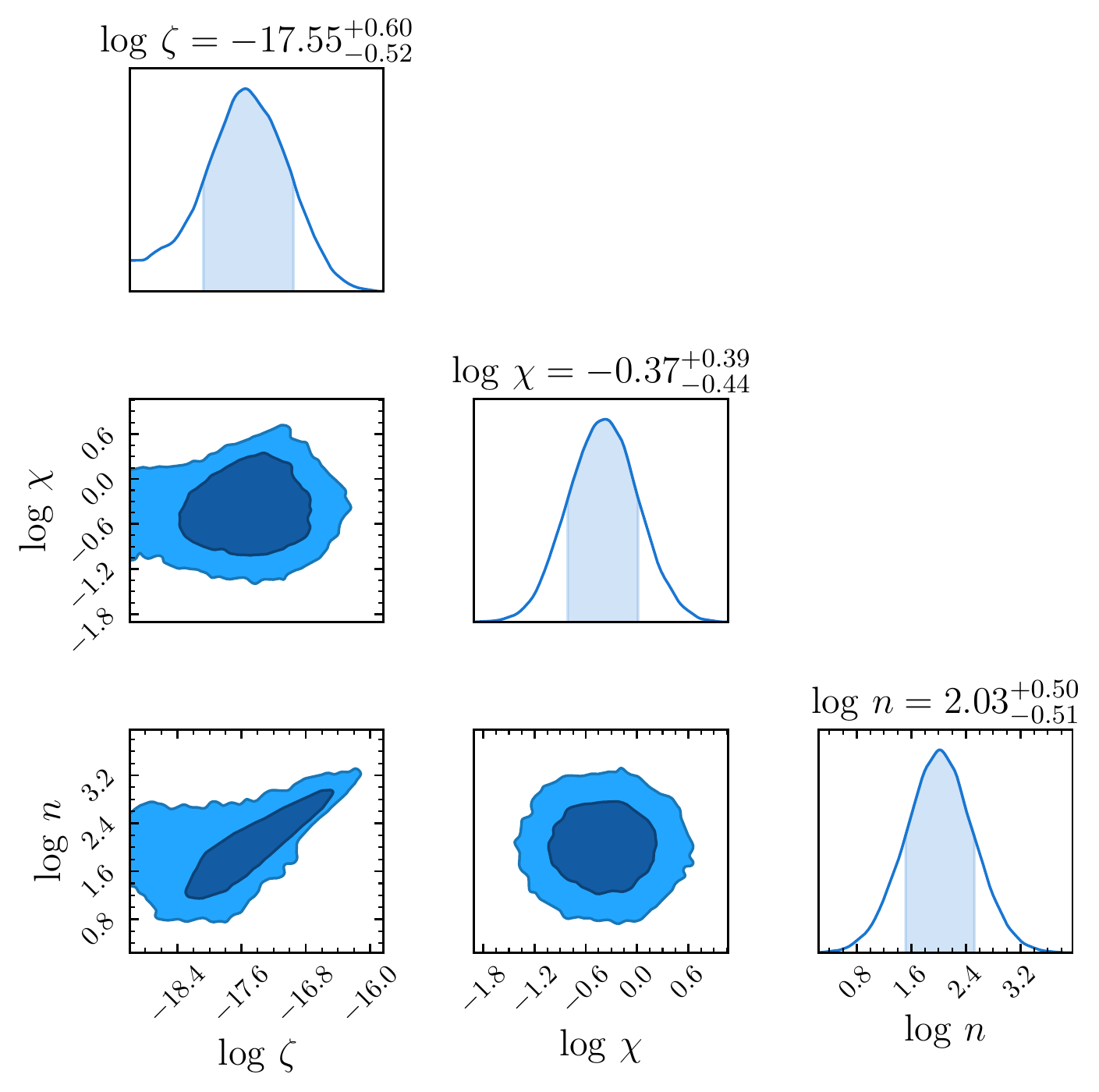}} 
{DLA at $z=2.588$ towards J\,2347$-$0051}
\end{minipage}

\caption{The marginalized posterior probability functions for CRIR, UV field strength, and number density (continued).
}
\label{fig: MCMC_results5}
\end{figure*}
\end{center}

\label{lastpage}

\end{document}